\documentclass{emulateapj}

\pdfoutput=1

\usepackage{graphicx}
\usepackage[percent]{overpic}
\usepackage{multirow}
\usepackage{color}
\usepackage{rotating}
\usepackage{amsmath}
\usepackage{mathtools}
\usepackage{physics}
\usepackage{tablefootnote}

\def\ltsima{$\; \buildrel < \over \sim \;$}
\def\simlt{\lower.5ex\hbox{\ltsima}}
\def\gtsima{$\; \buildrel > \over \sim \;$}
\def\simgt{\lower.5ex\hbox{\gtsima}}
\def\kms{{\rm\,km\,s^{-1}}}

\def\masyr{{\rm\,mas/yr}}
\def\kpc{{\rm\,kpc}}

\def\pc{{\rm\,pc}}

\def\magn{{\rm\,mag}}
\newcommand{\fmmm}[1]{\mbox{$#1$}}

\newcommand{\scnp}{\mbox{\fmmm{''}}}


\def\deg{^\circ}
\def\degg{\hbox{$\null^\circ$\hskip-3pt .}}

\def\s{\ifmmode \widetilde \else \~\fi}
\def\={\overline}

\def\spose#1{\hbox to 0pt{#1\hss}}

\def\lta{\mathrel{\spose{\lower 3pt\hbox{$\mathchar"218$}}
     \raise 2.0pt\hbox{$\mathchar"13C$}}}
\def\gta{\mathrel{\spose{\lower 3pt\hbox{$\mathchar"218$}}
     \raise 2.0pt\hbox{$\mathchar"13E$}}}
\def\Dt{\spose{\raise 1.5ex\hbox{\hskip3pt$\mathchar"201$}}}    
\def\dt{\spose{\raise 1.0ex\hbox{\hskip2pt$\mathchar"201$}}}    

\def\dotsfill{\leaders\hbox to 1em{\hss.\hss}\hfill}

\def\Gyr{{\rm\,Gyr}}

\def\ltsima{$\; \buildrel < \over \sim \;$}
\def\gtsima{$\; \buildrel > \over \sim \;$}
\def\lsim{\lower.5ex\hbox{\ltsima}}
\def\gsim{\lower.5ex\hbox{\gtsima}}

\slugcomment{Accepted by The Astrophysical Journal}

\shorttitle{Stellar Streams of the Inner Galaxy}
\shortauthors{Ibata, Malhan \& Martin}

\begin{document}

\title{The Streams of the Gaping Abyss:\\
A population of entangled stellar streams surrounding the Inner Galaxy}

\author{Rodrigo A. Ibata\altaffilmark{1}}
\author{Khyati Malhan\altaffilmark{1,2}}
\author{Nicolas F. Martin\altaffilmark{1,3}}

\altaffiltext{1}{Observatoire Astronomique, Universit\'e de Strasbourg, CNRS, 11, rue de l'Universit\'e, F-67000 Strasbourg, France; rodrigo.ibata@astro.unistra.fr}
\altaffiltext{2}{The Oskar Klein Centre for Cosmoparticle Physics, Department of Physics, Stockholm University, AlbaNova, 10691 Stockholm, Sweden}
\altaffiltext{3}{Max-Planck-Institut f\"ur Astronomie, K\"onigstuhl 17, D-69117 Heidelberg, Germany}

\begin{abstract}
We present the discovery of a large population of stellar streams that surround the inner Galaxy, found in the {\it Gaia} DR2 catalog using the new \texttt{STREAMFINDER} algorithm. Here we focus on the properties of eight new high-significance structures found at Heliocentric distances between $1$ and $10\kpc$ and at Galactic latitudes $|b|>20\deg$, named Slidr, Sylgr, Ylgr, Fimbulthul, Sv\"ol, Fj\"orm, Gj\"oll and Leiptr. Spectroscopic measurements of seven of the streams confirm the detections, which are based on Gaia astrometry and photometry alone, and show that these streams are predominantly metal-poor. The sample possesses diverse orbital properties, although most of the streams appear to be debris of inner-halo globular clusters. Many more candidate streams are visible in our maps, but require follow-up spectroscopy to confirm their nature. We also explain in detail the workings of the algorithm, and gauge the incidence of false detections by running the algorithm on a smooth model of the {\it Gaia} catalog.
\end{abstract}


\section{Introduction}
\label{sec:Introduction}

The advent of the second data release (DR2) of the {\it Gaia} mission \citep{2018A&A...616A...2L,2018A&A...616A...1G} has opened up an new panoramic window onto our Galaxy. The detailed astrometric measurements now show the motions of over a billion stars in our immediate environment, allowing us to deduce the kinematical structure of our host galaxy \citep{2018A&A...616A..12G}. One of the main aims of this endeavor is to probe the formation history of the Milky Way, by identifying the structures that formed in situ and those that were incorporated into it during the many galactic merging events that are expected to have contributed to its present ensemble.

One class of remnant structures that are ``fossils'' of the Galactic formation process are the so-called stellar streams, which are the remains of satellites that have been disrupted by the tidal forces of their host galaxy. They provide a very exciting new avenue to assess the distribution of the dark matter, by allowing us to directly measure accelerations. The basic reason for this is that a low-mass stream follows closely its progenitor's orbit, so a stream locus approximates an orbit in the Galaxy. Thus mapping the line of sight velocity and proper motion gradients along a stream allows us to uncover the acceleration that the stream is subject to. With several streams on different orbits it should be possible to build up a reliable three-dimensional map of the acceleration field throughout the region of the Galaxy traversed by such streams. These ideas can be extended to higher mass streams, by accounting for self-gravity and the consequent offset between the stream and the orbit of the progenitor \citep{2011MNRAS.417..198V}.

Streams can also reveal the small-scale granularity of the dark matter \citep{2002MNRAS.332..915I,2002ApJ...570..656J}, which is an issue of fundamental importance since one of the central predictions of $\Lambda$CDM cosmology is the existence of thousands of dark matter sub-halos in Milky Way sized galaxies \citep{Klypin:1999ej,Moore:1999ja}. The dark matter substructures heat the tidal tail, broadening the stream and also increasing the dispersion of the stream in the space of the integrals of motion. Gaps in the stream may also be induced by the close passage of these dark perturbers \citep{2012ApJ...748...20C,2016MNRAS.463..102E,2018ApJ...863L..20P}. 

The ideal tidal streams to search for dynamical heating and gaps are those that are old (and hence long), since these have the greatest cross-section for impacts from $\Lambda$CDM substructures and have existed for long enough in the potential to feel many close passages. Indeed, finding a single unheated long (i.e., old) low-mass halo stream would provide a severe challenge to $\Lambda$CDM, as it would provide incontrovertible proof that dark matter is smooth on small (sub-galactic) scales. Alternatively, obtaining positive proof of the heating action of dark matter lumps would require one either to find streams that do not stray into the Galactic disk where they may be heated by interactions with baryonic substructures \citep{2016MNRAS.463L..17A}, or to perform a very careful modelling of the stream population as an ensemble, and include the effects of non-axisymmetric perturbations such as that due to the Galactic bar \citep{2016MNRAS.460..497H,2016ApJ...824..104P,2017NatAs...1..633P}.

The necessary first step towards achieving these important goals of studying the dark matter as well as the structure, dynamics, composition, and assembly history of our Galaxy is to find samples of dynamically-cold stellar streams. While large sky photometric sky surveys like the Sloan Digital Sky Survey, Pan-STARRS1 and the Dark Energy Survey have been successfully trawled for these structures \citep{2006ApJ...643L..17G,2009ApJ...693.1118G,2016MNRAS.463.1759B,2018ApJ...862..114S}, there is a clear advantage to now try to search for stellar streams in the {\it Gaia} survey given its exquisite kinematic measurements. However, the streams of interest will be mixed in a complex fashion with the Milky Way's field populations over vast swathes of sky, so disentangling them from each other and from the Galaxy requires careful analysis. It was in view of these opportunities with {\it Gaia} (and other new surveys) that we developed an algorithm, which we called the {\tt STREAMFINDER}, to search for stream-like structures (\citealt{2018MNRAS.477.4063M}, hereafter Paper~I). 

In a previous contribution (\citealt{2018MNRAS.481.3442M}, hereafter Paper~II), we applied an updated version of the algorithm of Paper~I, which will be presented below, to undertake a first search for tidal streams in the {\it Gaia} DR2 catalog. Searching the high-latitude sky at $|b|>30\deg$, and at Heliocentric distances $d>5\kpc$, we reported the detection of five new streams (named {\it Gaia} 1 -- 5).

The aim of the present paper is to continue this search, extending the sky to $|b|>20\deg$ and to distances $d>0.5\kpc$. An early result of the present effort was the detection of the $75\deg$-long ``Phlegethon'' stream, which was presented in \citet[][hereafter Paper~III]{2018ApJ...865...85I}.

The {\it Gaia} DR2 survey is the principal dataset used here. We extinction-corrected this survey using the \citet{2011ApJ...737..103S} corrections to the \citet{Schlegel:1998fw} extinction maps, assuming the extinction ratios $A_G/A_V=0.85926$, $A_{G_{\rm BP}}/A_V=1.06794$ and $A_{G_{\rm RP}}/A_V=0.65199$, as listed on the web interface to the PARSEC isochrones \citep{2012MNRAS.427..127B}. Henceforth all magnitudes and colors refer to these extinction-corrected values. We also excised all stars in the {\it Gaia} DR2 catalog that lie close to certain compact structures. These included objects within two tidal radii of the globular clusters listed in \citet{2010arXiv1012.3224H}, objects within 7 half-light radii of the Galactic satellites listed in \citet{2012AJ....144....4M}, as well as within a $3\deg$ radius of M31 and $1\deg$ radius of M33. We also remove stars in appropriately-sized regions (typically $\sim 0.5\deg$) around the open clusters NGC~188, Berkeley~8, NGC~2204, NGC~2243, NGC~2266, Melotte~66, NGC~2420, NGC~2682 and NGC~6939. This pruned catalog is used both in the search for streams below, as well as in the construction of the contamination model.

The layout of the paper is as follows. In Section~\ref{sec:STREAMFINDER_improvements} we explain the new algorithm, paying particular attention to how it differs from the procedure described in Paper~I. Section~\ref{sec:Comparison} describes the artificial data that were processed to establish the incidence of false positive detections. The maps of the detected streams are presented in Section~\ref{sec:Results}, and then we describe each new high-significance stream individually in Section~\ref{sec:Streams}. In Section~\ref{sec:Discussion}, we discuss these results and draw the conclusions of our study.

\section{{\tt STREAMFINDER} improvements}
\label{sec:STREAMFINDER_improvements}

Although the original {\tt STREAMFINDER} concept of Paper~I worked well at high Galactic latitude, in regions of relatively uniform contamination, its shortcomings became evident when probing closer to the Galactic plane. There, the strong gradient in the background produced an unacceptably large population of false positives. We decided therefore to redesign the algorithm to undertake a full likelihood analysis, as described below.

Conceptually, we proceed in a similar manner to Paper~I, by considering, in turn, each star $i$ in the survey down to a chosen magnitude limit. The question we aim to answer is: if star $i$ forms part of a stream, does the maximum-likelihood stream solution, centered on $i$, imply that the putative stream is significant? To this end, we search for the most likely stream model $\mathcal{P}_{\rm stream}(\theta)$ and the corresponding stream fraction $\eta$, that maximizes:
\begin{equation}
{\ln} \mathcal{L} = \rm \sum_{\rm{data}} {\ln} \, [\eta \mathcal{P}_{\rm stream}(\theta) + (1 - \eta) \, \mathcal{P}_{\rm cont} ] \, ,
\label{eqn:likelihood}
\end{equation}
where $\theta$ are the stream fitting parameters, and the probability density function $\mathcal{P}_{\rm cont}$ is a model of the ``contamination'' (sometimes called a ``background'') from non-stream stars. By comparing to the likelihood of the no-stream case (when $\eta=0$), Equation~\ref{eqn:likelihood} easily allows us to measure the stream detection significance. In the analysis below, the vast majority of the {\it Gaia} survey stars are found to have $\eta=0$ (i.e., the most likely stream solution is to have no stream at all).

\subsection{Stream model}
\label{Sec:stream_model}

To build the stream model $\mathcal{P}_{\rm stream}(\theta)$, we use a simple leapfrog scheme to integrate an orbit both forwards and backwards from the star under consideration, under the influence of an acceleration field provided by the realistic Galactic density distribution of \citet{Dehnen:1998tk} (their model 1, which is axisymmetric and contains a bulge, disk, thick disk, interstellar medium, and halo components). The orbit defines the central line for a proposed stream. This line is projected into the space of the following six observed quantities in {\it Gaia} DR2: sky position $\alpha$, $\delta$, proper motions $\mu^*_\alpha (\equiv \mu_\alpha \cos(\delta))$, $\mu_\delta$, and color-magnitude position $G$, $G_{\rm BP}-G_{\rm RP}$, as described in Paper~I. To make this projection, we adopt the recent accurate determinations of the Galactocentric distance of $R_\odot=8.122\pm0.031\kpc$ \citep{2018A&A...615L..15G}, and of the circular velocity at the Solar neighborhood of $v_c(R_\odot)=229.0\pm0.2\kms$  \citep{Eilers:2018vh}. Given that $v_c(R_\odot)+V_\odot=255.2\pm5.1 \kms$ \citep{2014ApJ...783..130R}, we take the $V$-component of the peculiar velocity of the Sun to be $V_\odot=26.2\kms$ (which we keep fixed). The $U$ and $W$ components of the Sun's peculiar velocity are taken from \citet{2010MNRAS.403.1829S}. Henceforth, for convenience, we will drop the asterix superscript from $\mu^*_\alpha$.

With {\it Gaia} DR2 data we have access to very precise sky positions, and superb, but still noisy, proper motions. The line of sight velocities $v_{\rm los}$ of the survey stars are effectively all unknown. For distant halo stars, of which the majority are faint in {\it Gaia} (see \citealt{2017ApJ...848..129I}, their Figure~1), parallaxes are effectively uninformative, yet the distances can be constrained via the excellent {\it Gaia} photometry, given that plausible populations of stream stars should follow color-magnitude tracks and luminosity functions of plausible progenitors. For this purpose we choose the PARSEC single stellar population (SSP) models \citep{2012MNRAS.427..127B} of different metallicities and ages to serve as templates. Nevertheless, a given model may yield multiple distance solutions (as we discuss in Paper~I), while SSP models of different metallicity or age will also yield different distance solutions. 

The stream model is simply a smeared-out version of the orbit, blurred with Gaussian widths in each observed parameter in such a way as to display similar sky width, depth, velocity distribution and color-magnitude dispersion as known streams (or as the type of stream one wishes to detect). For the present work, we adopted a stream length of $L=20\deg$. For each star, we find the minimum angular distance from the stream, and calculate the corresponding perpendicular distance $s$, assuming that the star has the same line-of-sight distance $d$ as the orbit. We adopt a Gaussian stream model width of $w_s=100\pc$ (we use $w$ to denote model Gaussian dispersions and $\sigma$ to denote measurement uncertainties), and a velocity dispersion of $w_v = 3\kms$ in each kinematic dimension (which is converted into a model dispersion in proper motion $w_\mu$ given $d$).

The probability density model of the stream is then:
\begin{equation}
\mathcal{P}_{\rm stream}(\theta)  = \mathcal{P}_{\rm length} \times \mathcal{P}_{\rm width} \times \mathcal{P}_{\rm LF} \times \mathcal{P}_{\rm color} \times \mathcal{P}_\mu 
\end{equation}
where $\mathcal{P}_{\rm length}=1/L$ is the (uniform) probability of lying at the measured position along the stream, and $\mathcal{P}_{\rm width}= \mathcal{N}(\Delta s, w_s)$ is the Gaussian probability of lying at the measured distance $\Delta s$ perpendicular to the stream. $\mathcal{P}_{\rm LF}$ is the probability that the observed star has been drawn from the luminosity function of the adopted SSP model (placed at the stream distance $d$). $P_{\rm color} =  \mathcal{N}(\Delta (G_{\rm Bp} - G_{\rm Rp}), \sigma_{G_{\rm Bp} - G_{\rm Rp}})$ is the Gaussian probability that the measured color of the star is consistent with that expected of the SSP model (again at distance $d$) given the measured magnitude $G_0$ and the color uncertainty $\sigma_{G_{\rm Bp} - G_{\rm Rp}}$. Finally, 
\begin{align}
\begin{split}
\mathcal{P}_\mu & = {{1}\over{2\pi \sigma_{\mu_\alpha} \sigma_{\mu_\delta} \sqrt{1-\rho^2}}}  \times \\
& \exp(-{{1}\over{2(1-\rho^2)}} \Bigg[ {{\Delta^2_{\mu_\alpha}}\over{\sigma^2_{\mu_\alpha}}} + {{\Delta^2_{\mu_\delta}}\over{\sigma^2_{\mu_\delta}}}  
- {{2\rho \Delta_{\mu_\alpha} \Delta_{\mu_\delta}}\over{\sigma_{\mu_\alpha} \sigma_{\mu_\delta}   }}  \Bigg] )  \\
\end{split}
\end{align}
is the probability that the measured proper motions agree with those of the closest point on the orbit model (with measured offsets from the model of $\Delta_{\mu_\alpha}$ and $\Delta_{\mu_\delta}$). The Gaia proper motion uncertainties $\sigma_{\mu_\alpha}$ and $\sigma_{\mu_\delta}$ are thus taken into account along with their measured correlation $C \equiv {\tt pmra\_pmdec\_corr}$ (discussed in \citealt{2018A&A...616A...2L}), which is subsumed into the parameter
\begin{equation}
\rho = {{ C \, \sigma_{\mu_\alpha} \sigma_{\mu_\delta}}\over {\sqrt{ (\sigma^2_{\mu_\alpha} + w^2_\mu) (\sigma^2_{\mu_\delta} + w_\mu^2)}}} \, .
\end{equation}
This last equation can be derived by considering the convolution of the two-dimensional covariance matrix with an isotropic two-dimensional Gaussian of dispersion $w_\mu$. This allows us to incorporate the effect of the model dispersion. While we have included the measured correlation between $\mu_\alpha$ and $\mu_\delta$, we ignore the other elements of the ($5\times5$) astrometric covariance matrix, as they are much less important: correlations with sky position will be negligible given the size of $w_s$, and we can completely ignore the correlations of proper motions with parallax, as the latter is not used in the likelihood calculation.

\begin{figure}
\begin{center}
\includegraphics[angle=0, viewport= 135 40 692 317, clip, width=\hsize]{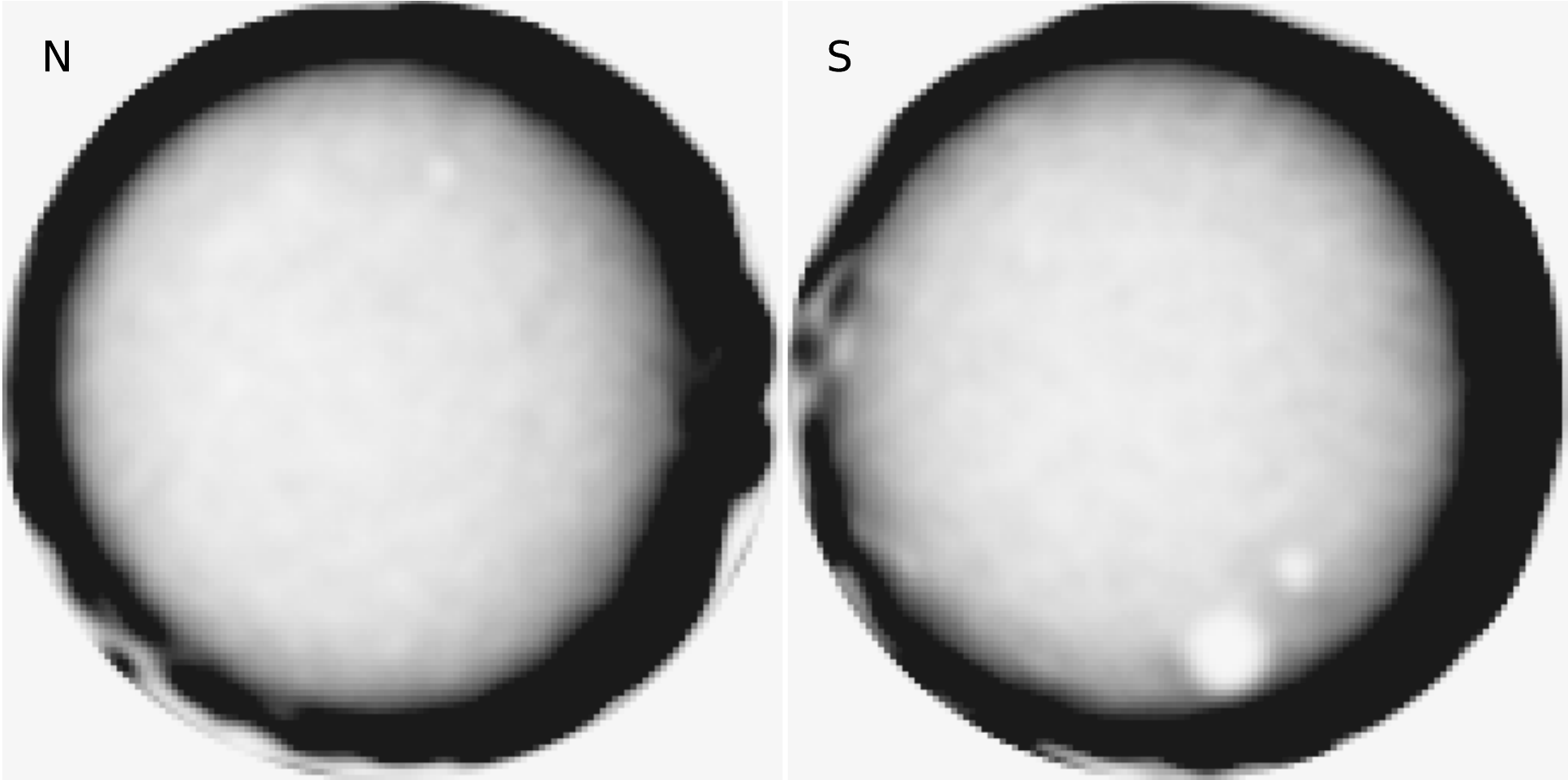}
\end{center}
\caption{Example of a slice through the ``fine'' spatial probability density function $P_{\rm fine}(\alpha,\delta,G,G_{\rm BP}-G_{\rm RP})$. The color-magnitude slice is centered at $G_{\rm BP}-G_{\rm RP}=1.1$, $G=18.1$, with bin width of $0.05 \, {\rm mag} \times 0.10 \, {\rm mag}$, respectively. The panels show a zenithal equal area projection of the counts (in linear scale) centered on the north Galactic pole (left) and south Galactic pole (right). These probability density distributions are constructed to follow the Galactic populations in a smooth manner.}
\label{fig:fine}
\end{figure}

\subsection{Contamination model}

To be useable in Eqn.~\ref{eqn:likelihood}, the contamination model $\mathcal{P}_{\rm cont}$ also needs to be a probability density function of the same six observables $(\alpha,\delta,\mu_\alpha,\mu_\delta,G,G_{\rm BP}-G_{\rm RP})$ as $\mathcal{P}_{\rm stream}$. By ``contamination'' what we mean here, of course, are the components of the Milky Way that are not stream-like. We could have used a Galactic synthesis model, such as the {\it Gaia} Universe Model Snapshot (GUMS; \citealt{2012A&A...543A.100R}), for this purpose, but we decided instead to build an empirical contamination model that we hoped would better trace the complexities of our Galaxy.

We smoothed-out spatially the {\it Gaia} catalogue by re-drawing each survey star 1000 times from a two-dimensional Gaussian in tangent-point coordinates using a dispersion of $2\deg$. The resulting catalogue was then binned into a 4-dimensional grid in position and color-magnitude. We adopted a spatial binning scheme in zenithal equal area (ZEA) Galactic polar projection (the projection used in the maps in Paper~II) with a bin size of $1.4\deg\times1.4\deg$. The color-magnitude intervals were chosen to be $0.05 \, {\rm mag} \times 0.10 \, {\rm mag}$. This binned array allows us to construct a probability density function in position and color-magnitude; since this will provide the fine spatial sampling of the contaminating population, let us call this $P_{\rm fine}(\alpha,\delta,G,G_{\rm BP}-G_{\rm RP})$. We present an example of a slice through this probability density function in Figure~\ref{fig:fine}.

\begin{figure}
\begin{center}
\includegraphics[angle=0, viewport= 5 5 400 400, clip, width=\hsize]{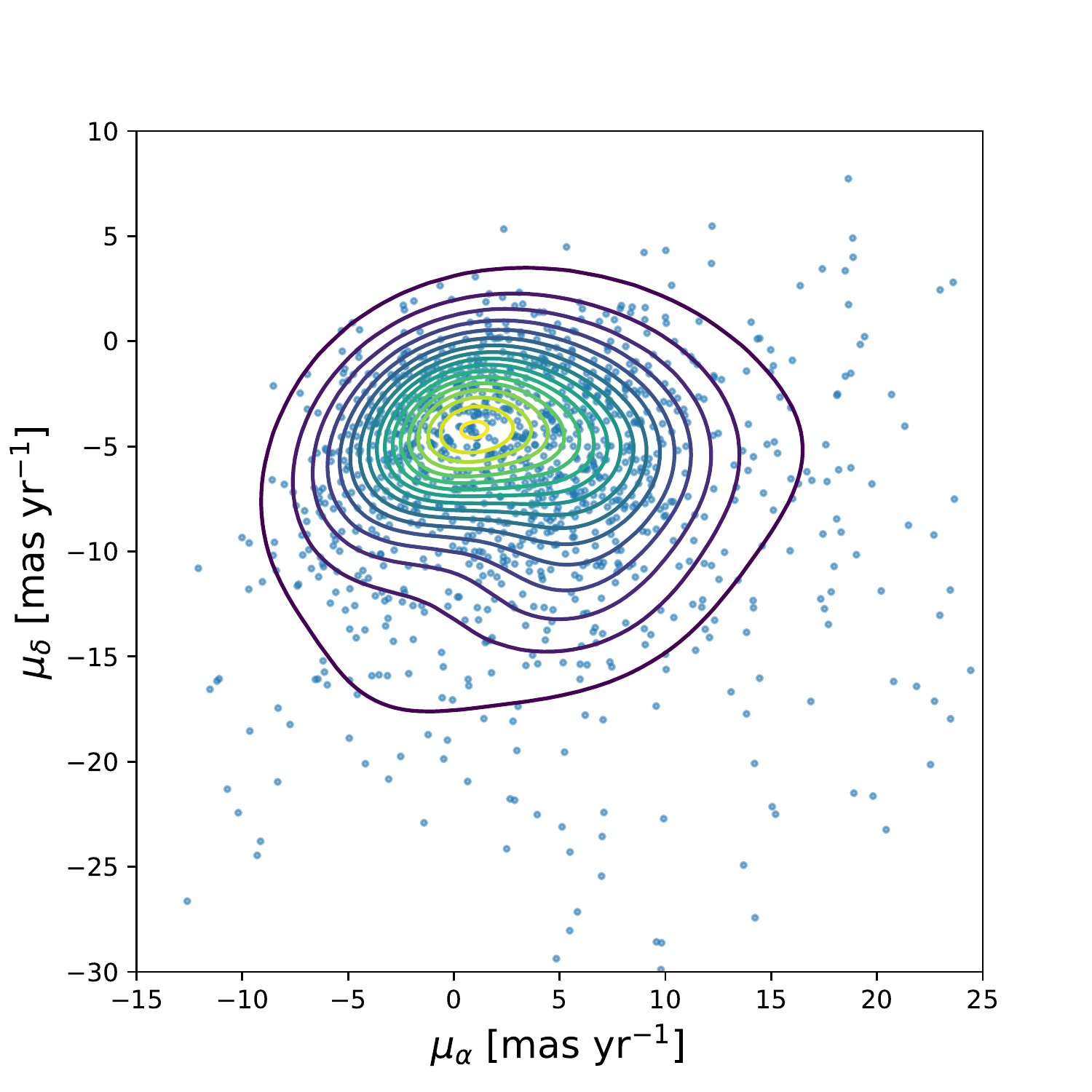}
\end{center}
\caption{Example of a slice through the Gaussian Mixture Model derived for a large $5.6\deg\times5.6\deg$ pixel centered at $\ell=105.2\deg$, $b=-60.8\deg$. The contours show the GMM model summed over the color-magnitude range $0.7<G_{\rm BP}-G_{\rm RP}<0.9$, $16<G<17$, and the dots show the data over the same range.}
\label{fig:GMM_mu}
\end{figure}

However, what we need to model are the correlations between color-magnitude and proper motion as a function of position on the sky. To this end we employed the Gaussian Mixture Model (GMM) decomposition implemented in the {\tt Armadillo} software package \citep{Sanderson:2016eq}. The GMM representation needs to have sufficient flexibility to reproduce the complex color-magnitude behavior of Galactic populations, while also following their proper motion behavior. By experimenting with the GMM software, we found that we could obtain a suitably smooth fit in the four dimensions of color-magnitude and proper motion with 100 GMM components, including cross-terms. To mitigate against varying depth due mainly to extinction, we limited the GMM fits to $G=20\magn$.

\begin{figure}
\begin{center}
\includegraphics[angle=0, viewport= 5 5 400 395, clip, width=\hsize]{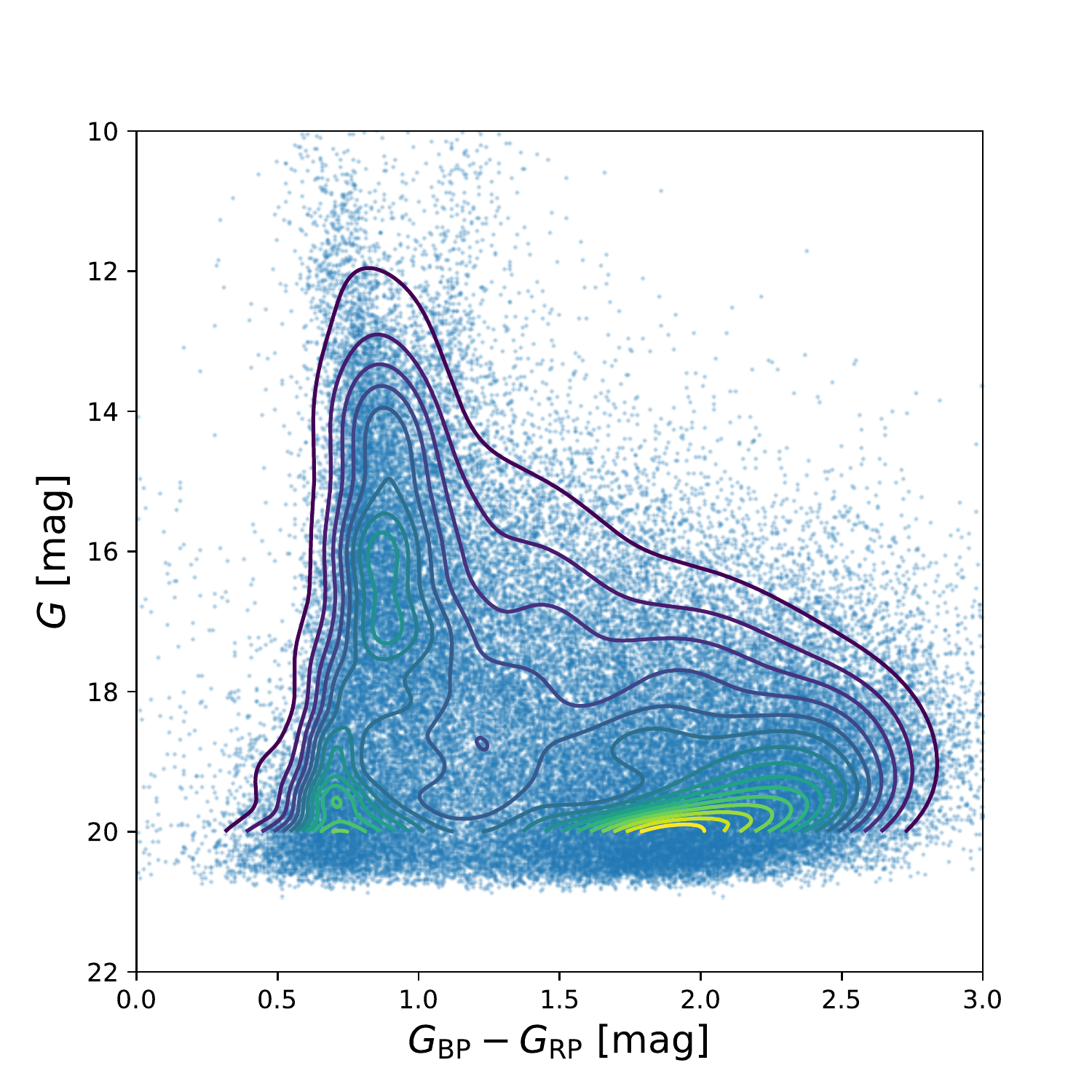}
\end{center}
\caption{Example of the color-magnitude behaviour of the Gaussian Mixture Model. We show the GMM model (contours) for the same sky region as shown previously in Figure~\ref{fig:GMM_mu}, marginalized over proper motion, along with the data (blue points) that were used to build it. The model is a 4-dimensional function built out of 100 Gaussian components, yet manages to represent the data in an apparently smooth way. In all of our hundreds of visual inspections of the GMM model over the sky a similar smooth behaviour was seen.}
\label{fig:GMM_CMD}
\end{figure}

In Figure~\ref{fig:GMM_mu} we show (for a representative high latitude field) the observed proper motion distribution extracted from a small color-magnitude box compared to the smooth GMM representation fit to the full color-magnitude sample. The model contours can be seen to provide a reasonable representation of the proper motion distribution. In Figure~\ref{fig:GMM_CMD} we have marginalized over proper motion to show the full color-magnitude distribution in the same spatial region as Figure~\ref{fig:GMM_mu}. Although it is obviously challenging to reproduce the color-magnitude behavior of stars in a typical Galactic field with Gaussian components, the software again achieves a visually-acceptable representation. 

For the GMM modelling we found it necessary to employ $4\times4$ larger spatial pixels of size $5.6\deg\times5.6\deg$, so that the spatial regions would contain sufficient stars ($>50000$ stars at the Galactic poles) for the GMM algorithm to yield smooth results. We label the probability density function derived in this way $P_{\rm GMM}(\mu_\alpha,\mu_\delta,G,G_{\rm BP}-G_{\rm RP} | \alpha,\delta)$, which is continuous and smooth in color-magnitude and proper motion by construction.

For each star, we cut through the GMM model at the value of its photometric $G,G_{\rm BP}-G_{\rm RP}$ measurements, to obtain the conditional probability $P_{\rm GMM}(\mu_\alpha,\mu_\delta | \alpha,\delta,G,G_{\rm BP}-G_{\rm RP})$. Clearly, this means that we have to normalize the function so that $\iint P_{\rm GMM} \mathrm{d} \mu_\alpha \, \mathrm{d} \mu_\delta = 1$.

The final contamination model is then:
\begin{align}
\begin{split}
\mathcal{P}_{\rm cont}(\alpha,\delta,\mu_\alpha,\mu_\delta,G,G_{\rm BP}-G_{\rm RP}) = & \\
P_{\rm fine}(\alpha,\delta,G,G_{\rm BP}-G_{\rm RP}) \times & \\
P_{\rm GMM}(\mu_\alpha,\mu_\delta | \alpha,\delta,G,G_{\rm BP}-G_{\rm RP}) \, \, . \\
\end{split}
\end{align}
This procedure may seem convoluted to the reader, but it has allowed us to build a Galaxy model that follows the density of the smooth populations with a relatively fine spatial and color-magnitude mesh, while simultaneously following the behavior of the populations in proper-motion. In contrast, our experiments of using the GMM method in six dimensions were not at all satisfactory: we could not both reproduce the complex structure in this six-dimensional space and also obtain a smooth model.

This contamination model passed hundreds of visual spot-checks, in which we compared the observations to the model at different locations on the sky in different two-dimensional slices through the $G,G_{\rm BP}-G_{\rm RP},\mu_\alpha,\mu_\delta$ parameter space. In these visual checks we ensured that the model was not under-fitting the data; we found occasionally that with $\sim 50$ GMM components or less, the model would only reproduce the disk dwarf sequence (the feature at $G_{\rm BP}-G_{\rm RP}\simgt 1.5$ in Figure~\ref{fig:GMM_CMD}). We also ensured that the model was not over-fitting the data, which would be visible as a lumpy model in color-magnitude space or in proper motion space. Despite these checks, we cannot of course be certain that our very complex contamination model provides a suitable representation of the Galaxy in this multi-dimensional parameter space. A particular worry is that artefacts in the contamination model could give rise to false positive detections. To address this issue in Section~\ref{sec:Comparison} below we will take recourse in a simulation of the Galaxy, and later show that the real {\it Gaia} data show stream-like structures unlike those found in the smooth simulation.

\subsection{Parameter fitting}

Within a given run of the {\tt STREAMFINDER} algorithm the free parameters are thus $\theta=[v_{\rm los}^{\rm model}, \mu_{\alpha}^{\rm model}, \mu_{\delta}^{\rm model}]$, and the stream fraction parameter $\eta$. The distance degeneracy is explored by re-running the algorithm with different plausible SSP models. We find the maximum value of ${\ln} \mathcal{L}$ by first sampling over $v_{\rm los}^{\rm model}$, $\mu_{\alpha}^{\rm model}$, and $\mu_{\delta}^{\rm model}$. Here, just as in Paper~I, we sample $v_{\rm los}^{\rm model}$ over the escape velocity ($\pm600\kms$) uniformly in steps of $20\kms$, and we also sample $\pm 1\sigma$ in each of the proper motion directions in three uniform steps. For a given proposed $\theta$, we hunt for the value of $\eta$ that yields the peak likelihood, using Brent's optimization method \citep{Press:1992vz}. The best trial solutions thus obtained are provided as starting input for a ``downhill simplex'' routine \citep{Press:1992vz} that refines $\theta$ and $\eta$ to maximize ${\ln} \mathcal{L}$, iterating until convergence is found to an absolute tolerance of $10^{-4}$ in ${\ln} \mathcal{L}$.

It can be appreciated from the above discussion that this is a computationally very intensive task, since at the location of each star we calculate many thousands of orbits, and calculate the stream likelihood by summing over all stars in the survey; the process being repeated until every star in the survey has been examined in this way. 

We decided not to use the {\it Gaia} parallaxes in the likelihood calculation since the algorithm was initially intended to detect distant halo streams for which we expected the parallax measurements to be too poor to be useful. In retrospect this was probably a mistake, which we may fix in future versions of the software. However it will likely be challenging to implement this upgrade, as it will require us to include another dimension in the contamination model, but one which is correlated with color and magnitude. In the meantime, although the parallaxes are not directly included in the likelihood calculation, we still use the measurements by simply filtering the {\tt STREAMFINDER} output to those solutions where the model distances are coherent with the measured {\it Gaia} parallaxes to within $2\sigma$. Although the heliocentric velocities of the streams are calculated internally within the algorithm, we do not yet include this additional observable dimension in the stream model; we intend to rectify this for the {\it Gaia} DR3 release which is expected to include the radial velocities of stars down to a limiting magnitude of $G \sim 16.5$.

\begin{figure}
\begin{center}
\includegraphics[angle=0, viewport= 131 38 760 325, clip, width=\hsize]{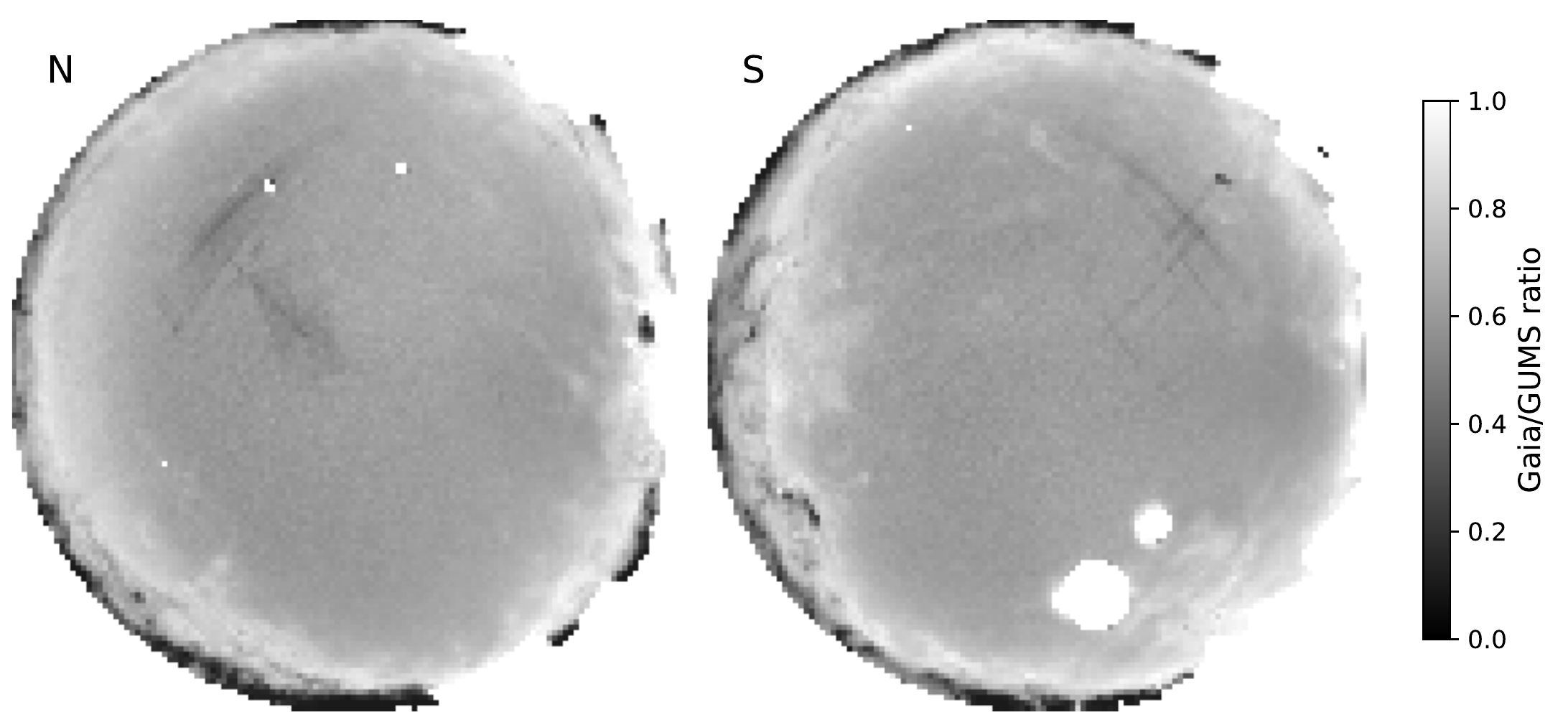}
\end{center}
\caption{Ratio of the number of stars in the {\it Gaia} catalog compared to the GUMS simulation. }
\label{fig:Gaia_to_GUMS}
\end{figure}

\begin{figure*}
\begin{center}
\vbox{
\hbox{
\includegraphics[angle=0, viewport= 55 23 600 589, clip, height=9cm]{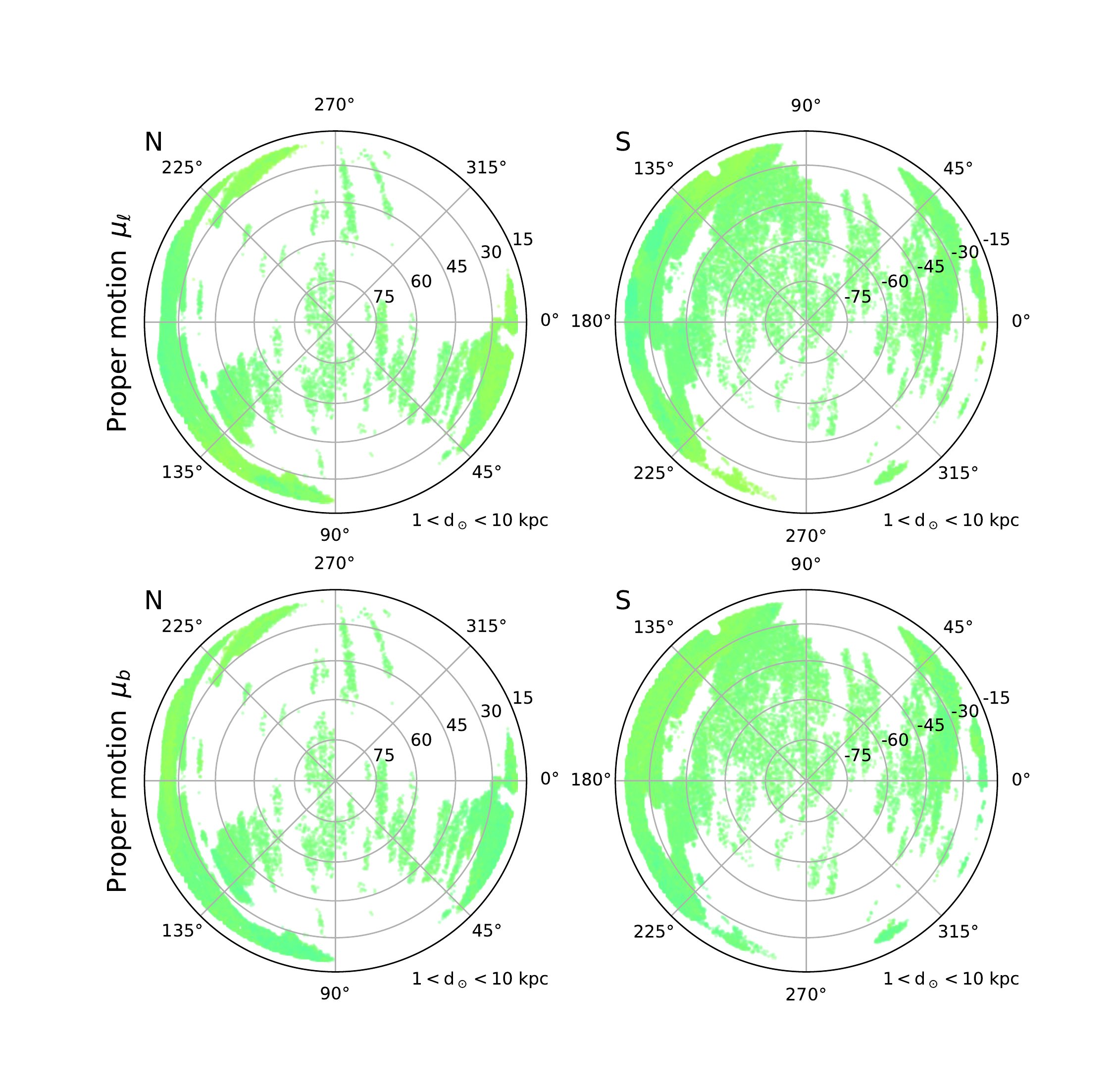}
\includegraphics[angle=0, viewport= 45 45 657 650, clip, height=9cm]{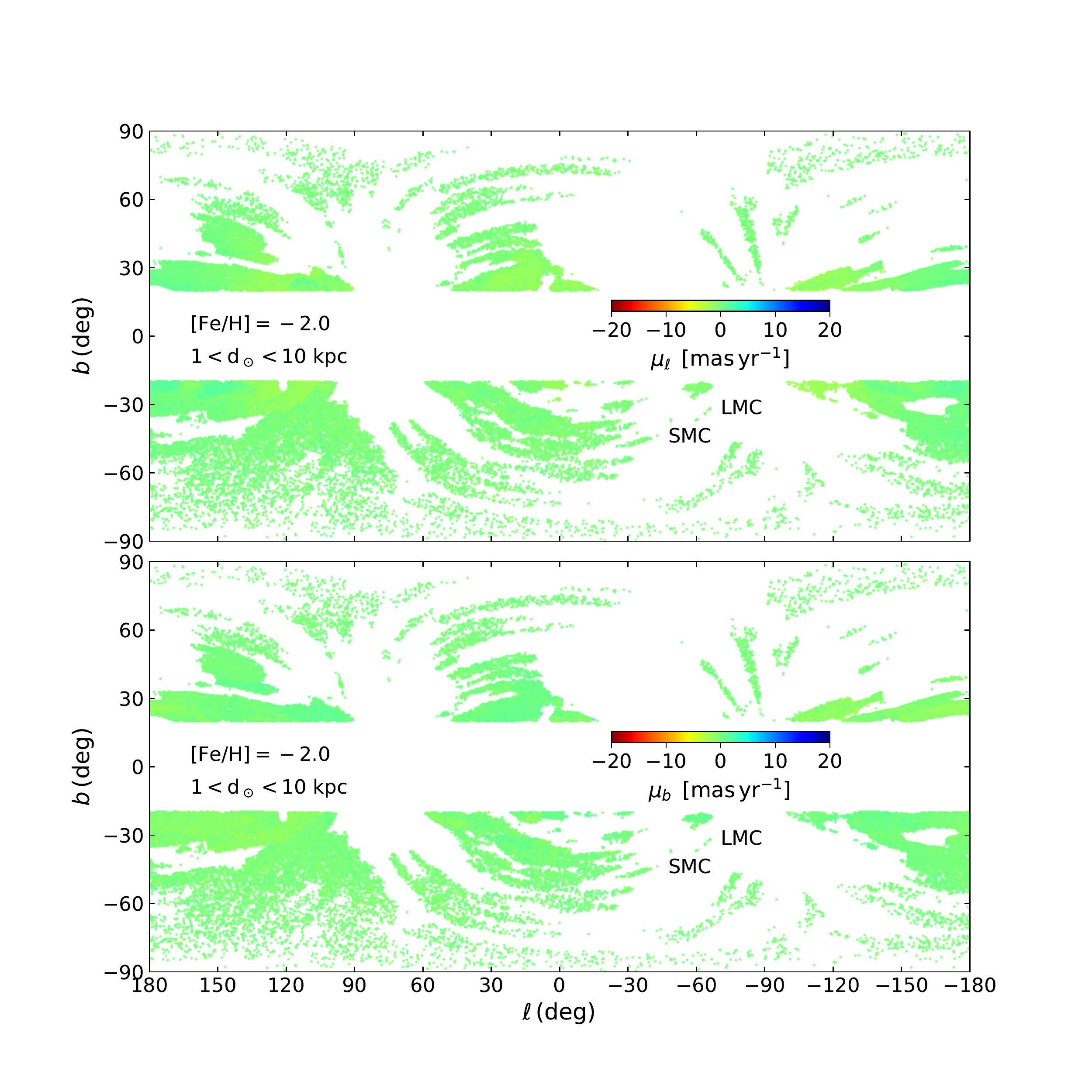}
}
\hbox{
\includegraphics[angle=0, viewport= 55 23 600 589, clip, height=9cm]{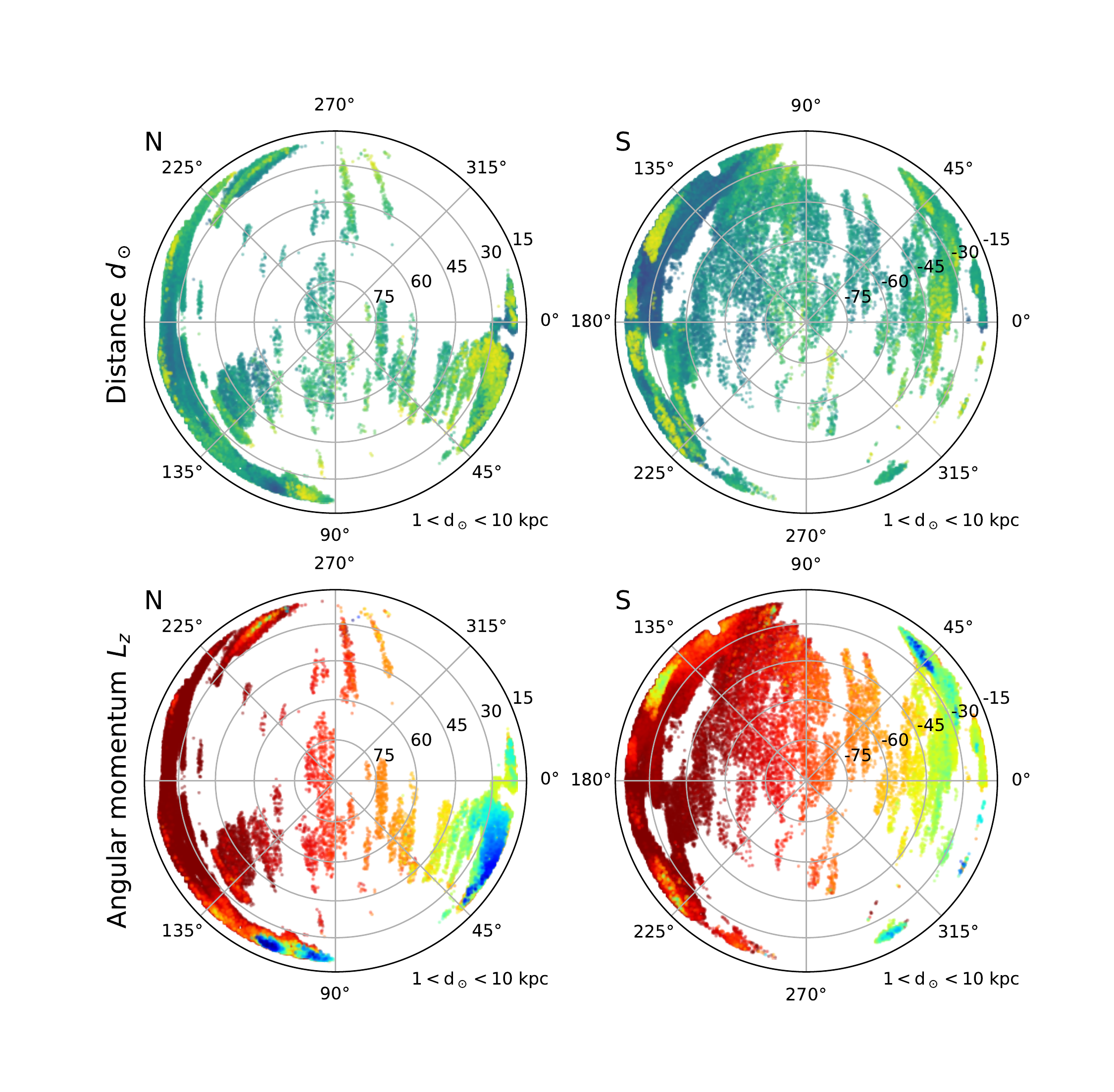}
\includegraphics[angle=0, viewport= 45 45 657 650, clip, height=9cm]{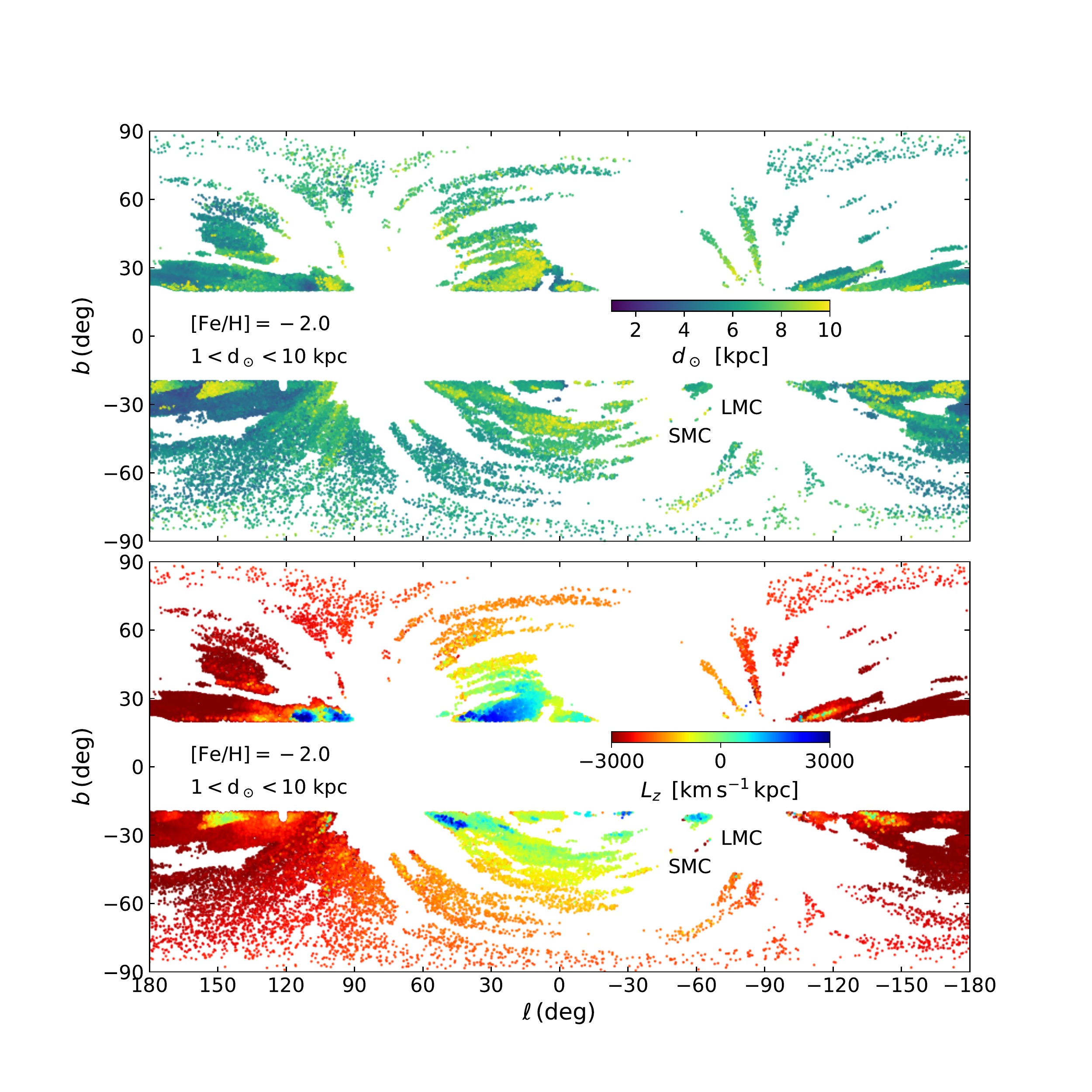}
}
}
\end{center}
\caption{Spatial distribution of stars that possess low proper motion ($\sqrt{\mu^2_\ell+\mu^2_b}<2\masyr$) and that are identified as stream-like with $>8\sigma$ confidence  in the distance interval $1<d_{\odot}<10\kpc$, assuming a metal-poor template with age=$12.5\Gyr$ and ${\rm [Fe/H]=-2.0}$. The first and second rows of panels show the distribution of proper motion in $\mu_\ell$ and $\mu_b$, respectively; the third row of panels shows the distance solution calculated by the algorithm with this stellar population template, and the bottom row of panels shows the $z$-component of angular momentum ($L_z$) of the best-fit orbit. The first and second columns of panels show Zenithal Equal Area projections centered on the north and south Galactic poles, respectively.} 
\label{fig:FeH_m2.0_fluff}
\end{figure*}

\begin{figure*}
\begin{center}
\vbox{
\hbox{
\includegraphics[angle=0, viewport= 55 23 600 589, clip,,height=9cm]{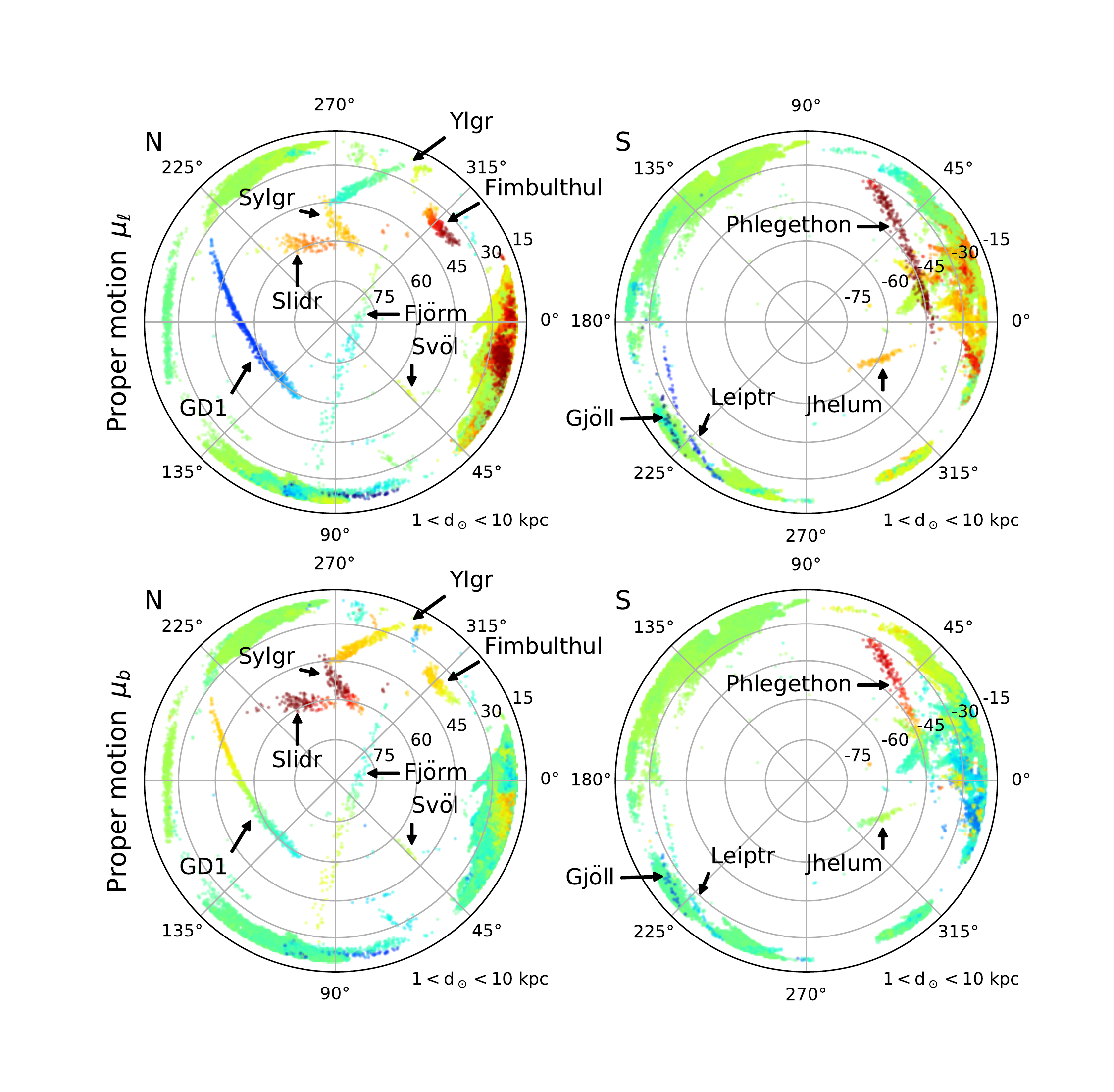}
\includegraphics[angle=0, viewport= 45 45 657 650, clip, height=9cm]{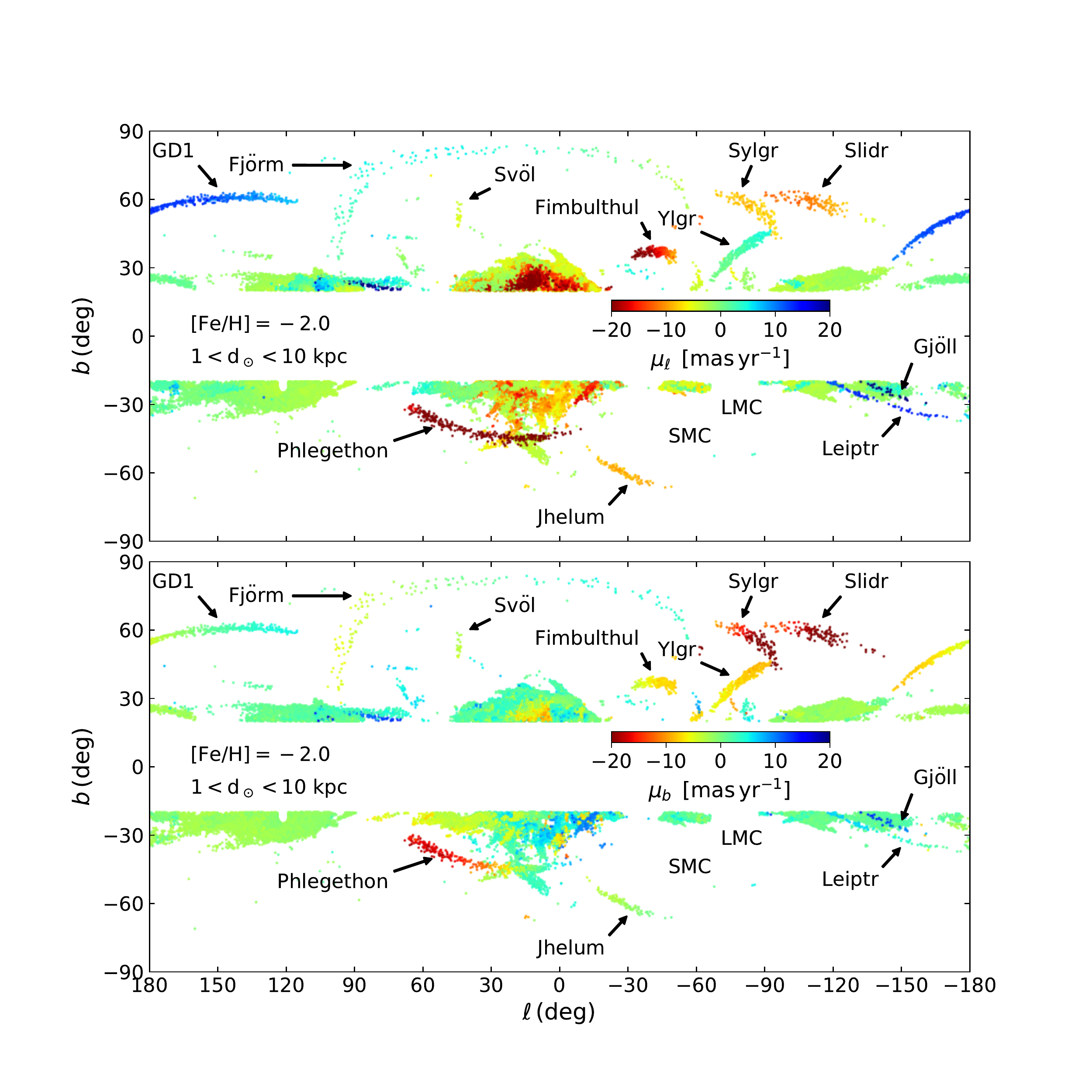}
}
\hbox{
\includegraphics[angle=0, viewport= 55 23 600 589, clip, height=9cm]{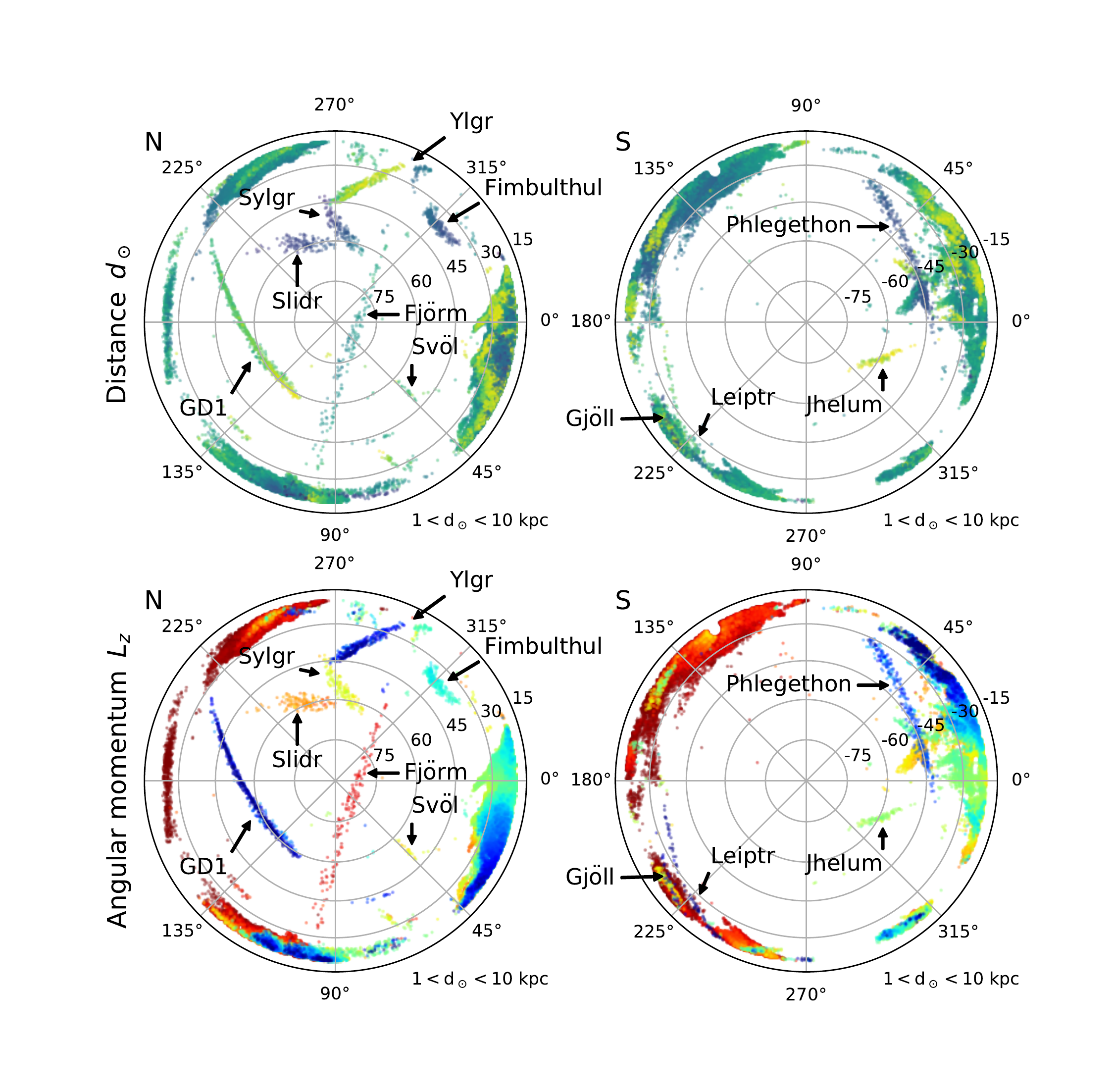}
\includegraphics[angle=0, viewport= 45 45 657 650, clip, height=9cm]{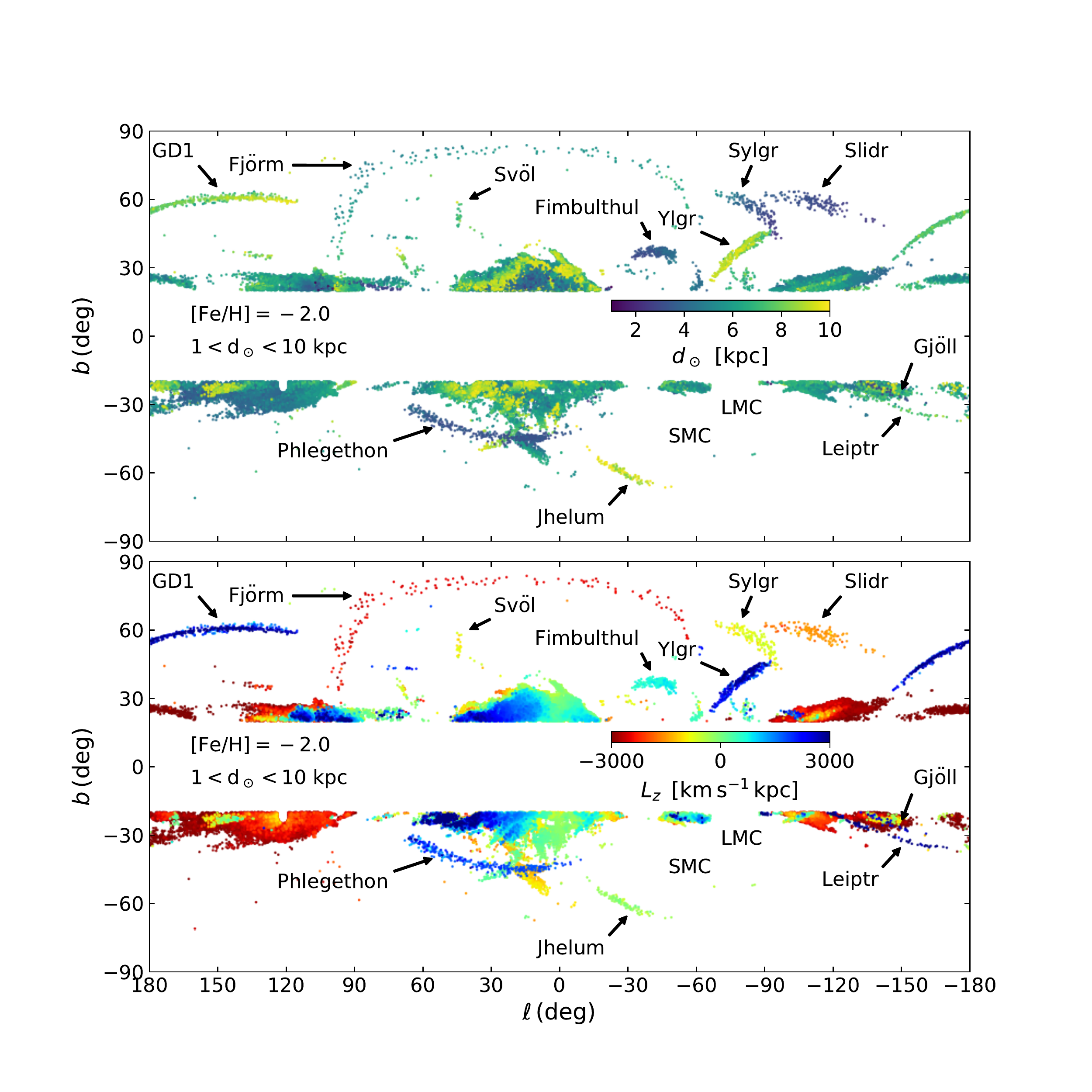}
}
}
\end{center}
\caption{As Figure~\ref{fig:FeH_m2.0_fluff}, but showing the stars with proper motion $\sqrt{\mu^2_\ell+\mu^2_b}>2\masyr$ that are identified as likely stream members using an $>8\sigma$ confidence threshold. In addition to the Phlegethon stream reported in Paper~III, we detect here the structures labelled ``Slidr'', ``Slygr'', ``Ylgr'',  ``Fimbulthul'', ``Sv\"ol'', ``Fj\"orm'', ``Gj\"oll'' and ``Leiptr''.}
\label{fig:FeH_m2.0}
\end{figure*}

\begin{figure*}
\begin{center}
\vbox{
\hbox{
\includegraphics[angle=0, viewport= 55 23 600 589, clip, height=9cm]{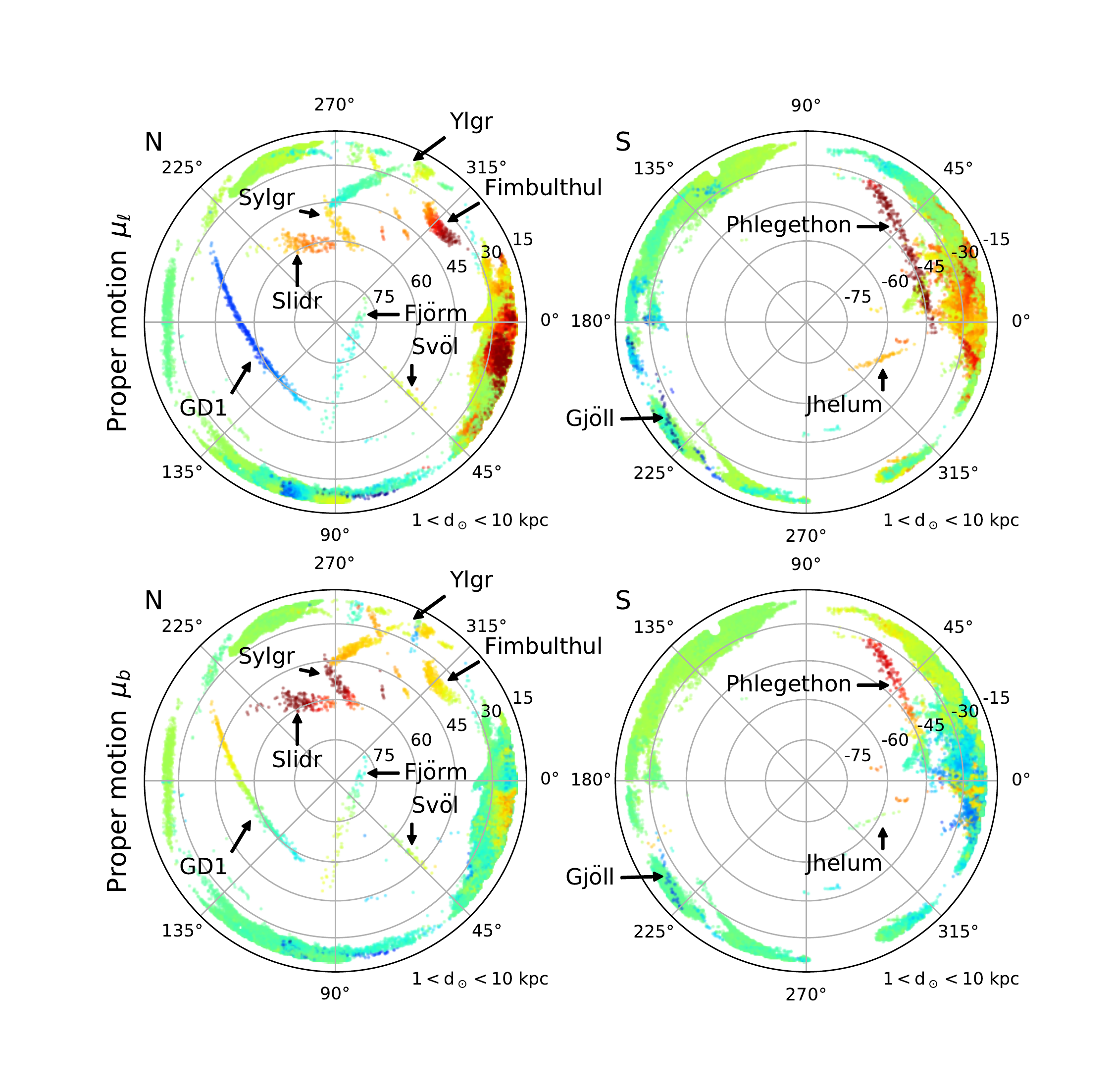}
\includegraphics[angle=0, viewport= 45 45 657 650, clip, height=9cm]{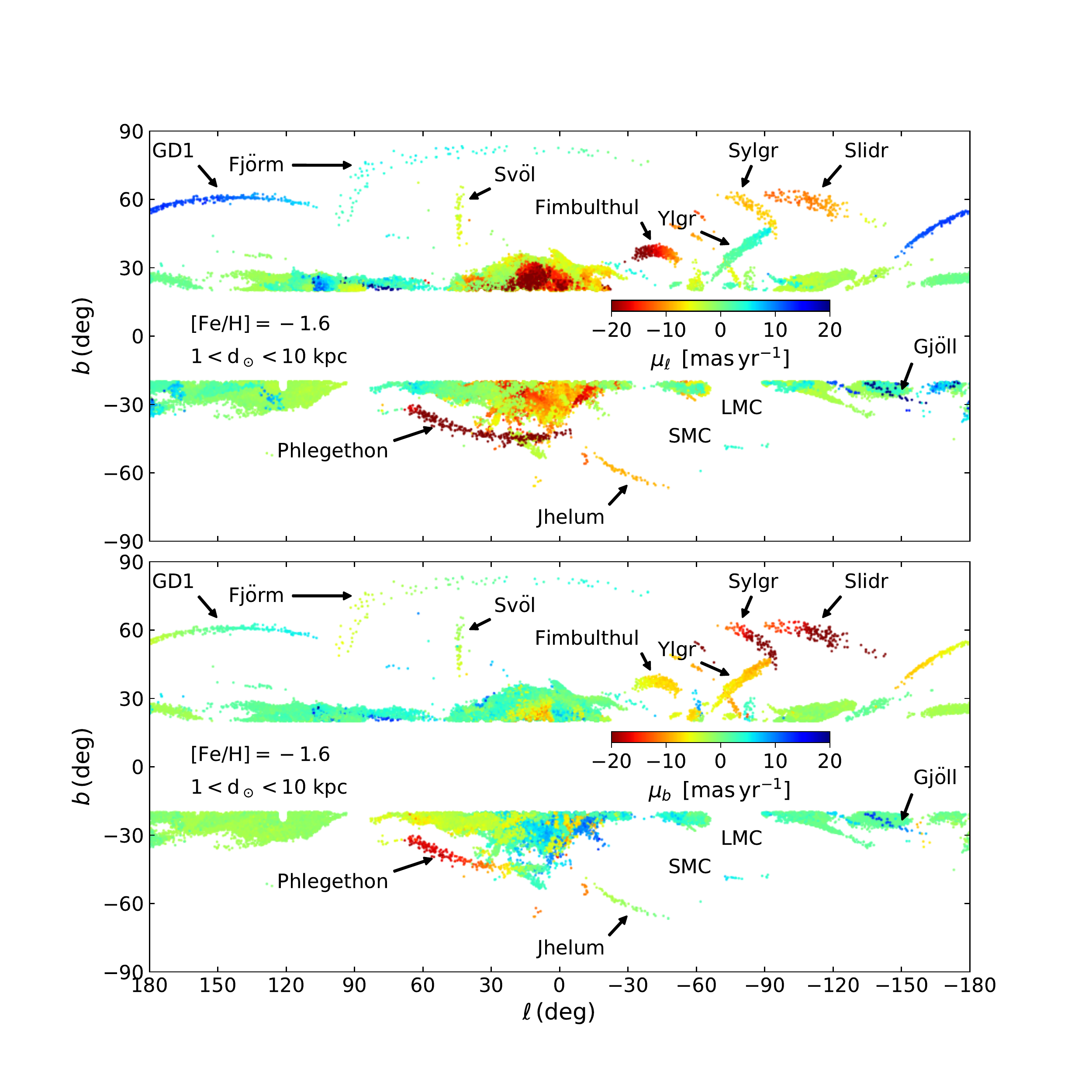}
}
\hbox{
\includegraphics[angle=0, viewport= 55 23 600 589, clip, height=9cm]{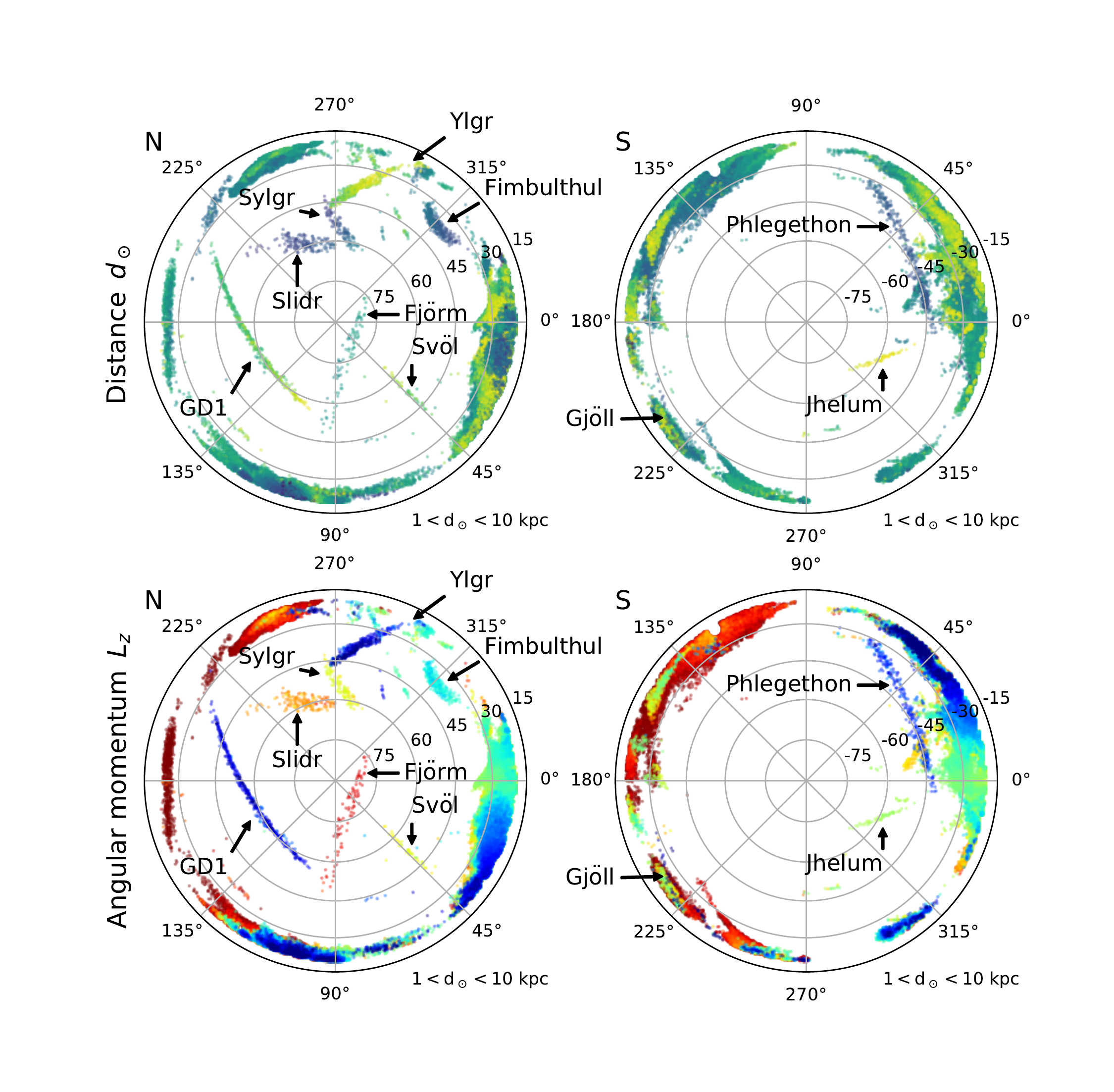}
\includegraphics[angle=0, viewport= 45 45 657 650, clip, height=9cm]{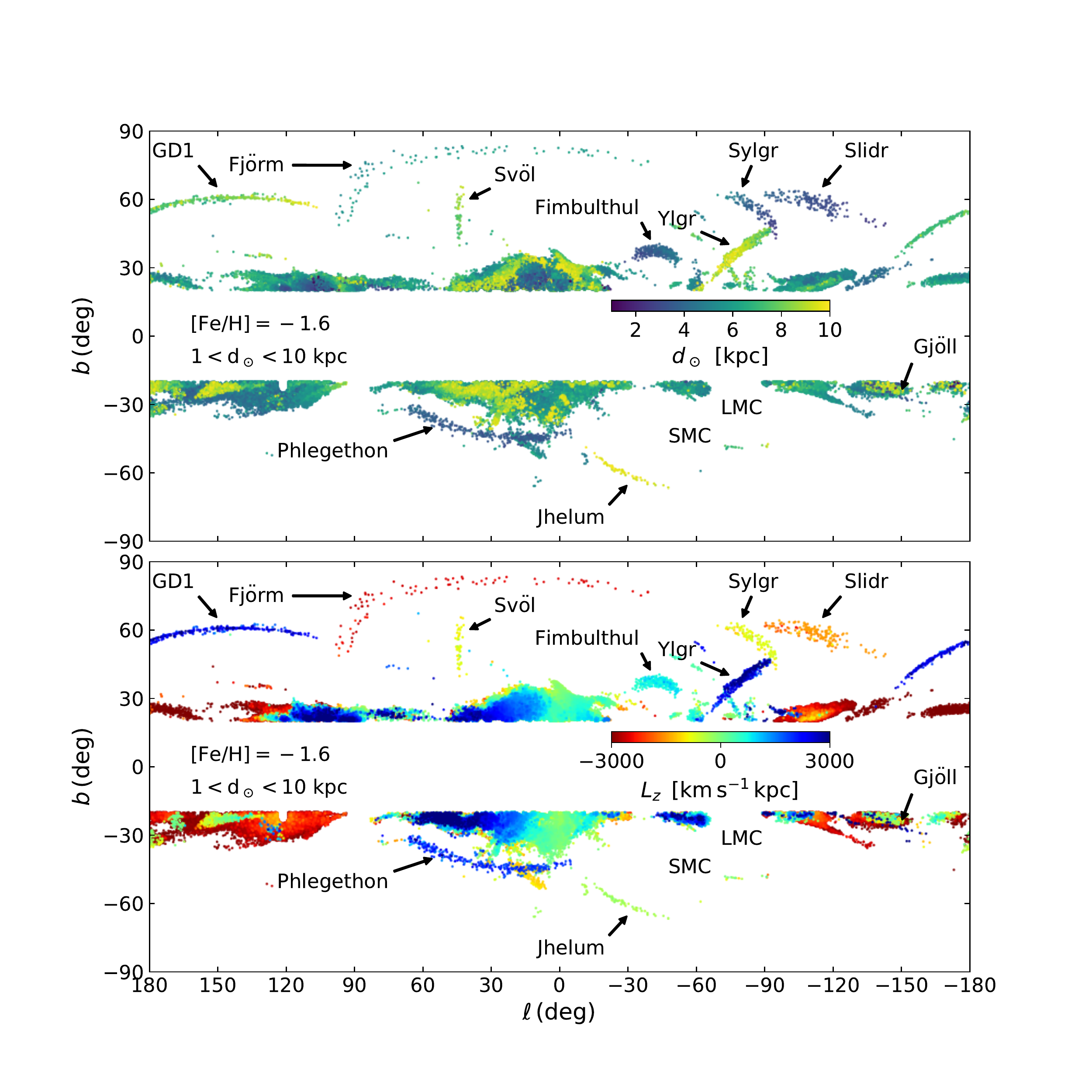}
}
}
\end{center}
\caption{As Figure~\ref{fig:FeH_m2.0}, but for stars with ${\rm [Fe/H]=-1.6}$.}
\label{fig:FeH_m1.6}
\end{figure*}

\begin{figure*}
\begin{center}
\vbox{
\hbox{
\includegraphics[angle=0, viewport= 55 23 600 589, clip, height=9cm]{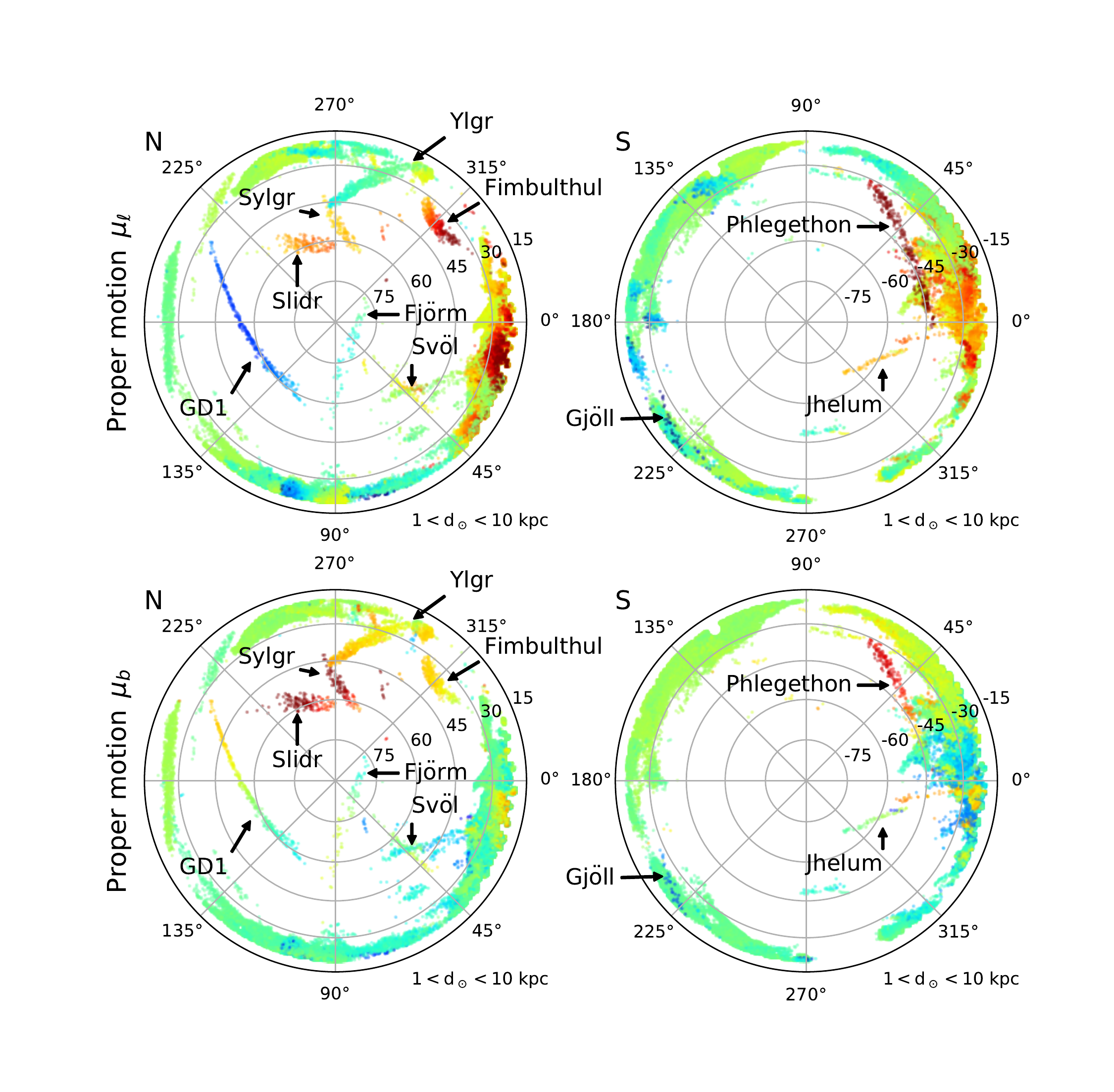}
\includegraphics[angle=0, viewport= 45 45 657 650, clip, height=9cm]{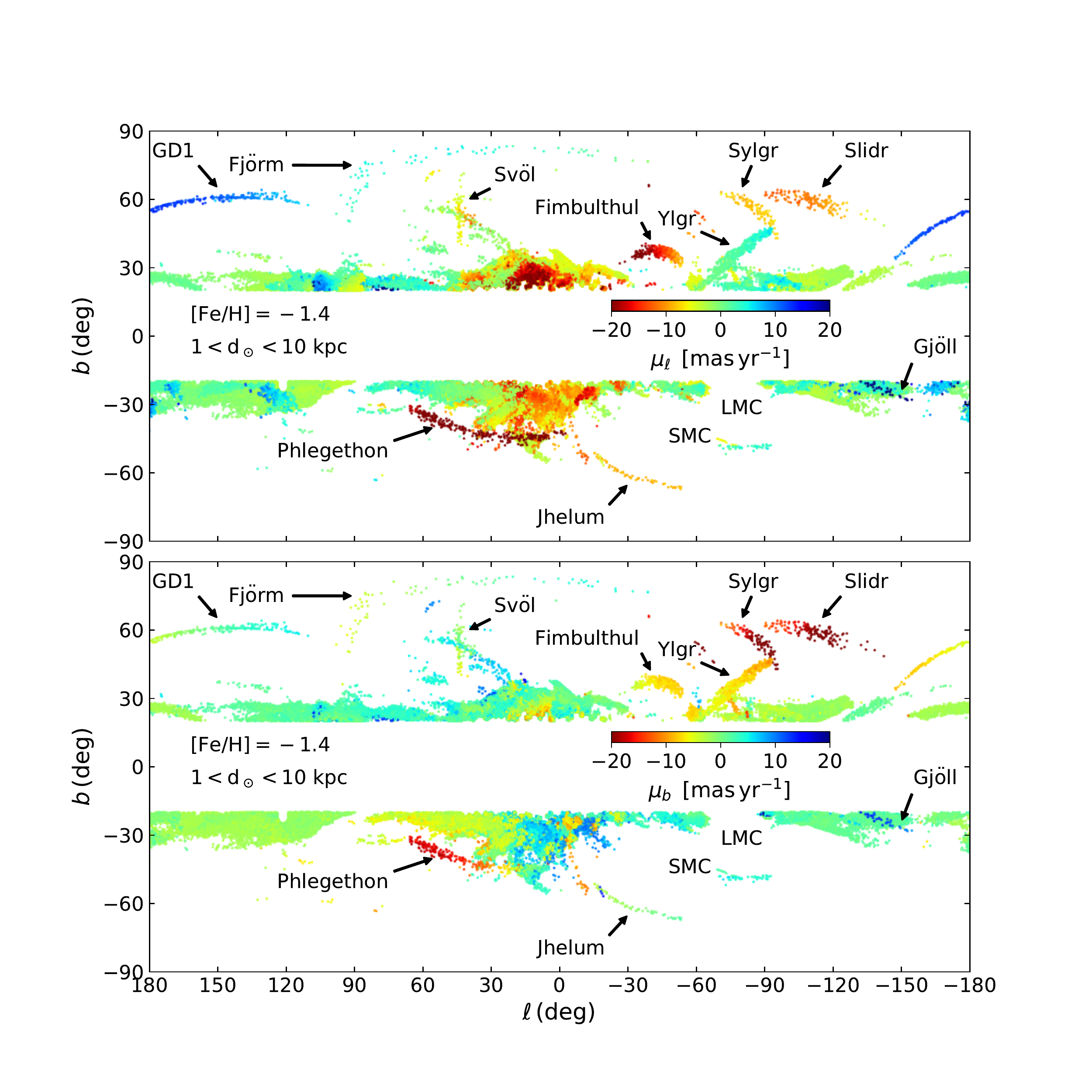}
}
\hbox{
\includegraphics[angle=0, viewport= 55 23 600 589, clip, height=9cm]{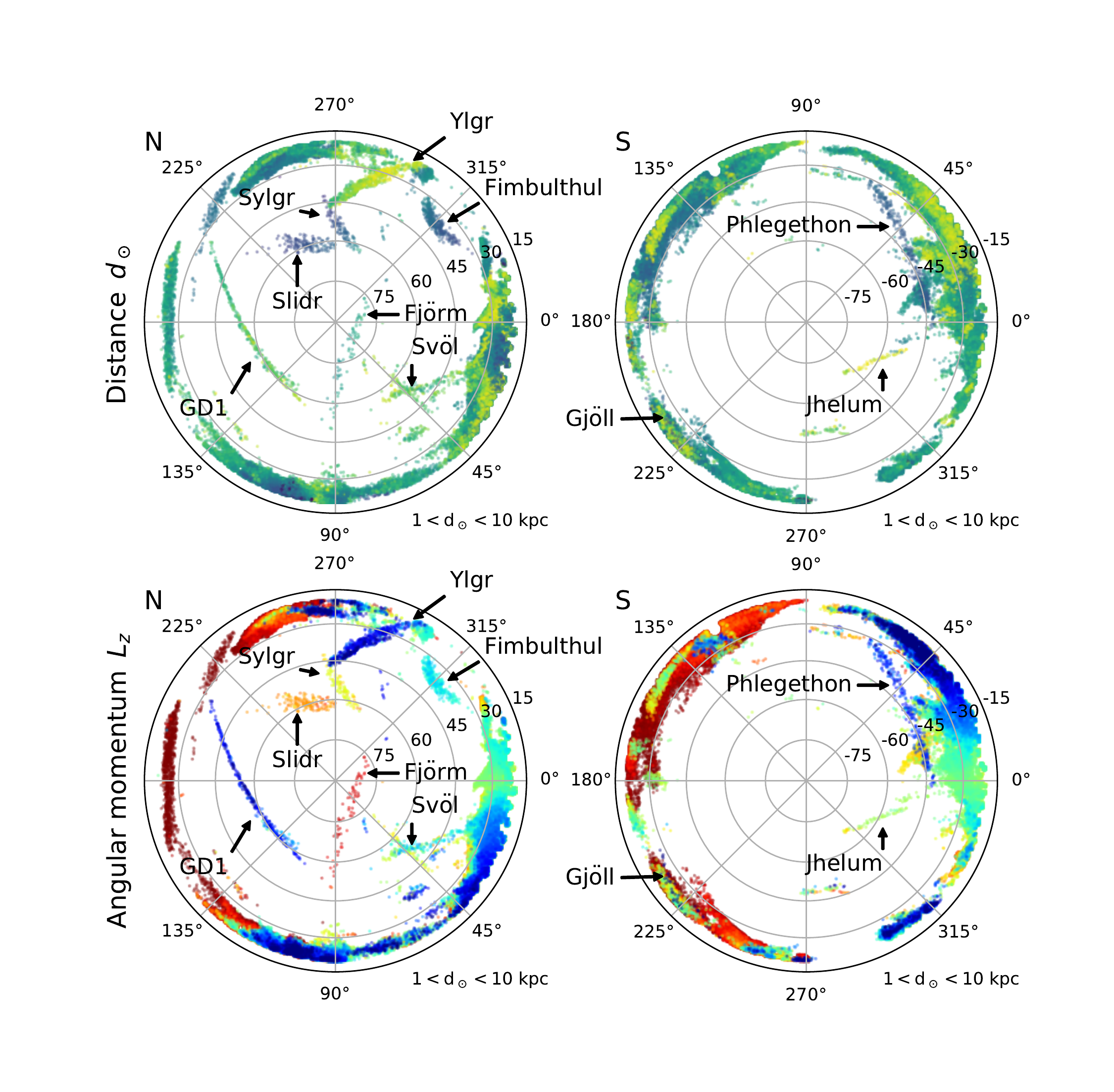}
\includegraphics[angle=0, viewport= 45 45 657 650, clip, height=9cm]{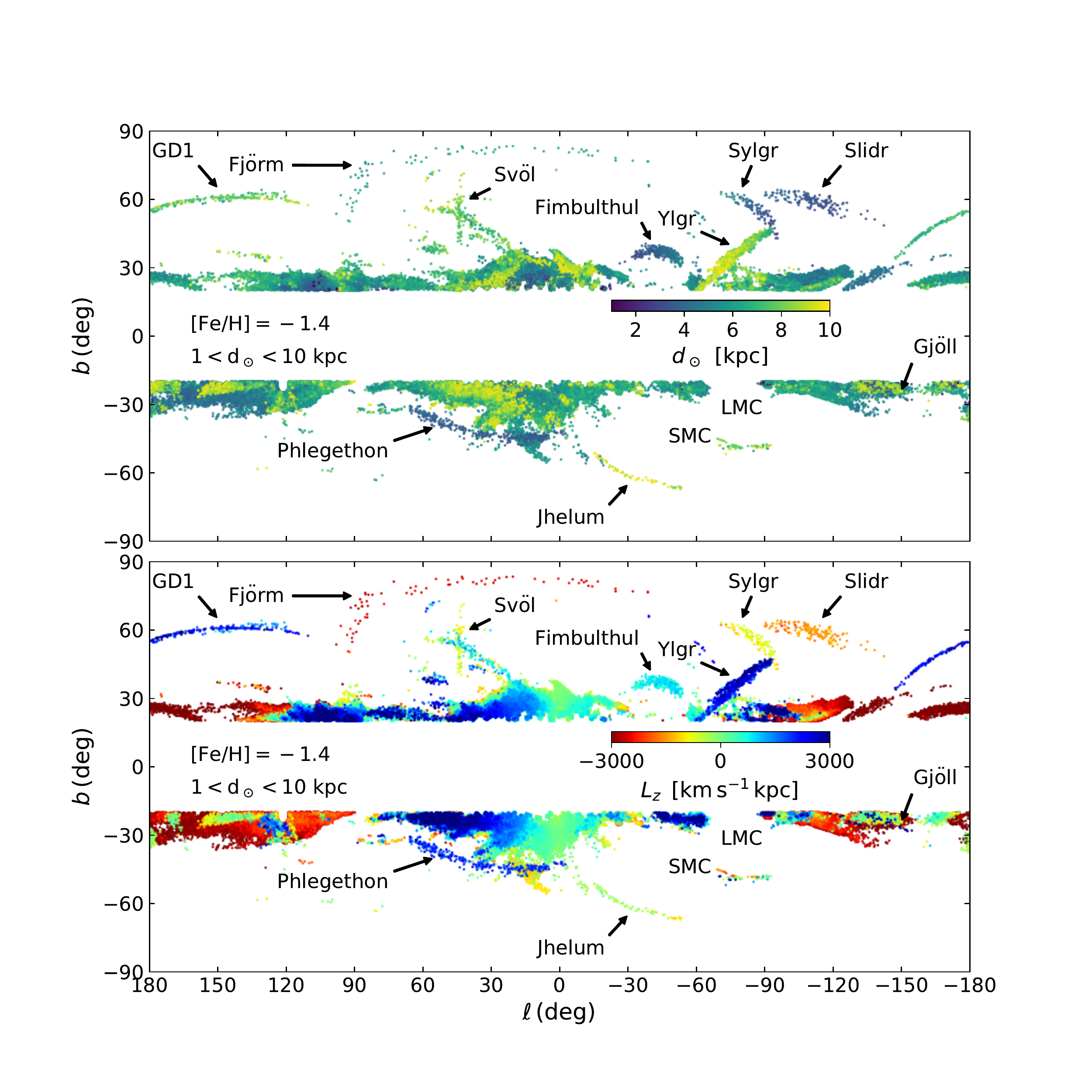}
}
}
\end{center}
\caption{As Figure~\ref{fig:FeH_m2.0}, but for stars with ${\rm [Fe/H]=-1.4}$.}
\label{fig:FeH_m1.4}
\end{figure*}

\section{Comparison data}
\label{sec:Comparison}

In Section~\ref{sec:Results} below we will present the maps of stream-like structures found by the {\tt STREAMFINDER} algorithm, in which many unexpected features are detected. It is natural to wonder whether these apparent streams are real or false positives created by our method of hunting for linear over-densities along orbits. In order to assess these concerns, we decided to also apply the algorithm to the {\it Gaia} Universe Model Snapshot (GUMS; \citealt{2012A&A...543A.100R}), which provides a realistic simulation of the stellar density and kinematics of the smooth components of the Milky Way. 

The GUMS simulation does not include an estimate of the observational uncertainties, however. So as to model the {\it Gaia} DR2 observations as closely as possible, we chose to directly reuse the uncertainties in the real data. This was implemented as follows. For every artificial star in GUMS, we searched for the spatially-closest real counterpart in {\it Gaia} within a $1\deg$ radius with a $G$-band magnitude within $0.1\magn$ of the GUMS star. If no such real star was found, the search was repeated doubling the magnitude difference criterion (repeated, if necessary, until a match was found). The proper motion and parallax uncertainties of the matched {\it Gaia} star were then assigned to the simulated GUMS star, together with the observed proper motion correlation, and these Gaussian uncertainties were used to resample the GUMS proper motions and parallaxes, thus making the properties of these parameters closer to the observations than in the original GUMS simulation. We apply exactly the same mask to remove stars around the locations of the Galactic satellites and globular clusters (so as to generate the same gaps in the survey, even though these satellites are not present in the simulation). 

Down to ${G_0=20}$, and at the Galactic latitudes of interest to the present study ($|b|\simgt 20\deg$), the number density of stars in the GUMS simulation is significantly higher than in the {\it Gaia} DR2 catalog. The {\it Gaia}$/$GUMS ratio is shown in Figure~\ref{fig:Gaia_to_GUMS}, and averages $0.6$ for the sky at $|b|> 20\deg$. The reason for the overestimate in the GUMS model is not clear to us, but we believe that the discrepancy is not  problematic for the present purposes, as long as we correct for it. To that end, we take the same $1\degg4\times1\degg4$ spatial bins as before (used in Figure~\ref{fig:Gaia_to_GUMS}), and simply randomly discard GUMS stars until we obtain the number density observed in {\it Gaia} DR2. This procedure has the added advantage of imprinting the effects of the actual {\it Gaia} scanning law on the simulated data, rendering it more realistic.

We then regenerate the contamination model $\mathcal{P}_{\rm cont}$ directly from the simulated data.

\section{Results}
\label{sec:Results}

The algorithm was run over metallicity values of ${\rm [Fe/H] = \{-2.0, -1.6, -1.4\}}$~dex, and we integrated orbits starting from {\it Gaia} stars with magnitudes ${G_0<19.5}$ and $|b|>20\deg$. The magnitude cut at ${G_0=19.5}$ was chosen so as to avoid edge-effects stemming from the faint limit of our GMM model at ${G_0=20.0}$ (see Figure~\ref{fig:GMM_CMD}). A single SSP model age of $12.5\Gyr$ was adopted. Since the allowed half-length of the orbits was set to $10\deg$, this meant that we actually examined data at all latitudes $|b|>10\deg$. The extension to lower latitude will be presented in future contributions, but this is computationally extremely expensive due to the high source density (the analysis presented here already took over a million CPU hours). 

Finally, following \citet{2018A&A...616A...2L}, we filtered the output stars to retain only those having flux excess $E (\equiv{\tt phot\_bp\_rp\_excess\_factor})$ in the range:
\begin{equation}
1+0.015 (G_{BP}-G_{RP})^2 < E < 1.3 + 0.06 (G_{BP}-G_{RP})^2 \, .
\end{equation}
This removes stars that have suspicious photometry, which amount to $\approx 10$\% of the sources in streams. We note, however, that the maps are qualitatively identical with or without this filter.

Inspection of the {\tt STREAMFINDER} output shows a clear dichotomy between the stream-like features with low proper motion ($\sqrt{\mu^2_\ell+\mu^2_b}<2\masyr$), and those of higher proper motion. The solutions with low proper motion form a relatively coherent pattern on sky, as we show in Figure~\ref{fig:FeH_m2.0_fluff}, which displays the stream-like features with $\sqrt{\mu^2_\ell+\mu^2_b}<2\masyr$ that are identified by the algorithm at distances between $1$ and $10\kpc$ when assuming a model of metallicity ${\rm [Fe/H]=-2.0}$ (other choices of the template metallicity result in similar sky distributions). The distance limits set the range that the algorithm is allowed to search over, and having only a narrow distance interval such as this helps speed up the computation. The top two rows of panels show the proper motion distributions in $\mu_\ell$ and $\mu_b$ (deliberately using the same color scale as will be used later in Figures~\ref{fig:FeH_m2.0}--\ref{fig:FeH_m1.4}), the third row shows the heliocentric distance distribution, while the bottom row shows the $z$-component of angular momentum $L_z$ of the stream solutions. We have set the detection threshold here, and in all subsequent maps, to $8\sigma$. These features are not present in the equivalent maps based on the GUMS comparison data, as we show in Figure~\ref{fig:FeH_m2.0_fluff_GUMS} in the Appendix, and so they do not appear to be an obvious artefact of the method. Nevertheless, these structures are not streams of low-mass accretions that we aim to identify here and constitute a contaminating population for the present work (we defer their analysis to a subsequent follow-up study). The bottom row of panels shows that they possess negative $L_z$ (meaning that they are prograde), and we found that we could eliminate this population by setting a higher detection threshold for prograde stars. However, in the following we choose to simply filter out the stars for which the most likely stream solutions have $\sqrt{\mu^2_\ell+\mu^2_b}<2\masyr$, as we can then present maps with a uniform detection threshold that does not depend on $L_z$. 

The complement to the sample in Figure~\ref{fig:FeH_m2.0_fluff} is shown in Figure~\ref{fig:FeH_m2.0}, which displays the stream solutions with $\sqrt{\mu^2_\ell+\mu^2_b}>2\masyr$ in the distance range between $1$--$10\kpc$ derived using the SSP template with age $12.5\Gyr$ and metallicity ${\rm [Fe/H]=-2.0}$. Figures~\ref{fig:FeH_m1.6} and \ref{fig:FeH_m1.4} show similar information but using a template with metallicity of ${\rm [Fe/H]=-1.6}$ and ${\rm [Fe/H]=-1.4}$, respectively.  The layout of the panels is identical to those in Figure~\ref{fig:FeH_m2.0_fluff}, and the color coding is identical as well. 

The well-studied GD-1 stream \citep{2006ApJ...643L..17G} is detected at high confidence (up to $20\sigma$ in portions of the stream) in all three metallicity maps, although its contrast decreases when using the higher metallicity templates. We note that the imposed upper distance limit of $10\kpc$ artificially limits this stream to $\ell \simgt 110\deg$. The only other previously-detected stream present in our maps is the Jhelum structure \citep{2018ApJ...862..114S}. The detection of GD-1 and Jhelum shows that the algorithm works as intended, and is able to recover low-surface brightness streams.

A large number of additional stream features are detected by the algorithm at this $8\sigma$ threshold. We have marked and named ten other streams that are distinct and coherent in all of the observed parameters, and clearly distinguishable from possible residual features of the Galactic disk or bulge. In addition to the ``Phlegethon'' stream, detected by our algorithm and reported in Paper~III, the new detections are ``Slidr'', ``Sylgr'', ``Ylgr'', ``Fimulthul'', ``Sv\"ol'', ``Fj\"orm'', ``Gj\"oll'' and ``Leiptr''. These names are taken from Norse mythology, and are eight of the eleven rivers that existed in the void (or gaping abyss) at the beginning of the world, with Slidr and Gj\"oll also being streams of the underworld (as is the Phlegethon in Greek mythology).

\begin{figure}
\begin{center}
\includegraphics[angle=0, width=\hsize]{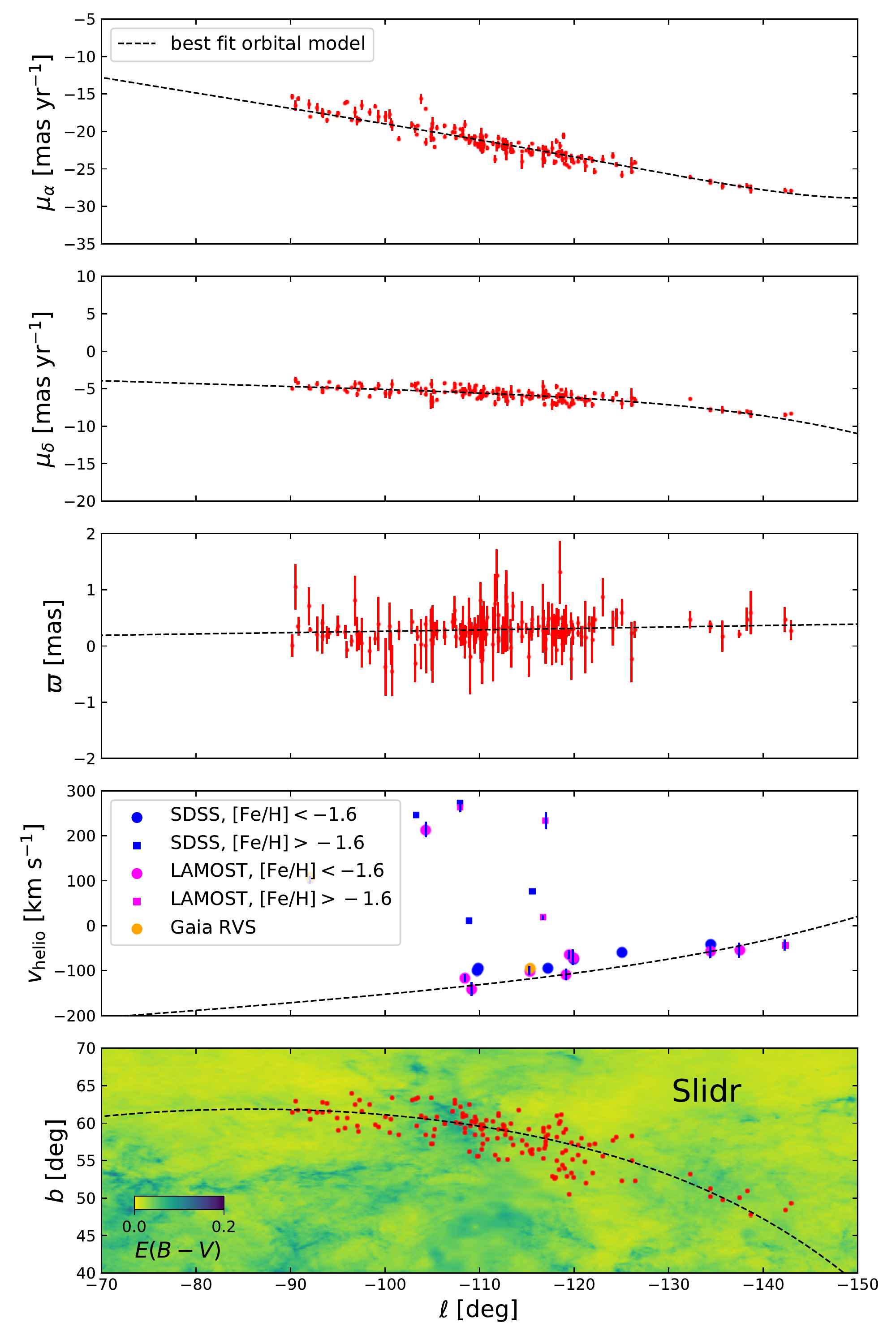}
\end{center}
\caption{Properties of the Slidr stream. From top to bottom, the panels show proper motion $\mu_\alpha$ and $\mu_\delta$, the parallax $\varpi$, the heliocentric line of sight velocity $v_{\rm helio}$ and Galactic latitude position of the stream stars, as a function of Galactic longitude $\ell$. The red points are the 156 members of the Gaia DR2 catalog with five-component astrometric solutions that are identified by the {\tt STREAMFINDER} software as stream stars using an $8\sigma$ threshold. The uncertainties on these points are shown with error bars. The large points in the fourth panel mark those stream members whose radial velocity has been measured by the SDSS, LAMOST or Gaia RVS surveys. As explained in the text, we fitted an orbit model to the Gaia data, without using any of the radial velocity information. The corresponding best-fit model is shown with a dashed line, and can be seen to give a reasonable representation of the parameter profiles. The radial velocity profile is effectively a prediction of the model, and can be seen to pass close to all but one of the metal poor stars with ${\rm [Fe/H]<-1.6}$, confirming the reality of the structure. Additionally, the bottom panel also shows the $E(B-V)$ extinction map, extracted from \citet{Schlegel:1998fw}, from which the reader may verify that the detected stream does not follow any obvious structure in the interstellar extinction.}
\label{fig:Slidr}
\end{figure}

In Figures~\ref{fig:FeH_m2.0_GUMS}--\ref{fig:FeH_m1.4_GUMS}, we reproduce exactly the same maps as for Figures~\ref{fig:FeH_m2.0}--\ref{fig:FeH_m1.4}, but for stream-like detections derived from the GUMS simulation. These false positives are found towards the bulge region of the simulation in areas of very high source density. Their distribution in proper motion is markedly different from the detections we report here.

\begin{figure}
\begin{center}
\includegraphics[angle=0, width=\hsize]{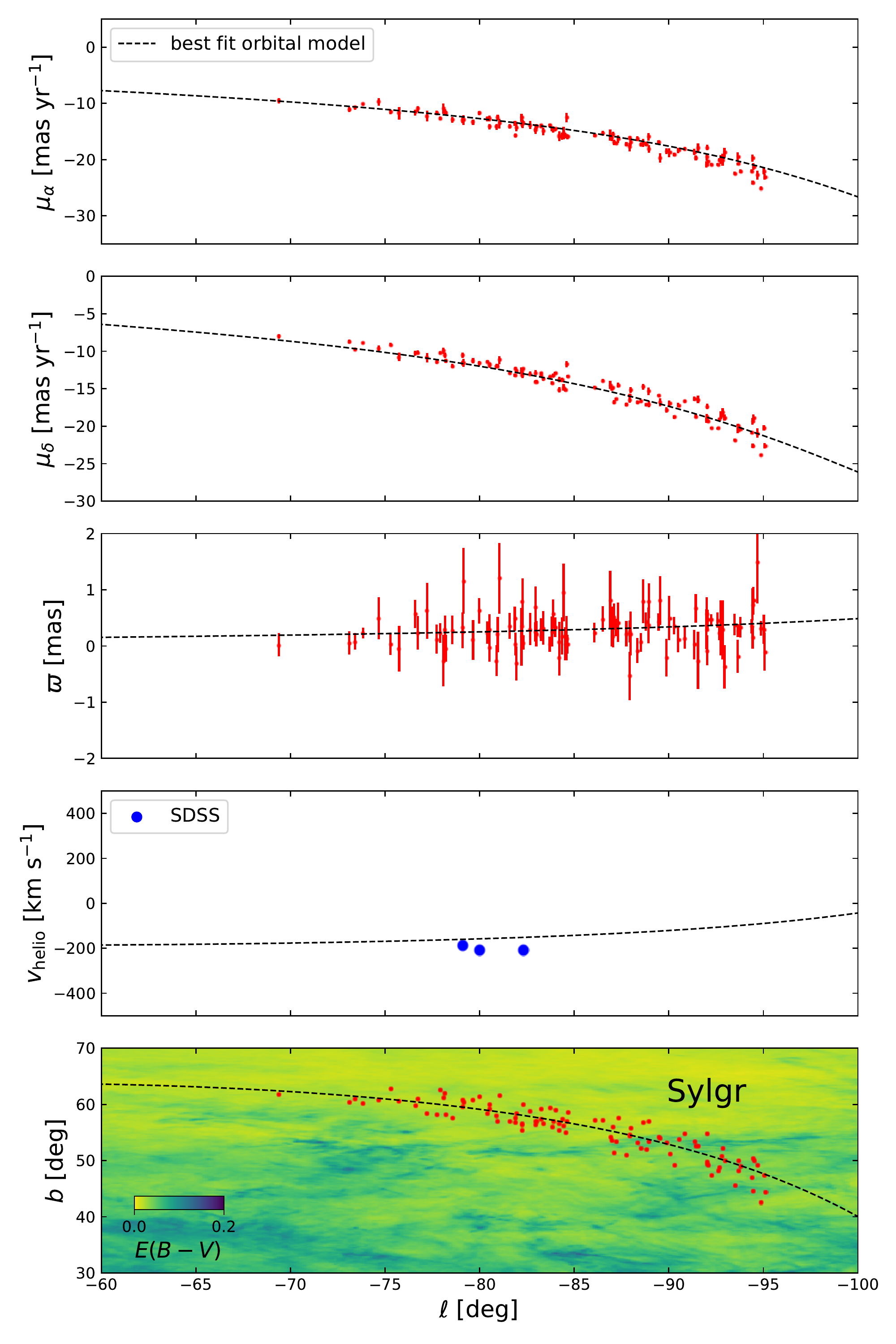}
\end{center}
\caption{As Figure~\ref{fig:Slidr}, but for the Sylgr stream. A total of 103 stars are identified as candidate members of this structure at $>8\sigma$ confidence.}
\label{fig:Sylgr}
\end{figure}

A number of other stream candidates can also be seen in these maps, especially in Figure~\ref{fig:FeH_m1.4}. They are however of somewhat lower significance than the eight new obvious structures we have named, or they have kinematic properties that make them not so easily distinguishable from the false positives in our processing of the GUMS catalog. 

We now proceed to examine each of the eight streams separately, first the streams in the northern Galactic hemisphere in order of increasing longitude in Figure~\ref{fig:FeH_m2.0}, and then we will proceed to examine the streams in the southern Galactic hemisphere, also in order of increasing longitude.

\begin{figure}
\begin{center}
\includegraphics[angle=0, width=\hsize]{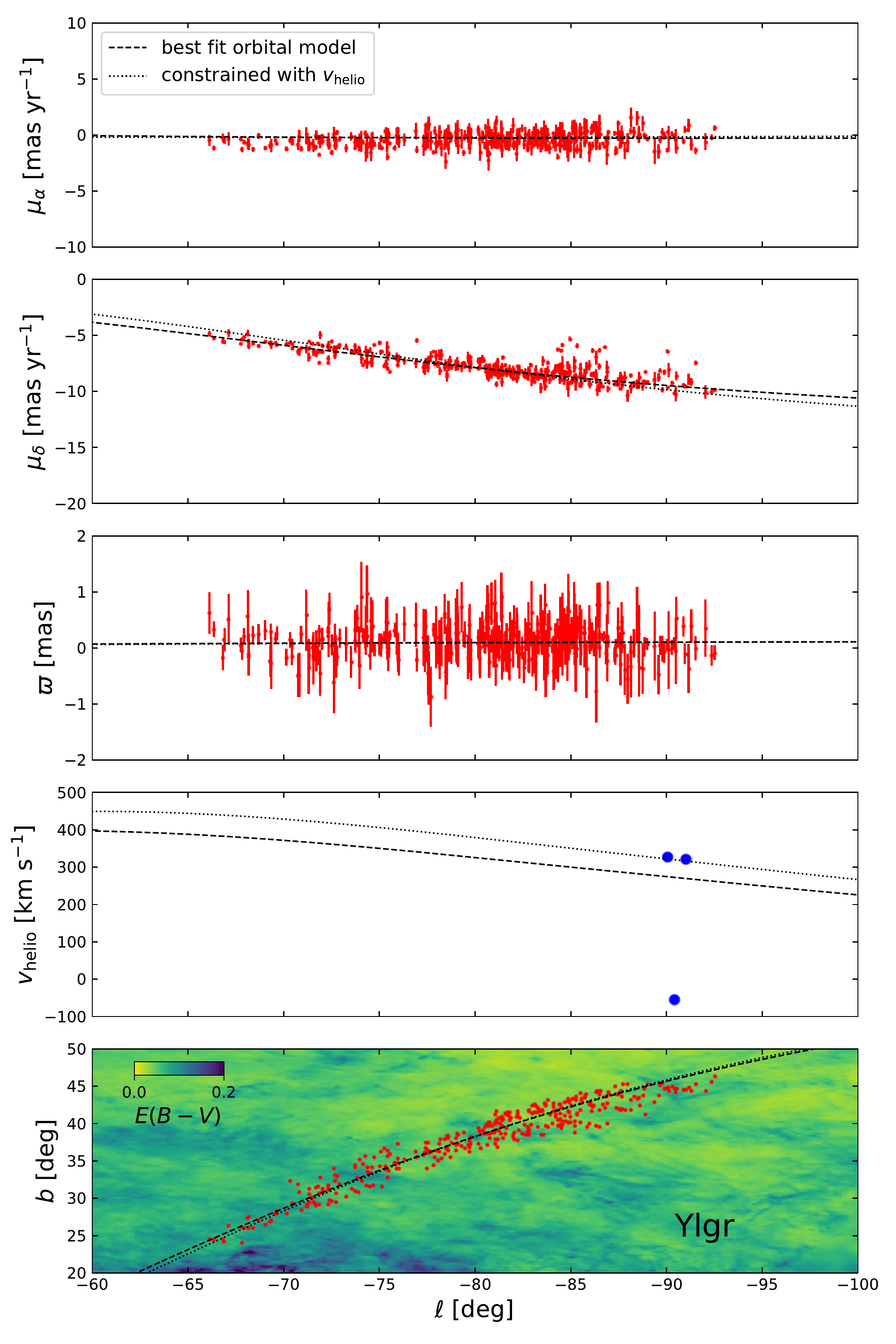}
\end{center}
\caption{As Figure~\ref{fig:Slidr}, but for the Ylgr stream, which contains 349 candidate members.}
\label{fig:Ylgr}
\end{figure}

\section{The Streams}
\label{sec:Streams}

\subsection{Slidr}

The Slidr stream can be perceived as a $32\deg$-long prograde structure in the northern hemisphere maps. Although it is detected with all of the chosen metallicity templates, it appears strongest in the ${\rm [Fe/H]=-1.6}$ map (Figure~\ref{fig:FeH_m1.6}). In Figure~\ref{fig:Slidr} we show the parameter correlations of stars of this structure that the algorithm identifies; these {\it Gaia} DR2 stars are displayed as red dots with their corresponding uncertainties. A total of 156 stars are detected at $>8\sigma$ confidence.

\begin{figure}
\begin{center}
\includegraphics[angle=0, width=\hsize]{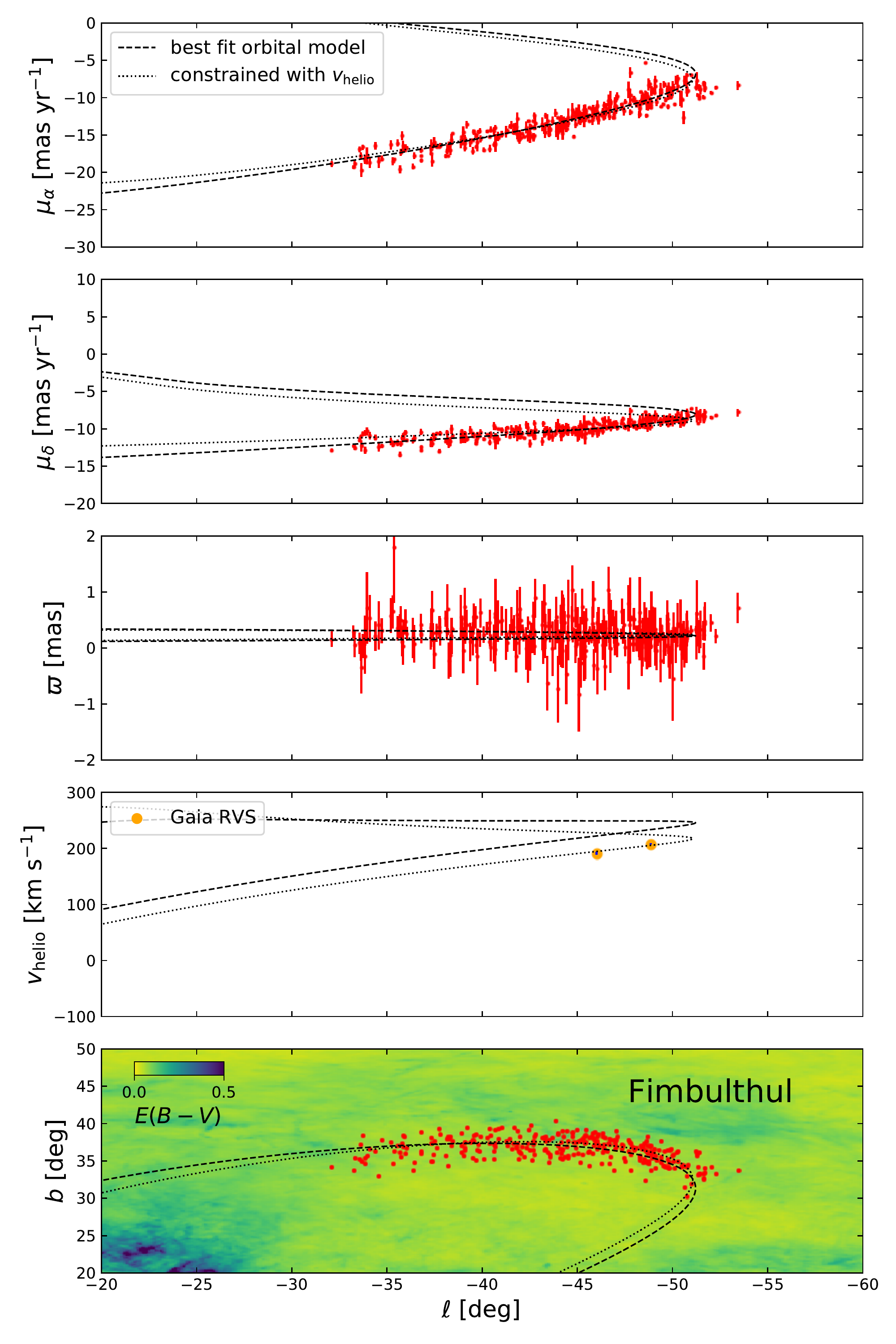}
\end{center}
\caption{As Figure~\ref{fig:Slidr}, but for the Fimbulthul stream. This $18\deg$-long structure contains 309 candidate members.}
\label{fig:Fimbulthul}
\end{figure}

\begin{figure}
\begin{center}
\includegraphics[angle=0, width=\hsize]{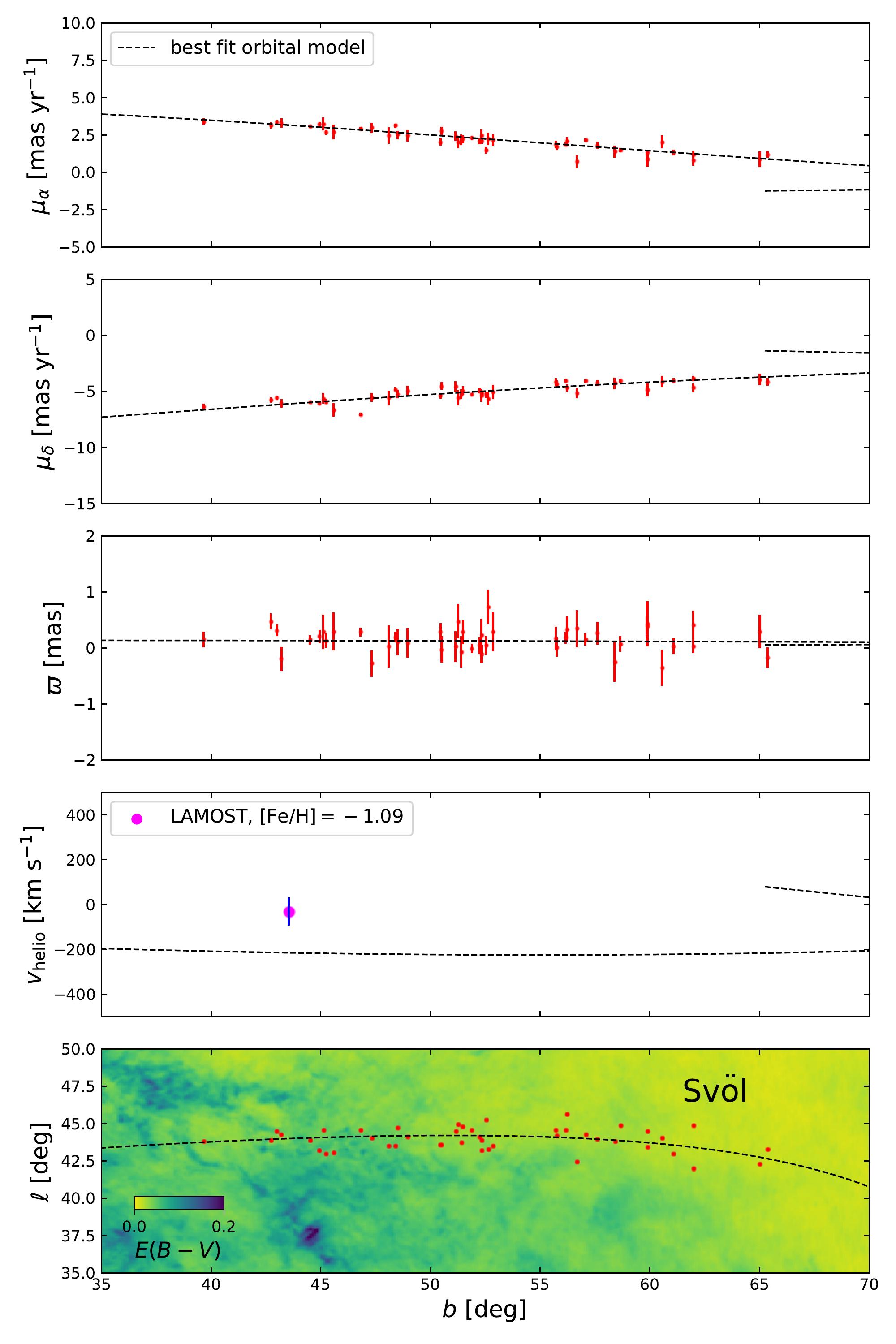}
\end{center}
\caption{As Figure~\ref{fig:Slidr}, but for the Sv{\" o}l stream, of which the algorithm finds 45 candidate members.}
\label{fig:Svol}
\end{figure}

\begin{figure}
\begin{center}
\includegraphics[angle=0, width=\hsize]{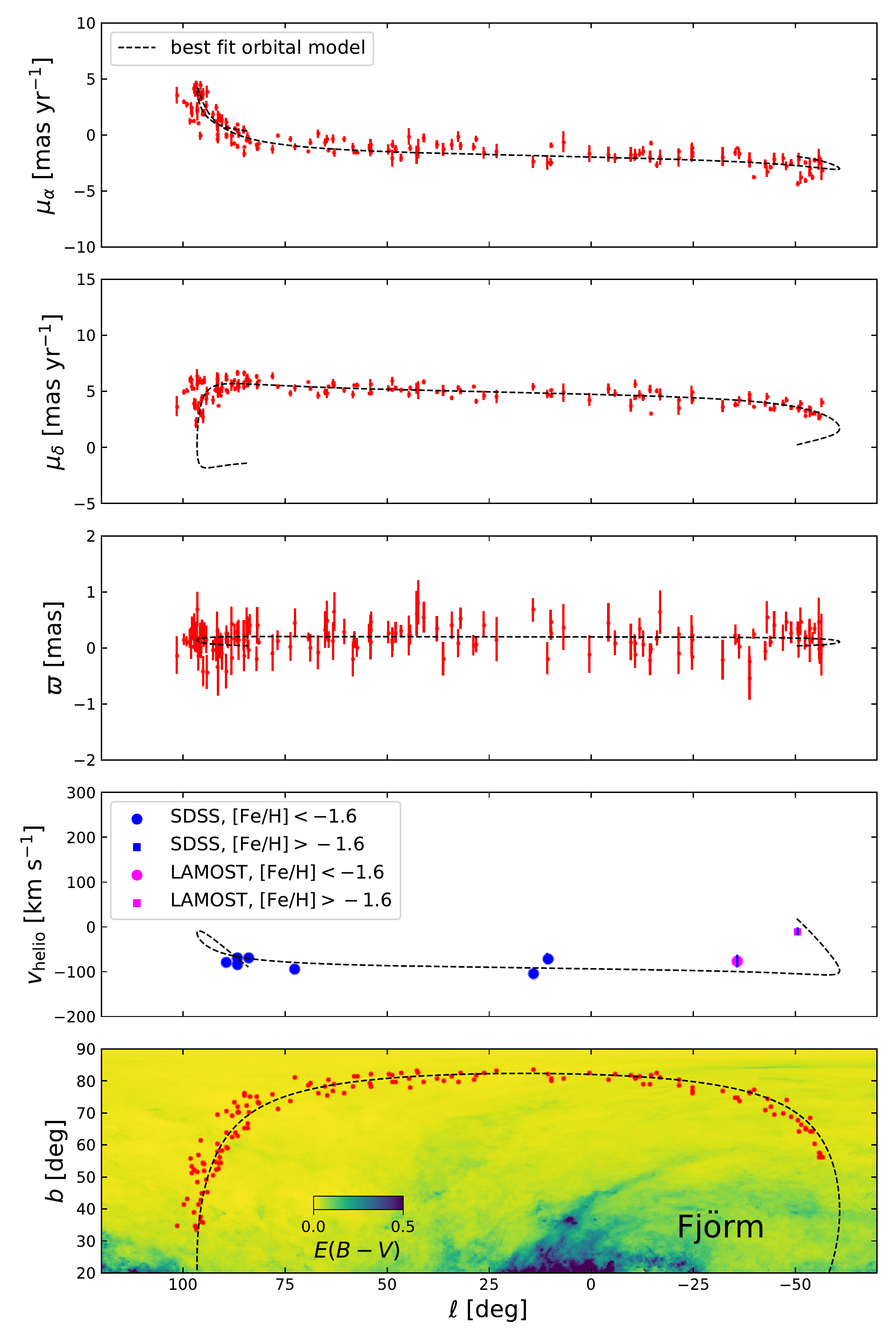}
\end{center}
\caption{As Figure~\ref{fig:Slidr}, but for the Fj{\" o}rm stream. The 148 candidate members of this structure span an arc of $87\deg$ in the northern sky.}
\label{fig:Fjorm}
\end{figure}

In order to obtain a global fit to these data, we reused the procedure developed in Paper~III to fit the ``Phlegethon'' stream. This re-fitting is necessary because the orbit fits computed by the {\tt STREAMFINDER} are optimized for speed rather than accuracy and reliability (for instance, by using the above-mentioned ``downhill simplex'' method, its solutions can get caught in local likelihood maxima), and furthermore the search routine only has access to data over the chosen stream search length $L$. The dedicated orbit fitting software implements a Markov Chain Monte Carlo (MCMC) exploration for the initial conditions of an orbit that best fits the data in position, proper motion and parallax. The orbit integrator and Galactic potential model are identical to those used in the {\tt STREAMFINDER}, and the parameter space is sampled using the affine-invariant ``walkers'' approach of \citet{Goodman:2010we}. Contrary to Paper~III, however, we did not attempt to fit any radial velocity measurements, partially because we do not have this information for all of the streams presented here, but more importantly because we desired to obtain predictions for the radial velocity gradient based on the Gaia astrometry alone. The algorithm maximises the likelihood of the orbit given the data, adopting the following probability density function for the stream:
\begin{equation}
\mathcal{P}_{\rm fit}(\theta)  = \mathcal{P}_{\rm width} \times \mathcal{P}_\mu \times \mathcal{P}_{\varpi} \, .
\end{equation}
The terms $\mathcal{P}_{\rm width}$ and $\mathcal{P}_\mu$ here are identical to those defined above in Section~\ref{Sec:stream_model} for the {\tt STREAMFINDER}. The last term is a simple Gaussian probability density function for the observed parallax $\mathcal{P}_{\varpi}= \mathcal{N}(\Delta \varpi, \sigma_\varpi)$, where we have ignored the intrinsic parallax dispersion, since the observational uncertainties completely dominate the dispersion of the data about the model. We ran the MCMC procedure for $1.1\times10^6$ iterations, and rejected the first $10^5$ (burn-in) steps. The best-fit model is shown as a dashed line in Figure~\ref{fig:Slidr}, and can be seen to give an excellent representation of the proper motions in $\mu_\alpha$ and $\mu_\delta$ (top two panels), the parallax (middle panel) and the position on the sky (bottom panel). The model distance to the structure (at $\alpha=171.043\deg$, $\delta=6\deg$) is $d=3.65\pm0.09\kpc$.

\begin{figure}
\begin{center}
\includegraphics[angle=0, width=\hsize]{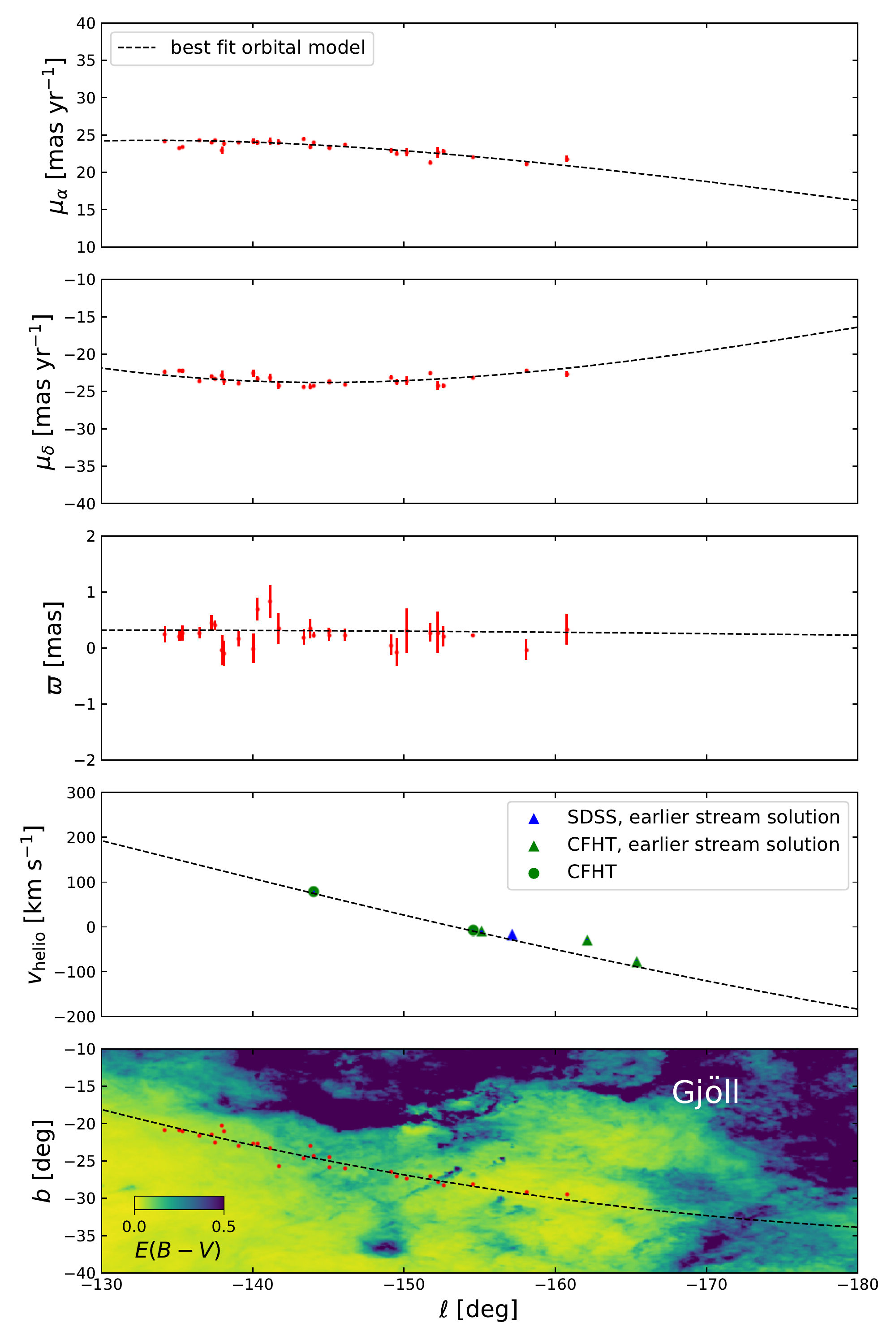}
\end{center}
\caption{As Figure~\ref{fig:Slidr}, but for the Gj\"oll stream. The {\tt STREAMFINDER} identifies 28 stars as being members of this system.}
\label{fig:Gjoll}
\end{figure}

In addition, the best-fit orbit gives a prediction of the heliocentric line of sight velocity of the structure (shown on the fourth panel). By chance, our Slidr sample contains ten stars previously observed in spectroscopy by the SDSS DR10 (\citealt{2009AJ....137.4377Y}, large blue points), 14 stars observed by LAMOST DR4 (\citealt{Cui:2012hd}, large magenta points), as well as two stars observed by the {\it Gaia} Radial Velocity Spectrometer (RVS; large orange points). The candidate stream members with radial velocity measurements are listed in Table~\ref{tab:RVs}. The two {\it Gaia} DR2 RVS stars happen to be a subset of the LAMOST sample. There is a clear separation in kinematic behavior between those stars that have spectroscopically-measured metallicities above and below ${\rm [Fe/H]=-1.6}$. We display the position of stars that are more metal-poor than this limit with large filled circles, while the more metal-rich stars are marked by squares. Apart from one outlier, all of the metal-poor subsample follows the predicted line of sight velocity gradient. The mean metallicity of the six stream members stars in the SDSS sample is ${\rm [Fe/H]=-1.80}$, essentially identical to the mean metallicity of ${\rm [Fe/H]=-1.84}$ of the eight LAMOST stars that are stream members.

This radial velocity confirmation firmly establishes the reality of the Slidr stream, and proves that the algorithm is able to uncover very low contrast structures that have been missed in previous searches. The presence of kinematic outliers here provides a reminder that the {\tt STREAMFINDER} produces samples of stream {\it candidates}. The $8\sigma$ selection threshold indicates that, at the position of each identified star, there is a stream present at $>8\sigma$ confidence that shares the kinematic properties of the star (and is consistent with its photometric properties and parallax). In particular, it does not mean that each stream candidate star identified by the {\tt STREAMFINDER} is a member with $>8\sigma$ confidence of the detected stream.

\begin{figure}
\begin{center}
\includegraphics[angle=0, width=\hsize]{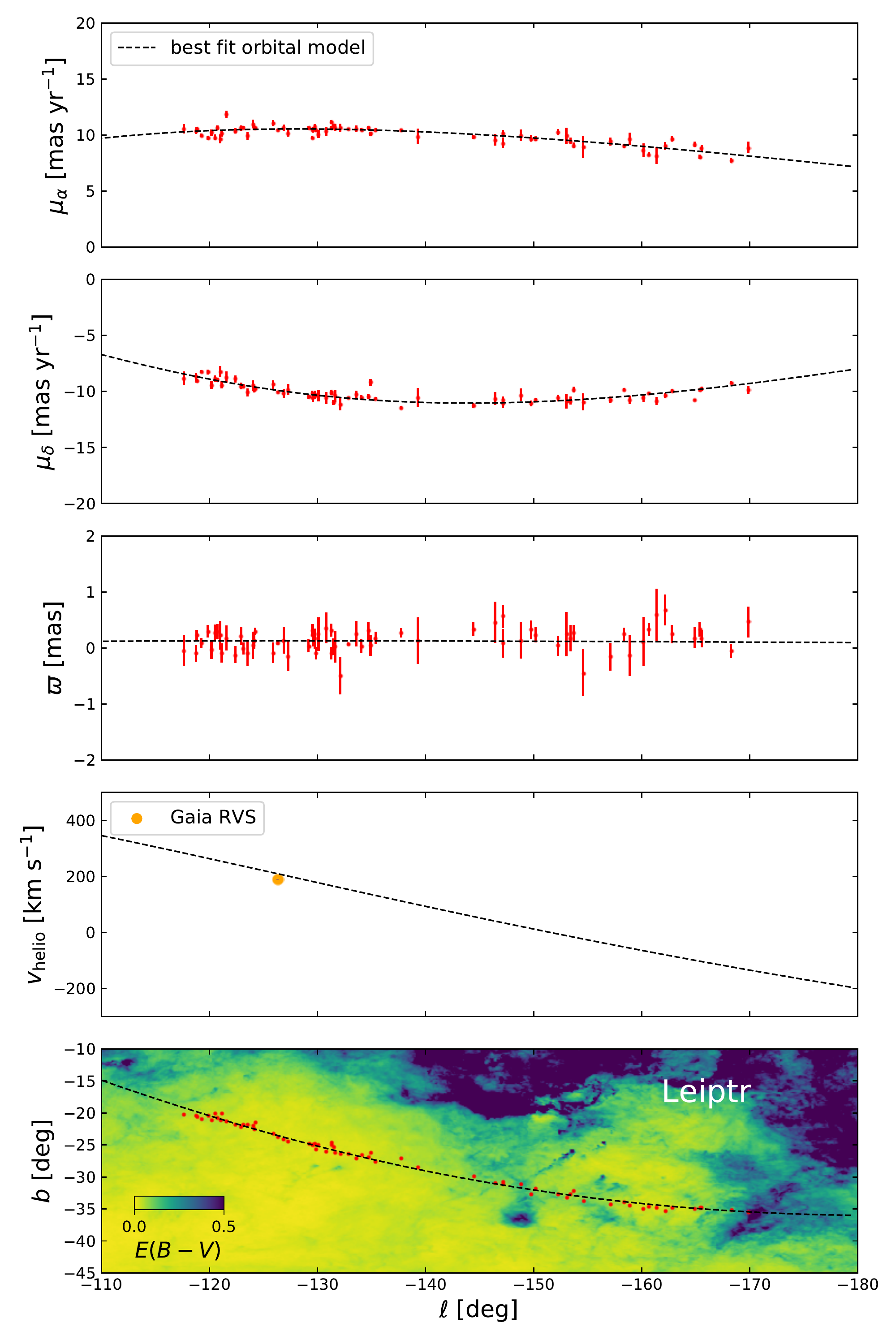}
\end{center}
\caption{As Figure~\ref{fig:Slidr}, but for the Leiptr stream, which contains 67 candidates.}
\label{fig:Leiptr}
\end{figure}

The orbit of Slidr is prograde but highly radial, with a pericenter at $4.45\pm0.08\kpc$ and apocenter at $42.4\pm3.2\kpc$. Its $z$-component of angular momentum is $L_z=-1458\pm9\kms \kpc$, according to the MCMC orbital fit (we adopt the usual convention that negative $L_z$ means prograde motion). Its orbit is shown in blue in Figure~\ref{fig:orbits}, and its position in the $L_z$ -- pericentric distance plane is displayed in Figure~\ref{fig:Lz}. These orbital parameters are listed in Table~\ref{tab:properties}, for easy comparison to the other streams.

The effect of sample contaminants on the derived orbits is hard to gauge. Here and in most cases below, the fact that the orbit as derived without radial velocities agrees with the small set of stars whose radial velocities have been measured suggests that the presence of contaminants is not problematic. Ultimately however, the candidate stream stars identified by the {\tt STREAMFINDER} should be followed-up with spectroscopy to eliminate any doubts and fully determine the orbit.

\subsection{Sylgr}

The Sylgr stream is a $24\deg$-long nearby structure that appears to have highest contrast in the ${\rm [Fe/H]=-1.6}$ map (Figure~\ref{fig:FeH_m1.6}), where it can be easily identified as a stream of 103 stars with highly negative $\mu_b$ in the northern hemisphere. The observed parameter gradients are shown in Figure~\ref{fig:Sylgr}. The result of the MCMC orbit-fitting procedure is shown with a dashed line, and gives a reasonable representation of the structure. Of the members we identify with the {\tt STREAMFINDER}, three have measured radial velocities in the SDSS; these are shown on the fourth panel of Figure~\ref{fig:Sylgr}. As before, we did not use the radial velocity information in the orbit-fitting procedure. All three of the SDSS stars lie close to the predicted radial velocity gradient of the stream, and they possess extremely low metallicities (${\rm \langle [Fe/H] \rangle=-2.7}$). It will be interesting to re-run the detection software in the region around this structure using a very metal-poor template.

\begin{figure}
\begin{center}
\includegraphics[angle=0, viewport= 4 10 420 855, clip, width=\hsize]{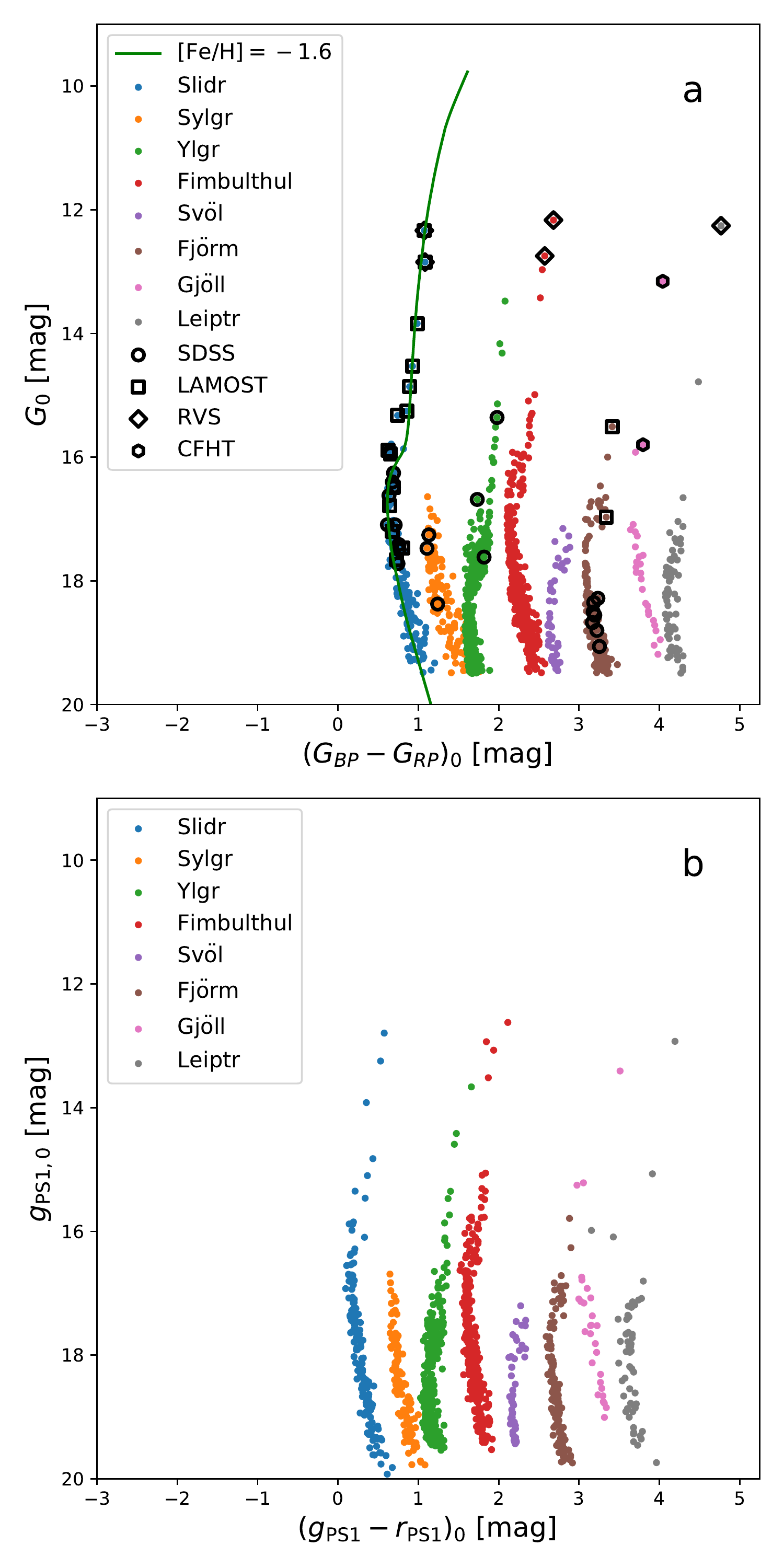}
\end{center}
\caption{Color-magnitude diagrams (CMDs) of the eight high-significance streams identified in this work. The Slidr system is shown without a color-shift, but each stream thereafter has been shifted by 0.5~mag in color so as to render the distributions visible. The PARSEC isochrone model with metallicity ${\rm [Fe/H]=-1.6}$ and age of $12.5\Gyr$ is superimposed, using the distance derived for the Slidr stream (corresponding to a distance modulus of 12.81). We also mark those stars with radial velocity measurements, as indicated in the legend. Panel (a) shows the original CMDs derived from Gaia DR2, while (b) displays the PanSTARRS1 DR1 (PS1) photometry \citep{2016arXiv161205560C} of the same stars. The consistency in the appearance of the Gaia and PS1 CMDs indicates that our detections are not due to spurious Gaia photometry.}
\label{fig:CMDs}
\end{figure}

\begin{figure*}
\begin{center}
\includegraphics[angle=0, viewport= 55 4 581 321, clip, width=\hsize]{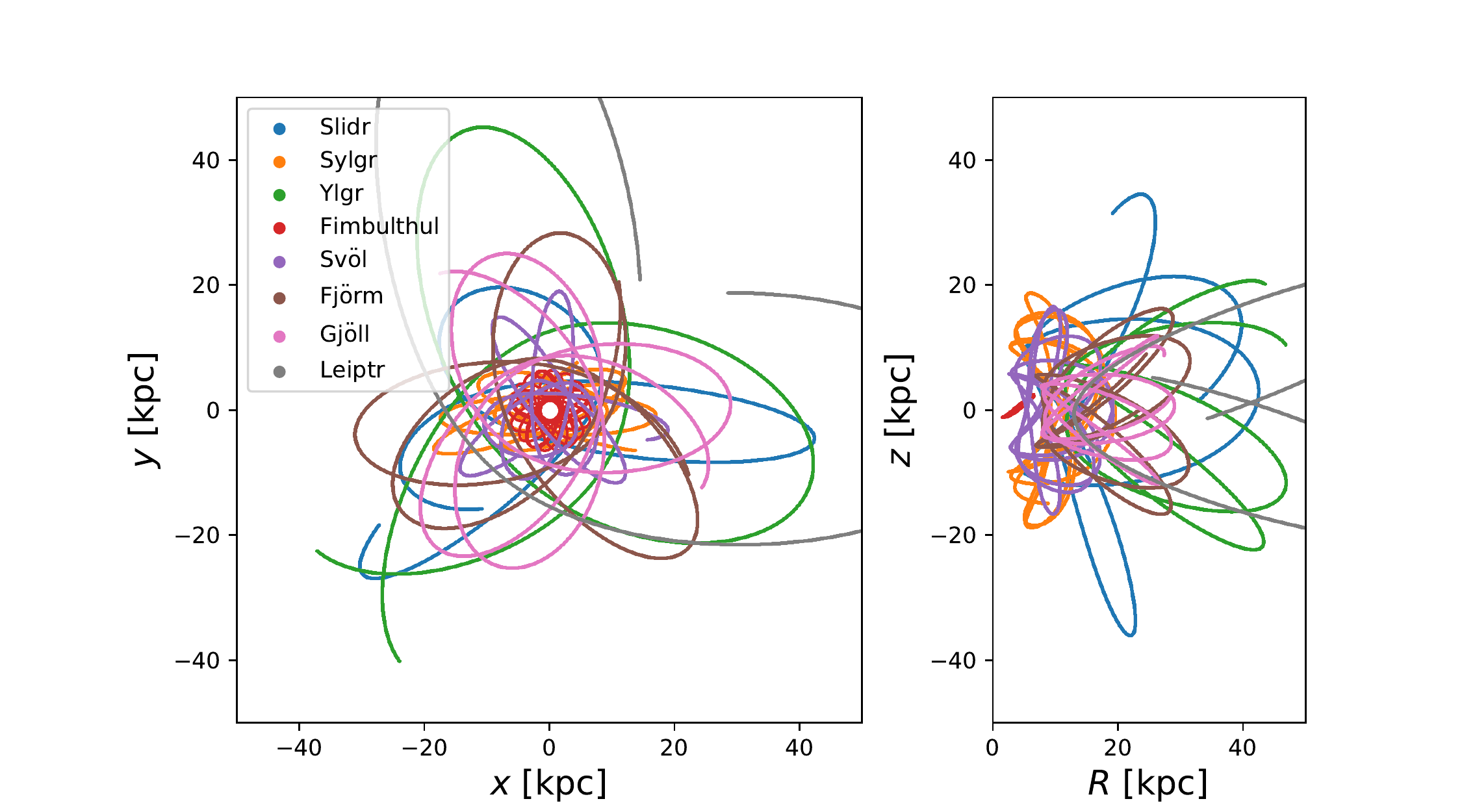}
\end{center}
\caption{Orbits of the eight streams, as derived by the MCMC orbit fitting algorithm. The best-fit orbit solutions presented in Section~\ref{sec:Streams} are shown in the $x$--$y$ plane (left) and the $R$--$z$ plane (right), and are integrated $1\Gyr$ into the past and $1\Gyr$ into the future. The position of the Sun in this coordinate system is $(x,y,z)=(-8.122,0,0.017)\kpc$.}
\label{fig:orbits}
\end{figure*}

The best-fit orbit shows that the structure is prograde ($L_z=-686\pm7\kms \kpc$), with a pericenter at $2.49\pm0.02\kpc$ and an apocenter at $19.5\pm0.5\kpc$. The CMD and orbital properties of the structure are shown with orange points in Figures~\ref{fig:CMDs}--\ref{fig:Lz}.

\subsection{Ylgr}

Figure~\ref{fig:Ylgr} shows the properties of the $30\deg$-long Ylgr stream. It is most clearly visible in the ${\rm [Fe/H]=-2.0}$ map (Figure~\ref{fig:FeH_m2.0}). The sample we identify contains 349 stars, three of which have been previously observed spectroscopically by the SDSS. All three of those stars are metal-poor (${\rm \langle [Fe/H] \rangle=-1.87}$) and the sample displays very small scatter ($0.04$~dex). The most likely stream fit without radial velocity information (dashed line) does not pass through the observed SDSS radial velocity measurements. If we include the radial velocity information into the likelihood, and assuming that the star with $v_{helio}=-53.6\kms$ is a contaminant, we obtain the dotted-line orbital fit. The orbit of the stream is highly retrograde ($L_z=3087\pm36\kms \kpc$), with a pericenter at $10.3\pm0.04\kpc$, and an apocenter at $30.6\pm0.9\kpc$. 

\subsection{Fimbulthul}

Fimbulthul is perhaps the most fascinating of the streams we have identified, and will be discussed in greater detail elsewhere (Ibata et al 2019, under review). A total of 309  stars over a $18\deg$ length of sky are identified as candidate members of this structure by the {\tt STREAMFINDER} algorithm. Two stars of this sample happen to have been measured by the {\it Gaia} RVS. The orbital fit without radial velocity anchors is shown with the dashed line in Figure~\ref{fig:Fimbulthul}. However, since this fit fails to account for the observed radial velocities, we decided to re-fit the structure including the RVS values as constraints. The new fit is shown with the dotted line, and provides a good description of the structure.

The revised orbit model has a pericentric distance of $1.50\pm0.01\kpc$, an apocenter distance of $6.53\pm0.01\kpc$ and is again highly retrograde ($L_z=490\pm1\kms \kpc$). These properties are very close to those of the prototypical massive cluster $\omega$ Centauri, and we show in Ibata et al (2019) that the Fimbulthul stream is indeed the long sought-for tidal tail of that globular cluster.

\subsection{Sv\"ol}

Sv\"ol is an $26\deg$-long stream seen in the northern hemisphere and that appears with highest contrast in the more metal-poor maps. The stars of this structure are displayed in Figure~\ref{fig:Svol}, to which we again applied our MCMC orbit-fitting procedure. Although one star of the sample has been observed by LAMOST, the relatively high metallicity of this object (${\rm [Fe/H]=-1.09\pm0.8}$) suggests that it may be a contaminant. 

Nevertheless, the orbit based purely on Gaia DR2 data is well-constrained: the pericenter lies at $2.53\pm0.13\kpc$, the apocenter is at $19.1\pm1.1\kpc$, and the stream is prograde, with $L_z=-793\pm30\kms \kpc$. This orbit is shown in purple in Figure~\ref{fig:orbits} and its angular momentum and pericenter are compared to other streams and globular clusters in Figure~\ref{fig:Lz}. As can be seen in Figure~\ref{fig:Lz}, and more quantitatively in Table~\ref{tab:properties}, the properties of Sv\"ol are extremely close to those of the Sylgr stream, which suggests the possibility that they are two fragments of a single structure. While we cannot rule out this possibility, we were unable to find trajectories that link the two features in less than one orbital period.

\subsection{Fj\"orm}

Fj\"orm is the longest of the streams presented here: it spans $87\deg$ in the northern Galactic hemisphere and passes close to the north Galactic pole. We identify 148 candidate members with the adopted $8\sigma$ threshold. Seven stars of this sample have spectroscopically-measured radial velocities and metallicities in the SDSS, and a further two in the LAMOST survey. 

In Figure~\ref{fig:Fjorm} we show the orbital fit based purely on the Gaia DR2 measurements. The spectroscopically-observed stars follow the expected velocity trend very well. All of the SDSS stars are very metal-poor, with an average metallicity of ${\rm \langle [Fe/H] \rangle=-2.20}$ and sample dispersion of $0.30$~dex. Of the two LAMOST stars, one has ${\rm [Fe/H]=-2.16\pm0.07}$, in excellent agreement with the SDSS subsample, while the other star is likely a contaminant since it is metal-rich (${\rm [Fe/H]=-0.70\pm0.18}$) and is also a velocity outlier.

The best stream fit has a pericenter at $6.93\pm0.02\kpc$, an apocenter at  $31.4\pm0.6\kpc$, and is prograde with $L_z-2373\pm7\kms \kpc$.

\subsection{Gj\"oll}

At first sight, the Gj\"oll stream (Figure~\ref{fig:Gjoll}) could perhaps be mistaken to be part of the Galactic Anticenter Stellar Structure (GASS), the broadly-distributed stellar population that is detected over most of the anticenter direction \citep{2014ApJ...791....9S} and that is probably the result of the warping and flaring of the outer disk. Our algorithm decomposes the GASS into numerous stream-like filaments. Even though it forms a $25\deg$-long arc within less than $30\deg$ of the Galactic plane in the Anticenter region, the Gj\"oll stream stands out as having very different kinematic properties to the GASS population, and at a Heliocentric distance of $3.4\pm0.1$, it is substantially closer.

In the first submission of the present work, we identified 57 stars as probable Gj\"oll stream members, and the resulting best-fit orbit is shown in Figure~\ref{fig:Gjoll}. One of those stars is present in the SDSS spectroscopic survey (blue triangle) and possesses a radial velocity that agrees well with the orbit prediction based on the earlier Gaia DR2 sample. 

We were awarded observing time at the Canada France Hawaii Telescope (CFHT) in semester 2018B to measure the radial velocities of our stream candidates, and we decided to follow up bright members of the Gj\"oll structure, which is particularly well-located on the sky for observations in October and November. We selected five stars from the original sample for spectroscopic follow-up with the very high-resolution ESPaDOnS instrument at the CFHT observatory. These stars are listed in Table~\ref{tab:CFHT}, where we also provide the date and CFHT odometer identification of these spectroscopic observations.

The data were reduced using the Libre-ESpRIT pipeline \citep{1997MNRAS.291..658D}, and we measured the velocities of the stars by cross-correlation to our observation of the radial velocity template HD182572 using the {\tt fxcor} command in PyRAF.

\begin{table*}
\caption{Gj\"oll stream stars observed with CFHT/ESPaDOnS.}
\label{tab:CFHT}
\begin{tabular}{rrccrcccc}
\hline
\hline
RA J2000 & Dec J2000 & $G_0$ & $(G_{\rm BP}-G_{\rm RP})_0$ & $v_{helio}$ & CFHT & Observation & Exposure & New \\
${\rm [deg]}$ & ${\rm [deg]}$ & ${\rm [mag]}$ & ${\rm [mag]}$ & $[\kms]$ & odometer & date & [s] & sample? \\
\hline
   65.581723 & -0.863916 & 14.026 &  0.974 & $-75.54\pm0.32$ & 2329638 & 2018-10-24 &  600 & N \\
   74.627326 & -6.423376 & 13.157 &  1.041 & $ -6.27\pm0.36$ & 2329640 & 2018-10-24 &  300 & Y \\
   69.793188 & -1.536265 & 15.753 &  0.841 & $-28.28\pm0.80$ & 2335537 & 2018-11-20 & 1200 & N \\
   72.848250 & -6.758777 & 16.396 &  0.645 & $ -9.79\pm1.20$ & 2335539 & 2018-11-20 & 1500 & N \\
   82.104259 &-13.340007 & 15.799 &  0.796 & $ 79.04\pm2.08$ & 2335541 & 2018-11-20 & 1200 & Y \\
\hline
\hline
\end{tabular}
\tablecomments{The positions correspond are the sky coordinates in Gaia DR2 of the sources identified as stream members according to the {\tt STREAMFINDER}. The final column lists whether the star is part of the final conservative sample of 28 stars generated during the revision of the manuscript.}
\end{table*}

With the more conservative parameter choices requested by the reviewers of the present manuscript and of the accompanying study of the Fimbulthul stream, the number of candidate members of Gj\"oll dropped to 28, and the previously cross-matched SDSS star is no longer a member of the new sample, and neither are three of the five stars we observed with ESPaDOnS. We tried to carefully track down the reason why these stars were lost to the final sample, but there are many causes: some have simply a lower stream likelihood than the adopted $8\sigma$ threshold, but others were removed by the updated filtering. Yet all of these stars follow the predicted trend derived from the earlier sample, as can be seen on the fourth panel of Figure~\ref{fig:Gjoll}. The confirmation of our radial velocity profile prediction proves that the algorithm works as intended and is capable of finding very low surface brightness streams, even close to the Galactic disk where the density of contaminating sources is high. 

We have not as yet used the ESPaDOnS spectra to measure the metallicity of the member stars, and we expect this to be challenging due to the very low signal to noise ratio of the spectra. However, we note that the SDSS star in the earlier sample has ${\rm [Fe/H]=-1.78\pm0.05}$, consistent with a metal-poor population.

This structure is strongly retrograde ($L_z=2721\pm159\kms \kpc$), which rules out a connection with the GASS. The orbit has pericenter $8.0\pm0.2\kpc$ and apocenter $31.9\pm4.4\kpc$. As can be seen in Figure~\ref{fig:orbits} (pink line), this stream stays close to the plane of the disk.

\subsection{Leiptr}

The $48\deg$-long Leiptr stream (Figure~\ref{fig:Leiptr}) also lies towards the Galactic anticenter, but is seen at higher Heliocentric distances ($> 5.5\kpc$). The largest sample of candidate members is found using the ${\rm [Fe/H]=-1.6}$ template. One of the stars we identify was also measured by the Gaia RVS, and again, the best-fit orbit passes close to the radial velocity of this star. The orbit has a pericenter distance of $12.8\pm0.6\kpc$, and an apocenter of $85\pm25\kpc$. The strongly retrograde nature of this structure is reflected in its very high $z$-component of angular momentum: $L_z=4689\pm314\kms \kpc$.

\begin{figure}
\begin{center}
\includegraphics[angle=0, width=\hsize]{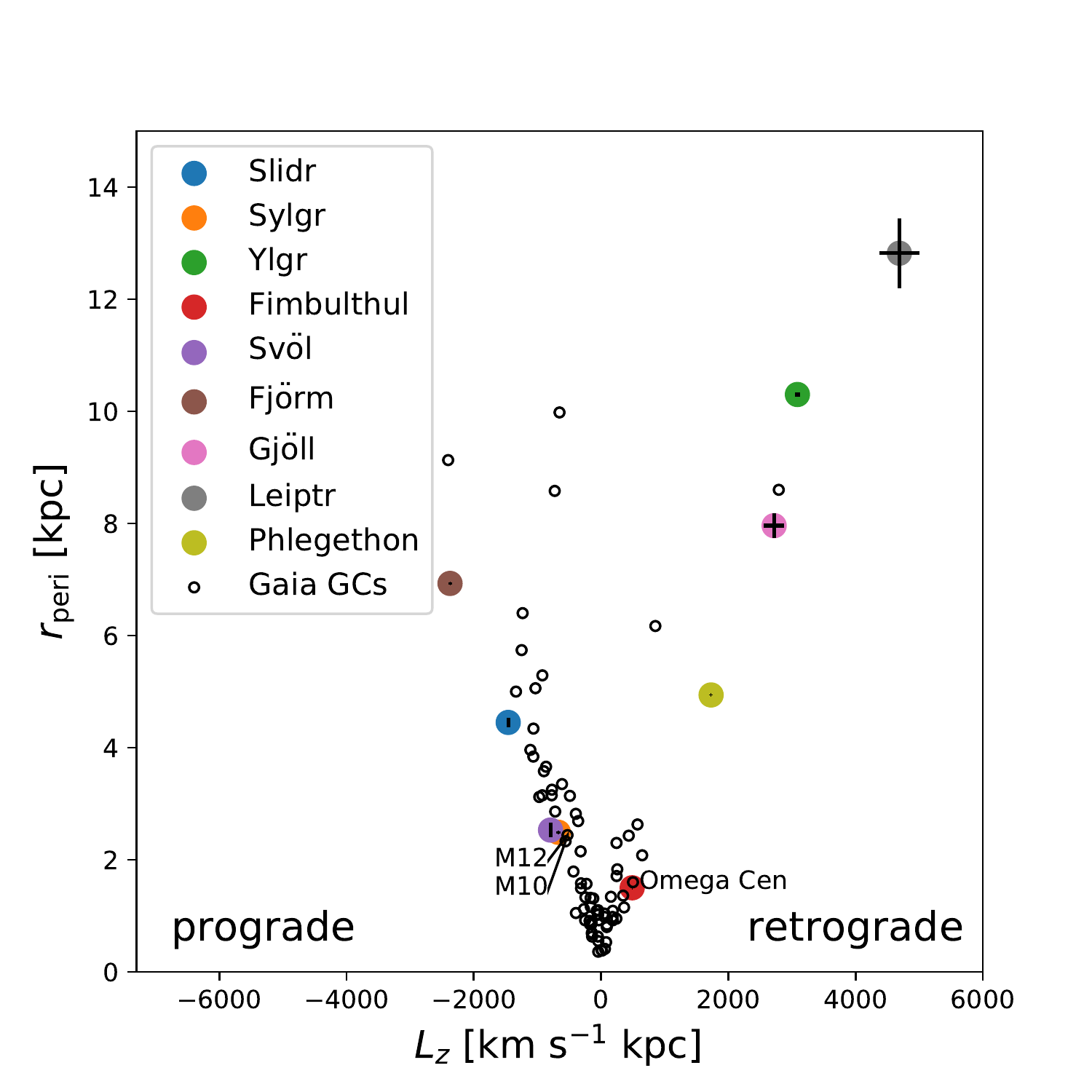}
\end{center}
\caption{The properties of the orbits of the detected streams in $L_z$ vs. pericenter distance. We have included the Phlegethon stream from Paper~III, shown in gray. The small dots represent the orbits of the Galactic globular clusters analysed by \citet{2018A&A...616A..12G}.}
\label{fig:Lz}
\end{figure}

\begin{table}
\caption{Orbital properties of the eight star streams presented here.}
\label{tab:properties}
\begin{tabular}{lcccc}
\hline
\hline
Name & $d$ & $r_{\rm Peri}$ & $r_{\rm Apo}$ & $L_z$ \\
           & $[\kpc]$ & $[\kpc]$ & $[\kpc]$ & $[\kms \kpc]$ \\ 
\hline
Slidr      & $3.65\pm0.09$ & $4.45\pm0.08$ & $42.4\pm3.2$      & $-1458\pm9$ \\
Sylgr      & $4.10\pm0.02$ & $2.49\pm0.02$ & $19.4\pm0.5$      & $ -671\pm7$ \\
Ylgr       & $9.53\pm0.26$ &$10.30\pm0.04$ & $30.6\pm0.9$      &  $3087\pm36$ \\
Fimbulthul & $4.22\pm0.01$ & $1.50\pm0.01$ & $6.53\pm0.01$     &   $490\pm1$ \\
Sv\"ol     & $7.76\pm0.21$ & $2.53\pm0.13$ & $19.1\pm1.1$      &  $-793\pm30$ \\
Fj\"orm    & $4.90\pm0.07$ & $6.93\pm0.02$ & $31.4\pm0.6$      & $-2373\pm7$ \\
Gj\"oll    & $3.38\pm0.10$ & $7.96\pm0.22$ & $31.9\pm4.4$      &  $2721\pm159$ \\
Leiptr     & $7.92\pm0.42$ & $12.8\pm0.6$  & $85.2\pm26$       &  $4689\pm314$ \\
\hline
\hline
\end{tabular}
\tablecomments{The tabulated $d$ is the heliocentric distance at the anchor point of the fit.}
\end{table}

\section{Discussion and Conclusions}
\label{sec:Discussion}

In this contribution, we have provided a detailed exposition of the workings of the updated {\tt STREAMFINDER} algorithm, which was used previously to detect 5 streams in Paper~II ({\it Gaia} 1--5) and the Phlegethon stream (Paper~III). Part of the necessary validation process consisted of building a Gaia-like mock catalog that is devoid of stream structures, based on the Gaia Universe Model Snapshot (GUMS) simulation. This mock was processed in an identical way to the real Gaia DR2 data. We thereby showed that we can limit the occurrence of false positives by setting a detection threshold based on our stream likelihood statistic.

In our stream search we used template stellar populations models of ancient metal-poor stars appropriate for tidal debris of disrupted globular clusters. Due to  computational expense, a single age ($12.5\Gyr$), and only three choices in metallicity (${\rm [Fe/H] = -1.4,-1.6}$ and $-2.0$) were used.  This rather coarse grid in metallicity will be refined in future work, and we will also explore different age choices. 

We detected a large number of candidate stream structures in the Gaia DR2 catalog between distances of $1$ and $10\kpc$. Many of these have only a small number of members ($\sim 10$) that pass the selected $8\sigma$ detection threshold, but they nevertheless appear to be plausible stream candidates. Examples of such cases include the features seen near ($\ell\sim-60\deg$, $b\sim 45\deg$) in Figure~\ref{fig:FeH_m1.6}. We also find that the most metal-rich map (Figure~\ref{fig:FeH_m1.4}) is highly structured around the bulge region and close to the disk, showing many dozens of stream-like features. Comparison to the false positives map derived from our processing of the GUMS simulation (Figure~\ref{fig:FeH_m1.4_GUMS}) suggests that some, but probably not most, of these features are artefacts. Follow-up spectroscopy is now needed to establish the reality of those candidates.

We selected eight high-significance streams (Slidr, Sylgr, Ylgr, Fimbulthul, Sv\"ol, Fj\"orm, Gj\"oll and Leiptr) out of our candidates, all of which contained (by chance) a small number of stars with extant radial velocity measurements. In five cases, our orbital models constructed from the sky position, proper motion and parallax information made predictions for the radial velocity trends of the streams that were close to the actual radial velocity measurements. In the two cases (Ylgr and Fimbulthul) where the default model failed, including the measured radial velocities in the fit produced a reasonable representation of the data. We cannot yet tell whether the radial velocity prediction for Sv\"ol is correct or not, as the single spectroscopically-observed star in that stream is very likely a contaminant.

We were also able to secure follow-up spectroscopy for the Gj\"oll stream using the ESPaDOnS high-resolution spectrograph at the CFHT. Of the 5 Gj\"oll stream star candidates that we observed based on a sample constructed before the first submission of the present work, all appear to be genuine stream members. This demonstrates the predictive power of the {\tt STREAMFINDER} algorithm. However, with the more stringent criteria imposed during the revision of this work, only two of the Gj\"oll stars observed with ESPaDOnS are retained in our final $8\sigma$ sample. This suggests that the present algorithm parameter settings are conservative, and that we will be able to find many more good stream candidates at lower detection thresholds.

We now switch to a brief discussion of the properties of the stream population. It is natural to include Phlegethon and the retrograde GD-1 stream \citep{2010ApJ...712..260K} into the current sample of streams, as they are also detected by our algorithm at $>8\sigma$ in the same $1$--$10\kpc$ distance range as the eight high-confidence  streams reported here (Jhelum is more distant at $\sim 13\kpc$, \citealt{2018ApJ...862..114S}).  It is perhaps initially surprising to notice that six out of these ten systems have retrograde orbits. In contrast, most globular clusters, which after all are the progenitors of such streams, have prograde orbits, as can be seen in Figure~\ref{fig:Lz}. We suspect that this overabundance of retrograde streams is due to a bias of our kinematic method, because prograde streams have lower contrast over the contaminating field star population. If this diagnosis is correct, as the astrometric accuracy of the Gaia data improve over the coming years, we should be able to pick out the prograde streams more and more cleanly. This argument would suggest that there should be many tens of streams with prograde orbits waiting to be detected out to the same distance as the retrograde streams found here (i.e. $\sim10\kpc$). Of course, it is also possible that dynamical effects such as interactions with spiral arms and the bar have preferentially perturbed the prograde streams, phase-mixing them more rapidly and hence rendering them undetectable with our method.

Most of the streams that we have identified do not have any obvious progenitor in the sense that there is no surviving globular cluster with the same pericenter, apocenter and $L_z$ (see Figure~\ref{fig:Lz}). The notable exception is Fimbulthul, which as we will discuss at greater length in an accompanying contribution, is the trailing arm of the tidal stream of $\omega$ Centauri. Figure~\ref{fig:Lz} shows that the Sylgr and Sv\"ol streams have similar pericenter and $L_z$ to the globular clusters M10 and M12; however, the apocenters of these clusters are $5.17\pm0.07\kpc$ and $4.88\pm0.04\kpc$, respectively, and so are inconsistent with being trivially related to Sylgr and Sv\"ol whose apocenters are $19.4\pm0.5\kpc$ and $19.1\pm1.1$, respectively. What seems more likely is that Sylgr and Sv\"ol are fragments of a now completely disrupted progenitor, and it is interesting to note in this context that the member stars of Sylgr are very metal-poor, and include an SDSS star with metallicity ${\rm [Fe/H]=-3.10\pm0.08}$. 

Our analysis is beginning to reveal the accretion events that contributed to the current state of our Milky Way, but that are now largely dissolved and as a consequence possess extremely low surface brightness. These streams are only detectable thanks to the exquisite astrometric information provided by Gaia DR2, which allows us to pick them out as coherently moving groups over vast swathes of sky. It is interesting to note that some metal-poor globular clusters may have formed prior to reionization \citep[see, e.g.,][and references therein]{2006ARA&A..44..193B}, and so mapping the properties of their stream remnants may allow us a new handle to probe the physics of that era. 

Follow-up studies to measure the radial velocities of the member stars are now essential to be able to employ these structures to measure the Galactic potential, and explore the their implications for the dark matter. Fj\"orm, the longest of these features, spans an $87\deg$-long arc in the northern sky, and on its own should be a powerful probe of the Galactic potential. With the continued improvements in the accuracy and depth of Gaia data expected over the coming years we foresee being able to greatly extend the detection horizon for faint stellar streams from ancient low-mass progenitors. Thus the prospects for this field appear very bright.

\acknowledgments

This work has made use of data from the European Space Agency (ESA) mission {\it Gaia} (\url{https://www.cosmos.esa.int/gaia}), processed by the {\it Gaia} Data Processing and Analysis Consortium (DPAC, \url{https://www.cosmos.esa.int/web/gaia/dpac/consortium}). Funding for the DPAC has been provided by national institutions, in particular the institutions participating in the {\it Gaia} Multilateral Agreement. 

We thank the staff of the Canada-France-Hawaii Telescope for taking the ESPaDOnS data used here, and for their continued support throughout the project. Based on observations obtained at the Canada-France-Hawaii Telescope (CFHT) which is operated by the National Research Council of Canada, the Institut National des Sciences de l'Univers of the Centre National de la Recherche Scientique of France, and the University of Hawaii.

Funding for SDSS-III has been provided by the Alfred P. Sloan Foundation, the Participating Institutions, the National Science Foundation, and the U.S. Department of Energy Office of Science. The SDSS-III web site is http://www.sdss3.org/.

SDSS-III is managed by the Astrophysical Research Consortium for the Participating Institutions of the SDSS-III Collaboration including the University of Arizona, the Brazilian Participation Group, Brookhaven National Laboratory, Carnegie Mellon University, University of Florida, the French Participation Group, the German Participation Group, Harvard University, the Instituto de Astrofisica de Canarias, the Michigan State/Notre Dame/JINA Participation Group, Johns Hopkins University, Lawrence Berkeley National Laboratory, Max Planck Institute for Astrophysics, Max Planck Institute for Extraterrestrial Physics, New Mexico State University, New York University, Ohio State University, Pennsylvania State University, University of Portsmouth, Princeton University, the Spanish Participation Group, University of Tokyo, University of Utah, Vanderbilt University, University of Virginia, University of Washington, and Yale University.

Guoshoujing Telescope (the Large Sky Area Multi-Object Fiber Spectroscopic Telescope LAMOST) is a National Major Scientific Project built by the Chinese Academy of Sciences. Funding for the project has been provided by the National Development and Reform Commission. LAMOST is operated and managed by the National Astronomical Observatories, Chinese Academy of Sciences.

PyRAF is a product of the Space Telescope Science Institute, which is operated by AURA for NASA.
 
This work has been published under the framework of the IdEx Unistra and benefits from a funding from the state managed by the French National Research Agency as part of the investments for the future program. RAI and NFM gratefully acknowledge support from a ``Programme National Cosmologie et Galaxies'' grant.


\begin{table*}
\caption{Stream candidate stars with measured radial velocity in the SDSS, LAMOST and Gaia-RVS catalogues.}
\label{tab:RVs}
\begin{tabular}{rrccrcccc}
\hline
\hline
RA J2000 & Dec J2000 & $G_0$ & $(G_{\rm BP}-G_{\rm RP})_0$ & $v_{helio}$ & ${\rm [Fe/H]}$ & Source & Stream & Probable\\
${\rm [deg]}$ & ${\rm [deg]}$ & ${\rm [mag]}$ & ${\rm [mag]}$ & $[\kms]$ & & & & member?\\
\hline
  166.127838 &  4.167013 & 17.722 &  0.746 &$ -98.05\pm 4.17$& $-1.78$ & S & Slidr & Y \\
  166.144402 &  4.184200 & 17.093 &  0.615 &$ -93.08\pm 3.04$& $-1.88$ & S & Slidr & Y \\
  166.883060 &  4.226261 & 17.413 &  0.758 &$  11.81\pm 2.45$& $-1.46$ & S & Slidr & N \\
  171.380846 &  4.785477 & 16.401 &  0.677 &$ 246.79\pm 3.27$& $-1.60$ & S & Slidr & N \\
  160.048762 &  6.708863 & 15.948 &  0.652 &$ -74.18\pm 2.93$& $-1.83$ & S & Slidr & Y \\
  164.598178 &  7.083380 & 16.256 &  0.691 &$  76.21\pm 2.91$& $-1.57$ & S & Slidr & N \\
  171.452492 &  8.174430 & 16.485 &  0.689 &$ 272.06\pm 2.58$& $-1.13$ & S & Slidr & N \\
  165.533067 &  9.028086 & 16.621 &  0.634 &$ -94.34\pm 4.21$& $-1.86$ & S & Slidr & Y \\
  157.866653 &  9.050309 & 17.496 &  0.760 &$ -58.27\pm 3.85$& $-1.72$ & S & Slidr & Y \\
  153.073899 & 13.038558 & 17.095 &  0.711 &$ -41.80\pm 3.20$& $-1.71$ & S & Slidr & Y \\
  176.467662 &  2.183338 & 12.846 &  1.083 &$ 101.12\pm 7.56$& $-1.20$ & L & Slidr & N \\
  158.456446 &  5.152103 & 14.861 &  0.892 &$ -64.11\pm10.05$& $-1.71$ & L & Slidr & Y \\
  171.674974 &  5.826564 & 16.783 &  0.640 &$ 213.89\pm17.43$& $-1.68$ & L & Slidr & N \\
  169.298167 &  6.306373 & 13.840 &  0.985 &$-116.06\pm 9.26$& $-1.69$ & L & Slidr & Y \\
  160.048762 &  6.708863 & 15.948 &  0.652 &$ -69.98\pm17.13$& $-2.13$ & L & Slidr & Y \\
  164.515970 &  6.835322 & 12.336 &  1.074 &$ -99.75\pm10.14$& $-1.77$ & L & Slidr & Y \\
  169.580920 &  7.111003 & 17.198 &  0.673 &$-140.68\pm15.56$& $-1.74$ & L & Slidr & Y \\
  171.452492 &  8.174430 & 16.485 &  0.689 &$ 264.07\pm11.84$& $-1.37$ & L & Slidr & N \\
  165.351172 &  8.748006 & 17.658 &  0.730 &$ 233.61\pm18.56$& $-1.38$ & L & Slidr & N \\
  166.425204 &  9.396366 & 14.529 &  0.927 &$  18.71\pm 4.90$& $-0.25$ & L & Slidr & N \\
  165.450935 & 10.228027 & 15.254 &  0.857 &$-108.02\pm12.99$& $-1.76$ & L & Slidr & Y \\
  153.969672 & 13.511505 & 15.326 &  0.742 &$ -55.47\pm17.18$& $-1.95$ & L & Slidr & Y \\
  152.074217 & 14.758481 & 15.889 &  0.621 &$ -54.09\pm16.97$& $-2.00$ & L & Slidr & Y \\
  149.086722 & 16.938353 & 17.467 &  0.803 &$ -43.31\pm12.45$& $-1.48$ & L & Slidr & Y \\
  176.467662 &  2.183338 & 12.833 &  1.076 & $108.73\pm1.90$ &         & G & Slidr & N \\
  164.515970 &  6.835322 & 12.321 &  1.067 & $-93.26\pm2.16$ &         & G & Slidr & Y \\
\hline
  180.587088 & -0.344169 & 17.257 &  0.630 &$-208.56\pm 5.12$& $-2.38$ & S & Slygr & Y \\
  182.320384 & -0.241553 & 18.375 &  0.740 &$-184.84\pm 5.81$& $-2.58$ & S & Slygr & Y \\
  182.105642 &  0.411199 & 17.473 &  0.608 &$-205.63\pm 3.28$& $-3.10$ & S & Slygr & Y \\
\hline
  169.596593 &-12.042540 & 15.359 &  0.979 &$ 327.50\pm 1.91$& $-1.88$ & S & Ylgr & Y \\
  168.852187 &-11.994377 & 17.616 &  0.818 &$ 322.58\pm 5.34$& $-1.92$ & S & Ylgr & Y \\
  169.349924 &-11.958677 & 16.684 &  0.733 &$ -53.65\pm 3.05$& $-1.82$ & S & Ylgr & N \\
\hline
  200.291401 &-26.516432 & 12.131 &  1.164 & $207.99\pm1.17$ &         & G & Fimbulthul & Y \\
  202.470324 &-24.439468 & 12.716 &  1.058 & $191.37\pm2.62$ &         & G & Fimbulthul & Y \\
\hline
  239.801474 & 26.730392 & 17.458 &  0.885 &$ -31.06\pm62.67$& $-1.08$ & S & Sv\"ol & N \\
\hline
  200.725566 & 23.984631 & 19.061 &  0.754 &$ -70.33\pm12.53$& $-2.09$ & S & Fj\"orm & Y \\
  199.536720 & 24.934613 & 18.282 &  0.735 &$-104.11\pm 5.18$& $-2.07$ & S & Fj\"orm & Y \\
  201.073852 & 32.632881 & 18.800 &  0.722 &$ -94.45\pm 7.99$& $-1.67$ & S & Fj\"orm & Y \\
  206.874238 & 40.968247 & 18.674 &  0.667 &$ -68.97\pm 7.67$& $-2.07$ & S & Fj\"orm & Y \\
  209.218215 & 41.462402 & 18.503 &  0.676 &$ -69.67\pm 6.61$& $-2.61$ & S & Fj\"orm & Y \\
  208.598025 & 42.246064 & 18.555 &  0.696 &$ -84.00\pm 6.50$& $-2.44$ & S & Fj\"orm & Y \\
  207.328617 & 42.609333 & 18.350 &  0.682 &$ -78.93\pm 7.64$& $-2.45$ & S & Fj\"orm & Y \\
  195.329241 &  4.878705 & 16.970 &  0.839 &$  -9.97\pm 8.76$& $-0.70$ & L & Fj\"orm & N \\
  198.416415 & 13.004091 & 15.508 &  0.913 &$ -76.30\pm14.25$& $-2.16$ & L & Fj\"orm & Y \\
\hline
  72.0846775 & -5.176483 & 18.721 &  0.859 & $-15.18\pm4.38$ & $-1.78$ & S & Gj\"oll & 
  Y \footnote{Although a likely stream member, changes to the algorithm made during the revision of this work caused this star to no longer appear in the $8\sigma$ stream candidate sample.} \\
\hline
  89.1120223 &-28.189130 & 12.241 &  1.259 & $190.98\pm0.58$ &         & G & Leiptr & Y \\
\hline
\hline
\end{tabular}
\tablecomments{The Source column lists the provenance of the radial velocity and metallicity information: S=SDSS, L=LAMOST and G=Gaia RVS. The positions correspond to the sky coordinates in Gaia DR2 of the sources identified as stream members according to the {\tt STREAMFINDER}. Cross-matches with SDSS and LAMOST were performed with a $1\scnp$ search radius.}
\end{table*}

\newpage
\section*{Appendix -- Maps of false positives in GUMS}

\begin{figure*}
\begin{center}
\vbox{
\hbox{
\includegraphics[angle=0, viewport= 55 23 600 589, clip, height=9cm]{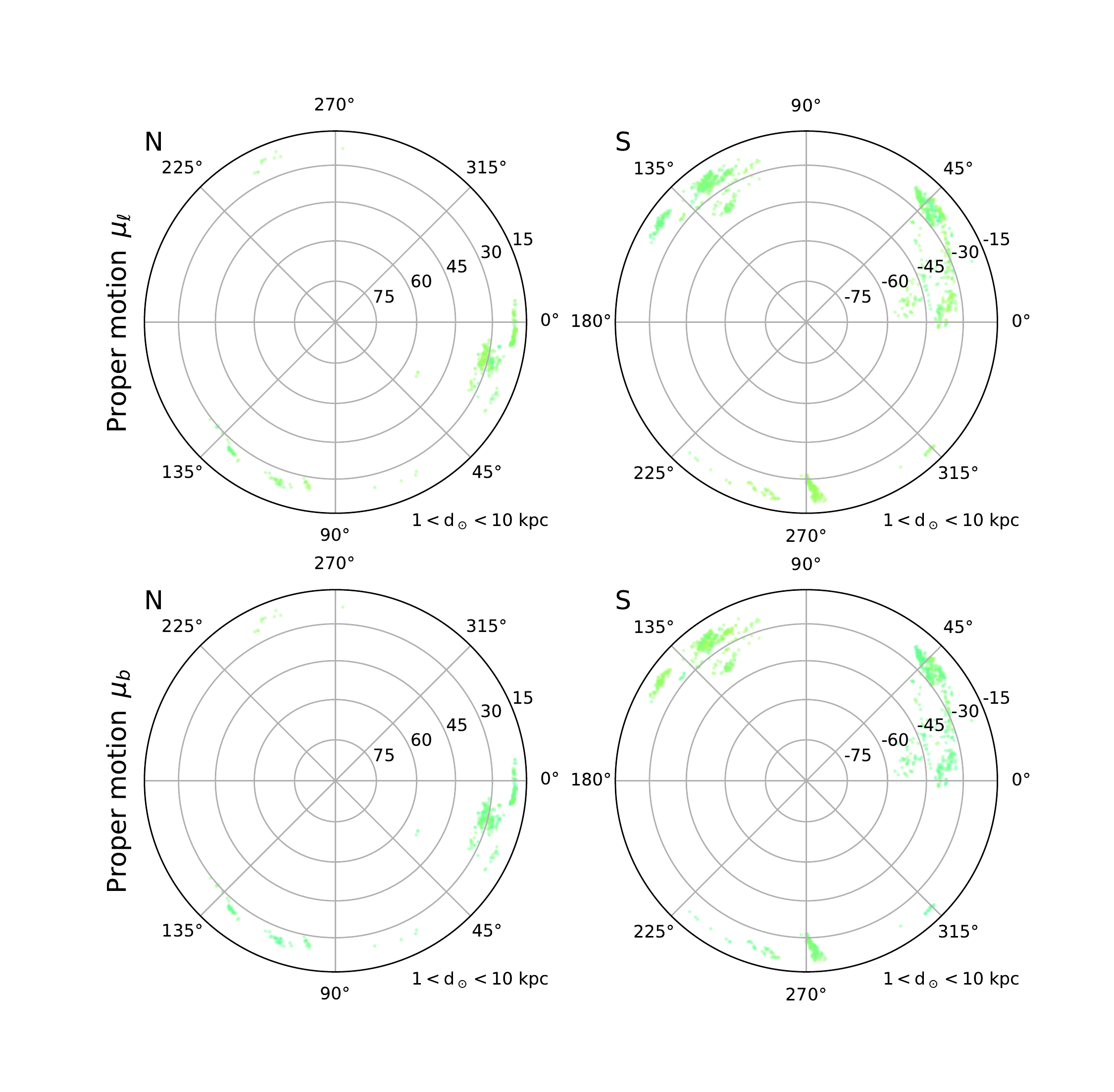}
\includegraphics[angle=0, viewport= 45 45 657 650, clip, height=9cm]{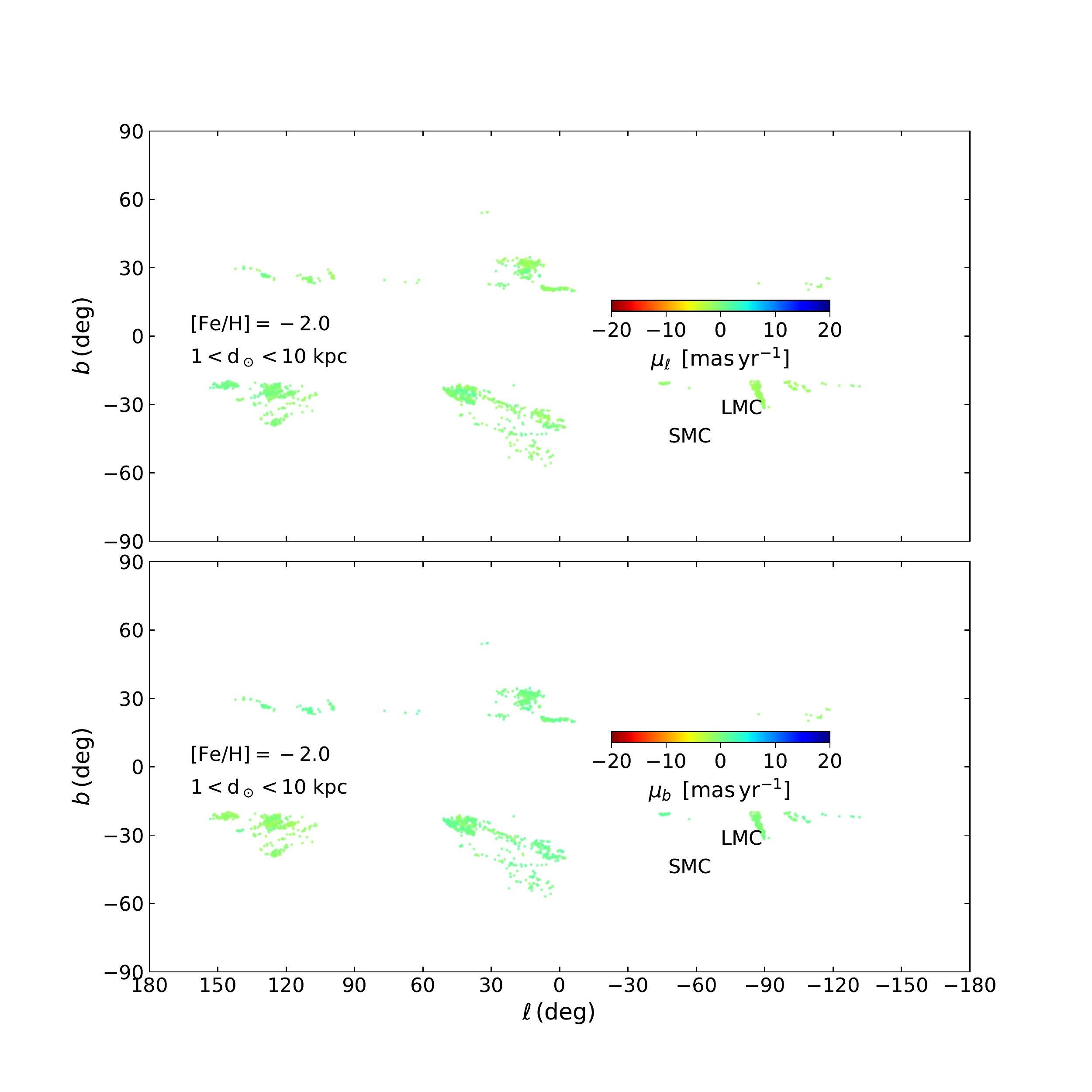}
}
\hbox{
\includegraphics[angle=0, viewport= 55 23 600 589, clip, height=9cm]{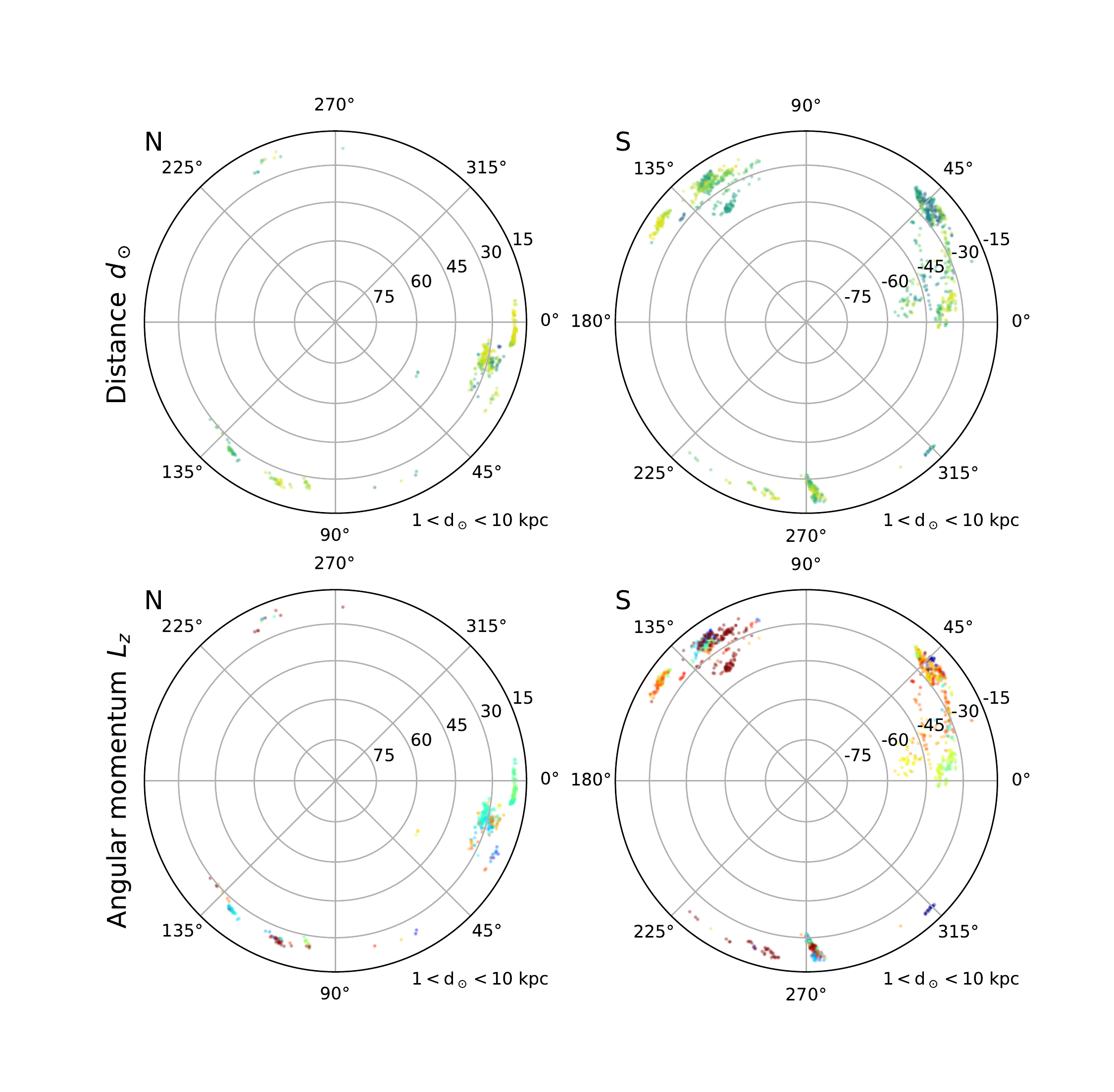}
\includegraphics[angle=0, viewport= 45 45 657 650, clip, height=9cm]{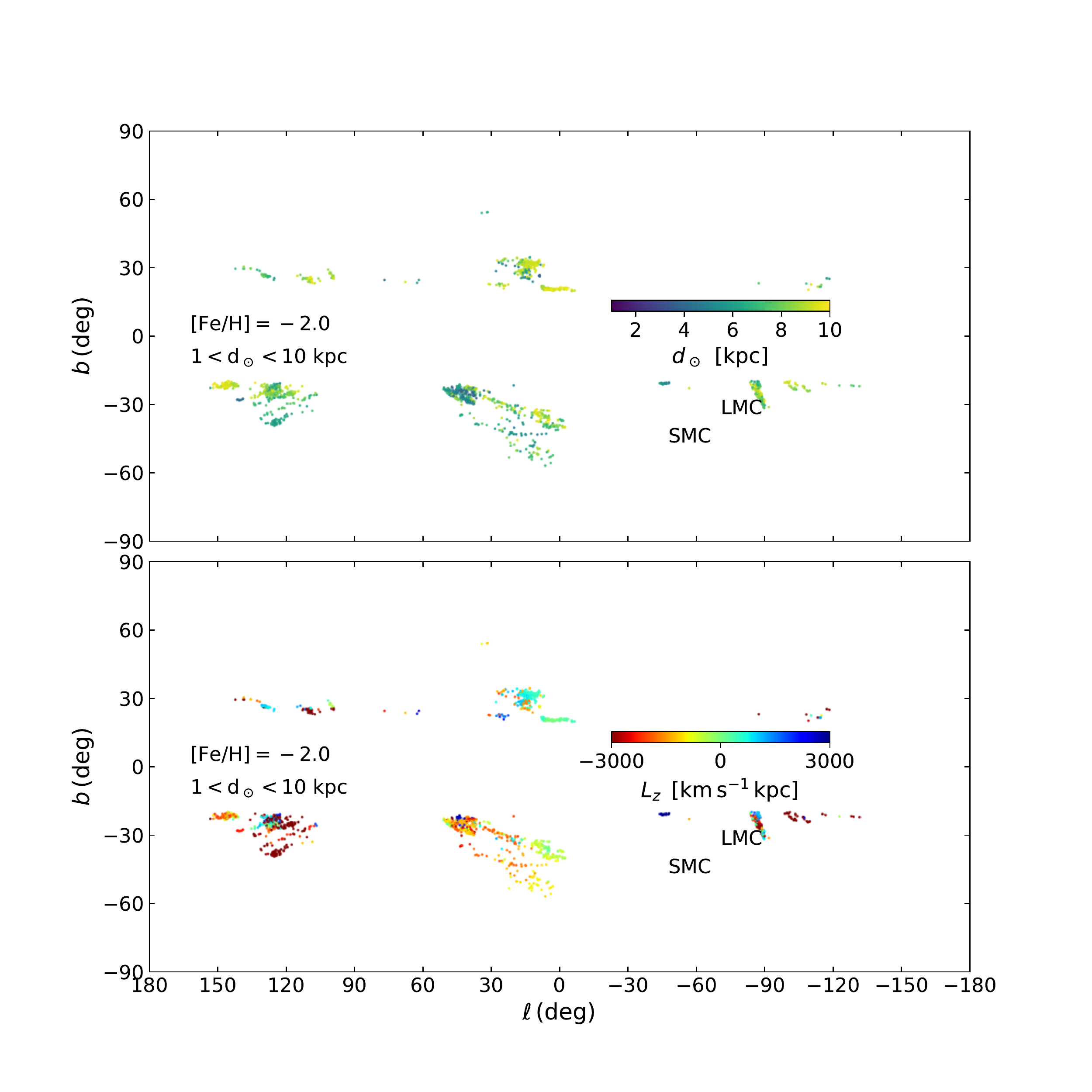}
}
}
\end{center}
\caption{Map of false positive stream-like detections with low proper motion ($\sqrt{\mu^2_\ell+\mu^2_b}<2\masyr$), derived by applying the {\tt STREAMFINDER} to the GUMS simulation. The layout of the panels is identical to that of Figure~\ref{fig:FeH_m2.0_fluff}.}
\label{fig:FeH_m2.0_fluff_GUMS}
\end{figure*}

\begin{figure*}
\begin{center}
\vbox{
\hbox{
\includegraphics[angle=0, viewport= 55 23 600 589, clip, height=9cm]{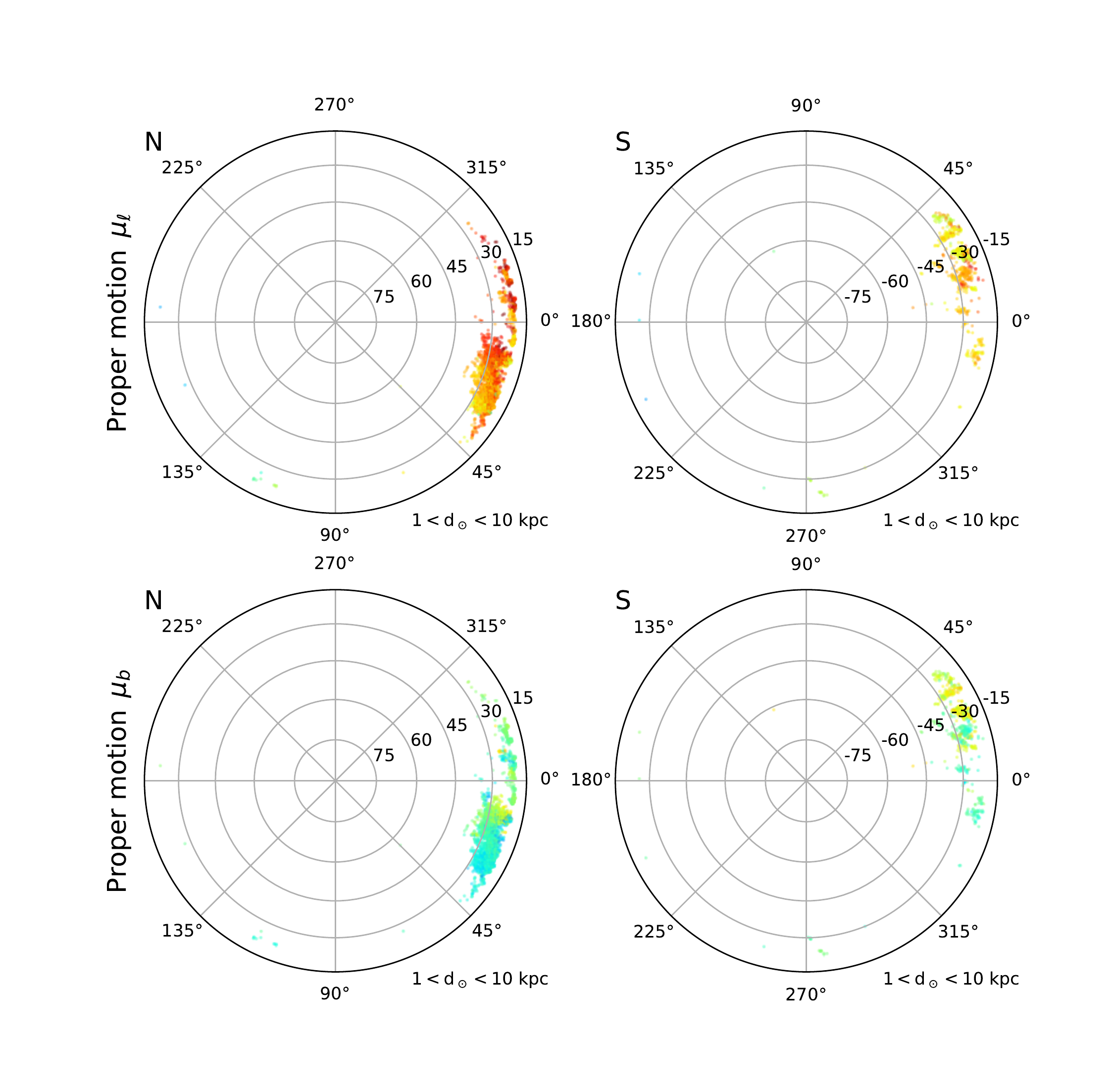}
\includegraphics[angle=0, viewport= 45 45 657 650, clip, height=9cm]{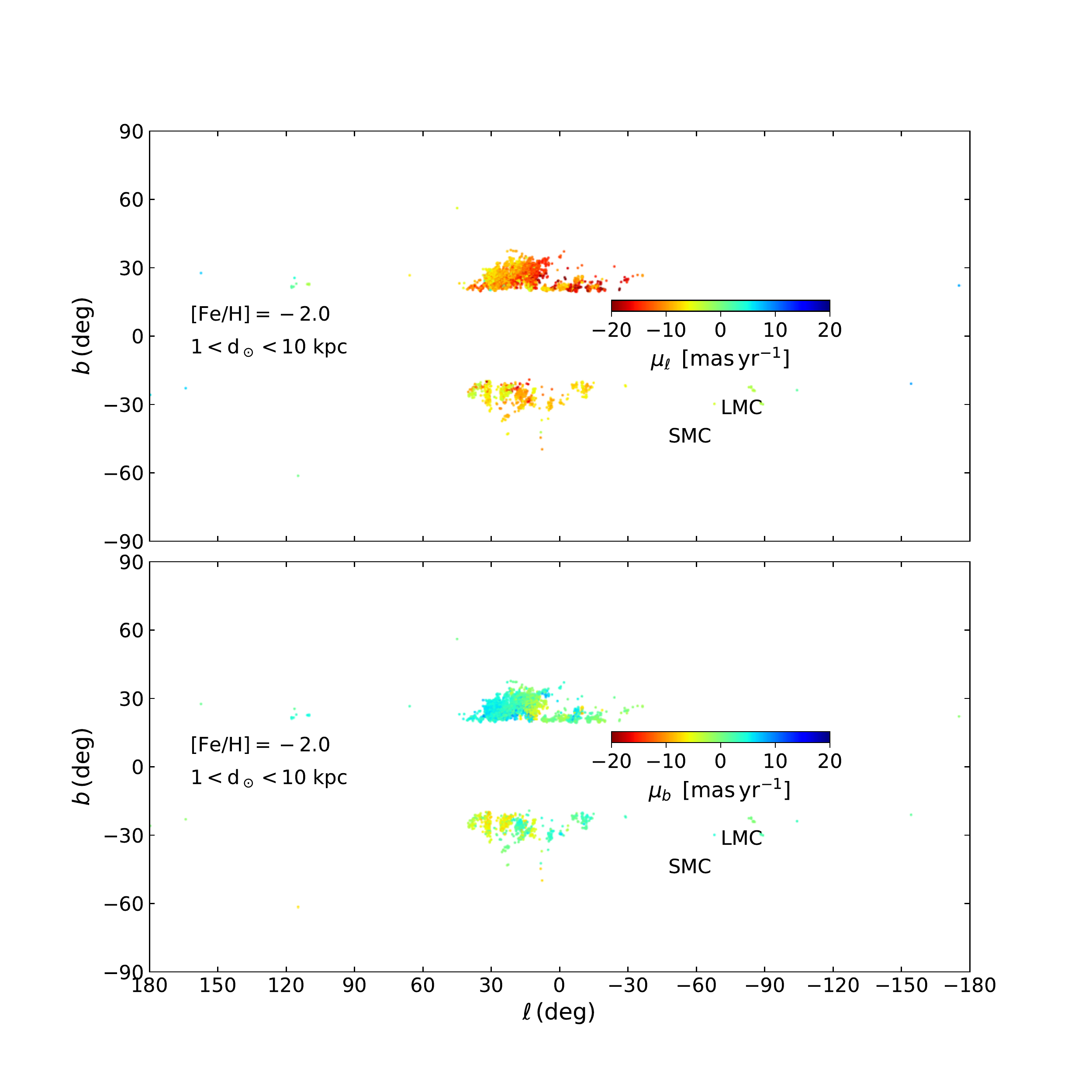}
}
\hbox{
\includegraphics[angle=0, viewport= 55 23 600 589, clip, height=9cm]{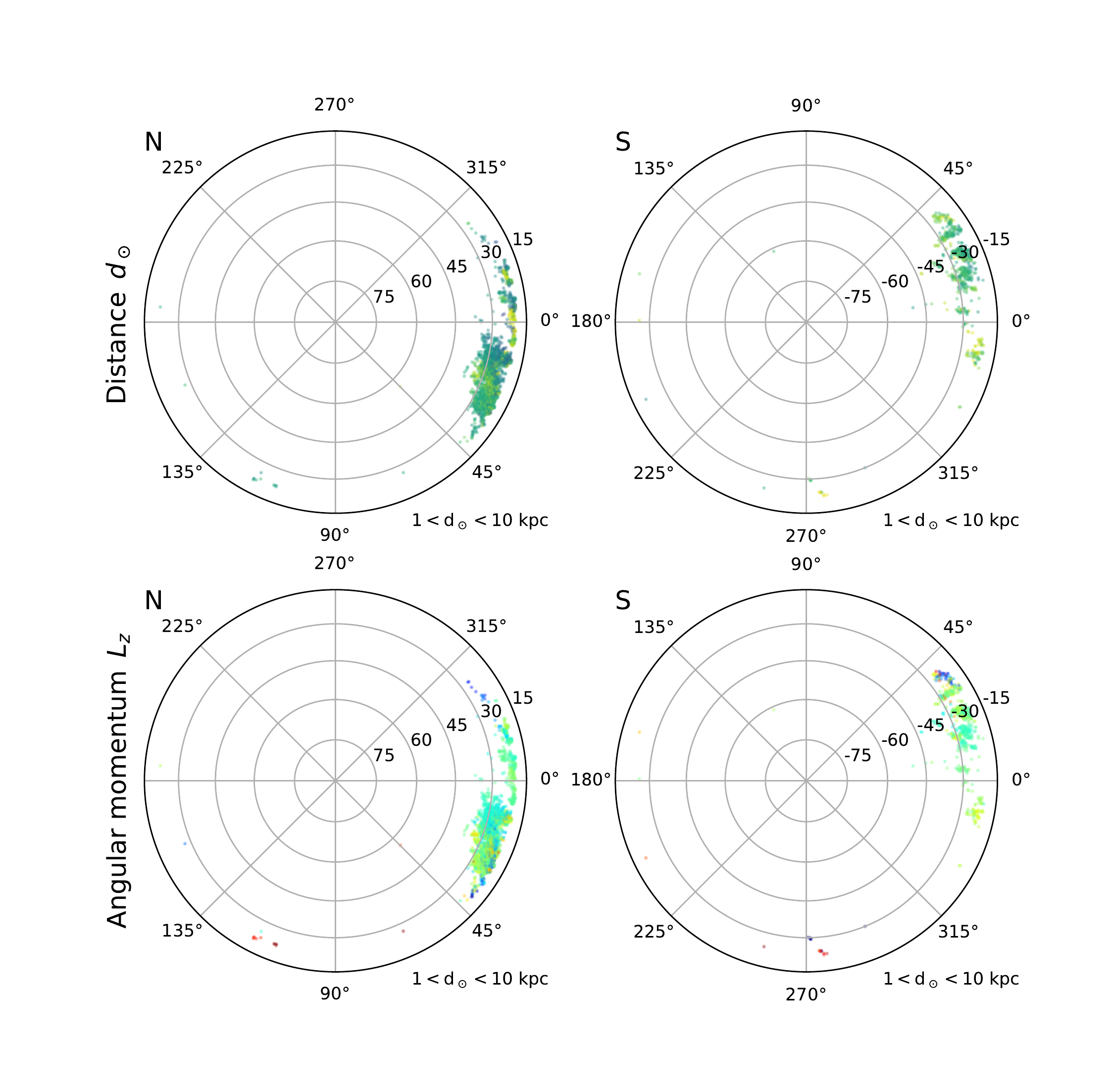}
\includegraphics[angle=0, viewport= 45 45 657 650, clip, height=9cm]{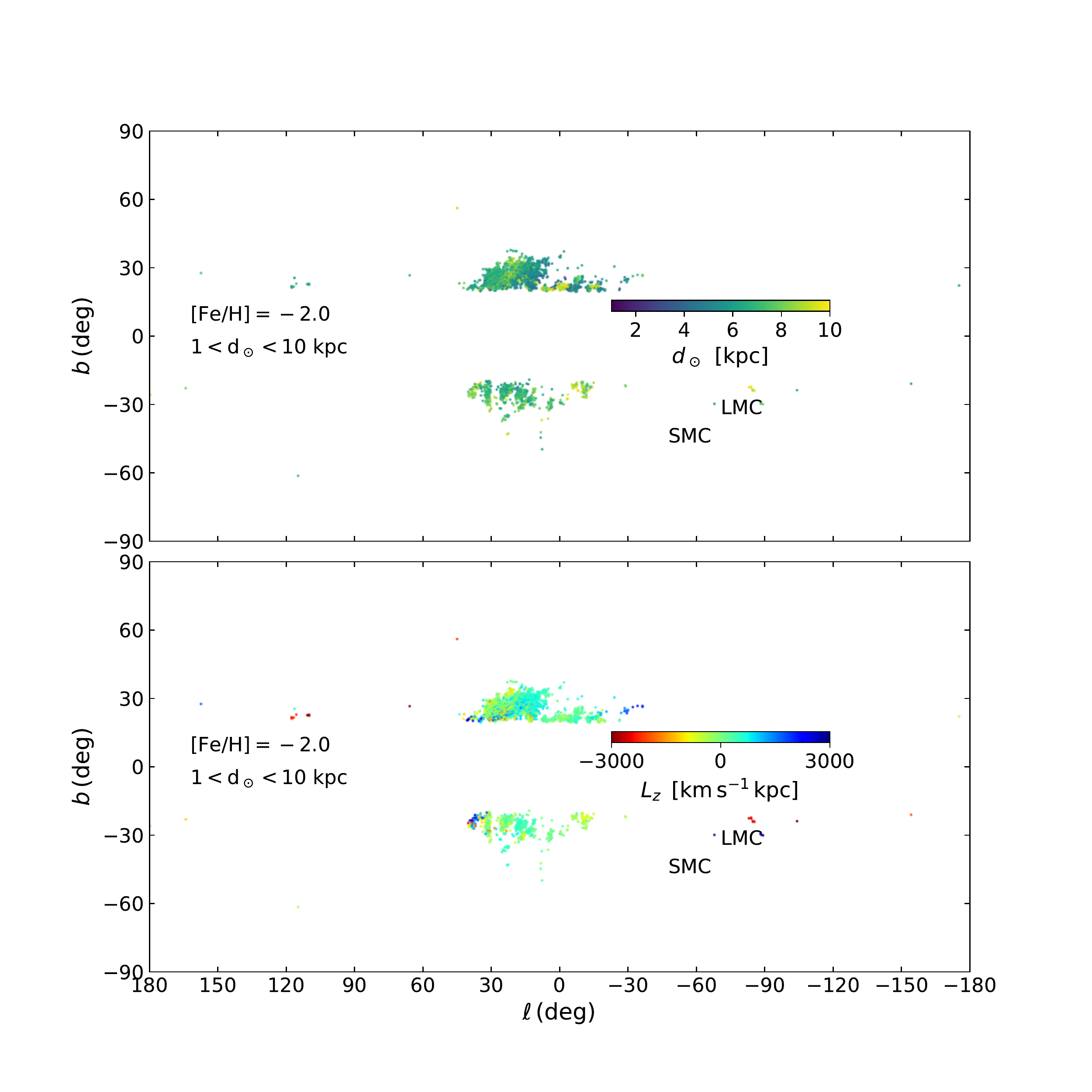}
}
}
\end{center}
\caption{Map of false positive stream-like detections in the GUMS simulation, using an SSP template with ${\rm [Fe/H]=-2.0}$. The layout of the panels is identical to that of Figure~\ref{fig:FeH_m2.0}, to which this figure should be compared.}
\label{fig:FeH_m2.0_GUMS}
\end{figure*}

\begin{figure*}
\begin{center}
\vbox{
\hbox{
\includegraphics[angle=0, viewport= 55 23 600 589, clip, height=9cm]{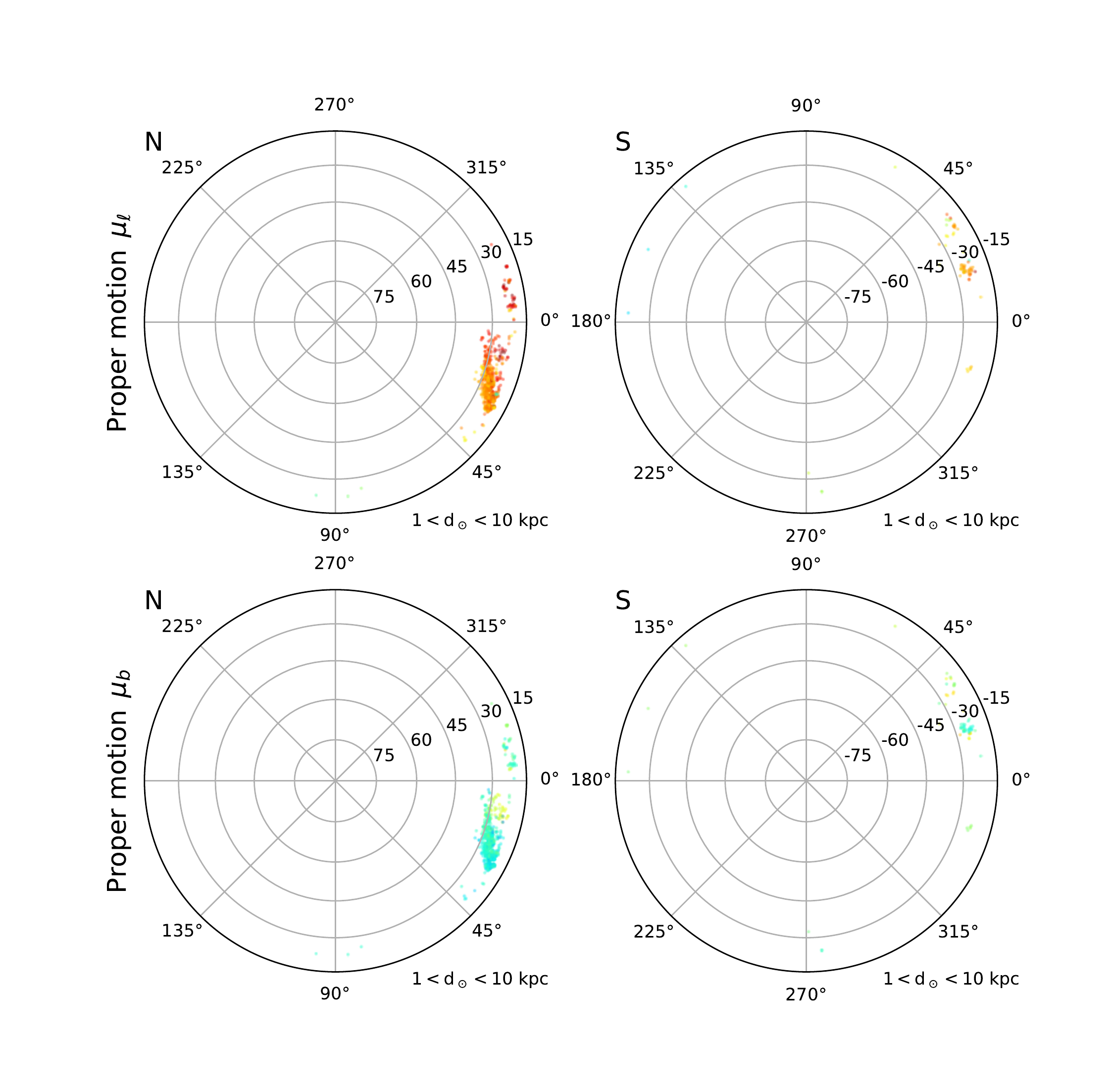}
\includegraphics[angle=0, viewport= 45 45 657 650, clip, height=9cm]{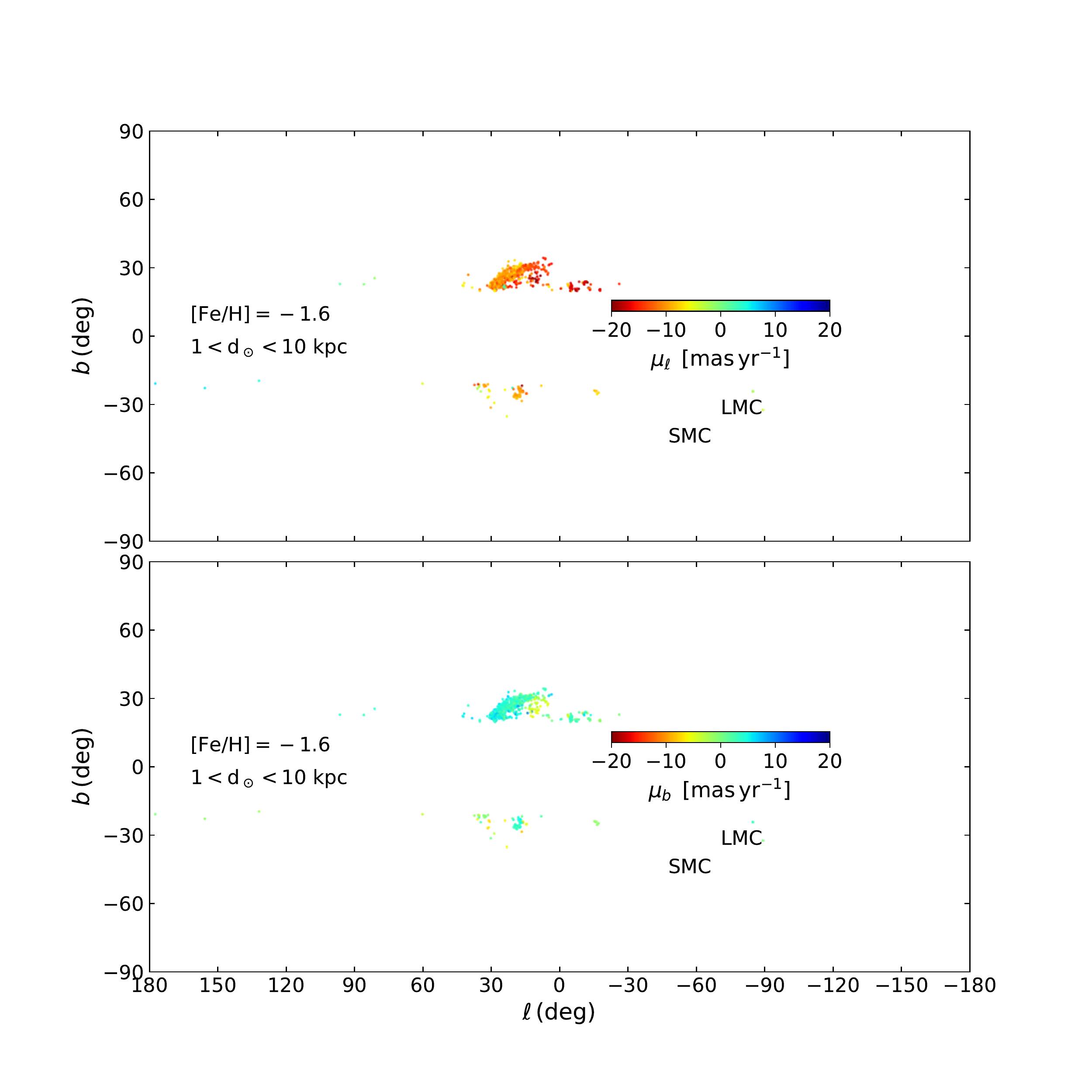}
}
\hbox{
\includegraphics[angle=0, viewport= 55 23 600 589, clip, height=9cm]{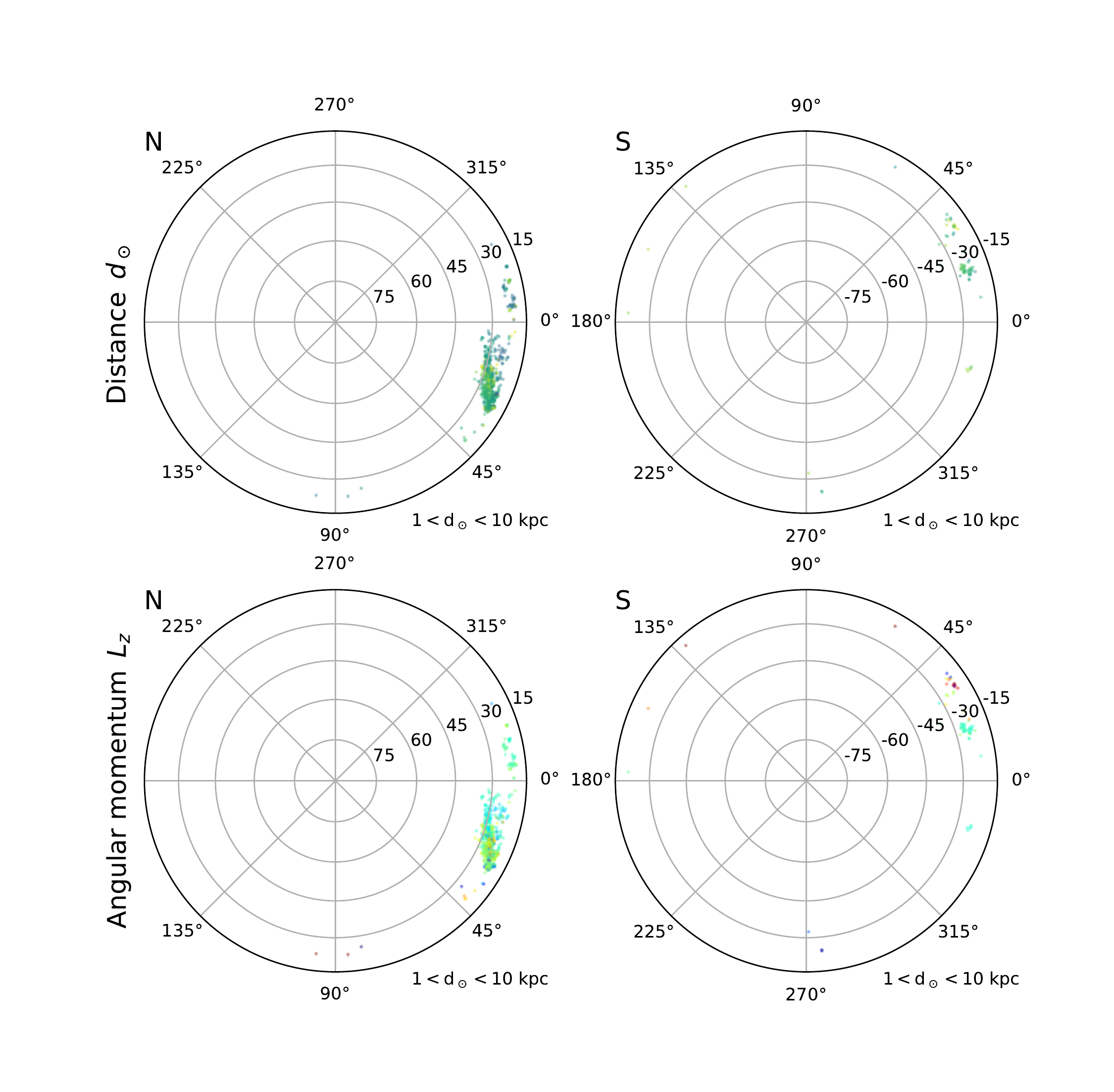}
\includegraphics[angle=0, viewport= 45 45 657 650, clip, height=9cm]{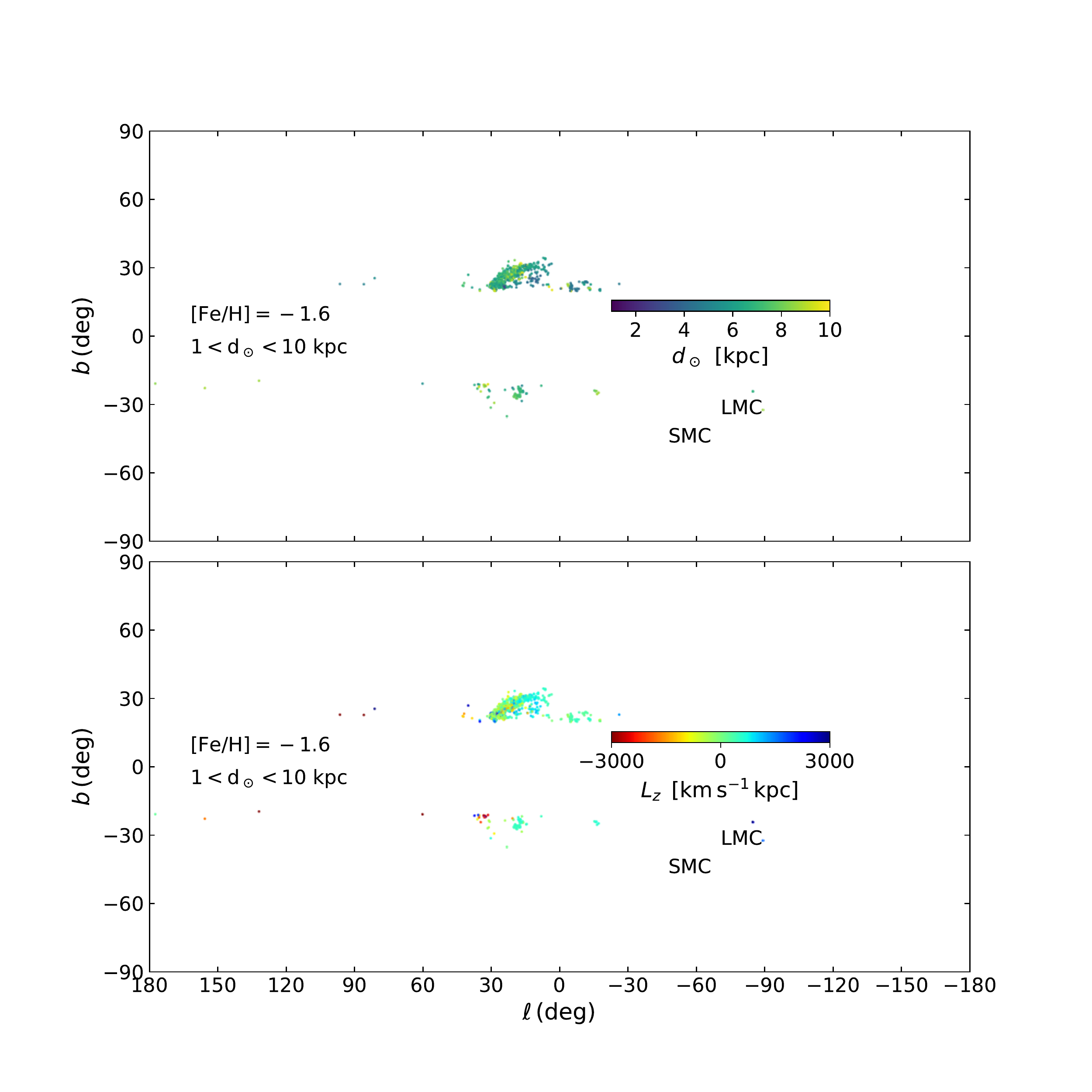}
}
}
\end{center}
\caption{As Figure~\ref{fig:FeH_m2.0_GUMS}, but for ${\rm [Fe/H]=-1.6}$ (cf. Figure~\ref{fig:FeH_m1.6}).}
\label{fig:FeH_m1.6_GUMS}
\end{figure*}

\begin{figure*}
\begin{center}
\vbox{
\hbox{
\includegraphics[angle=0, viewport= 55 23 600 589, clip, height=9cm]{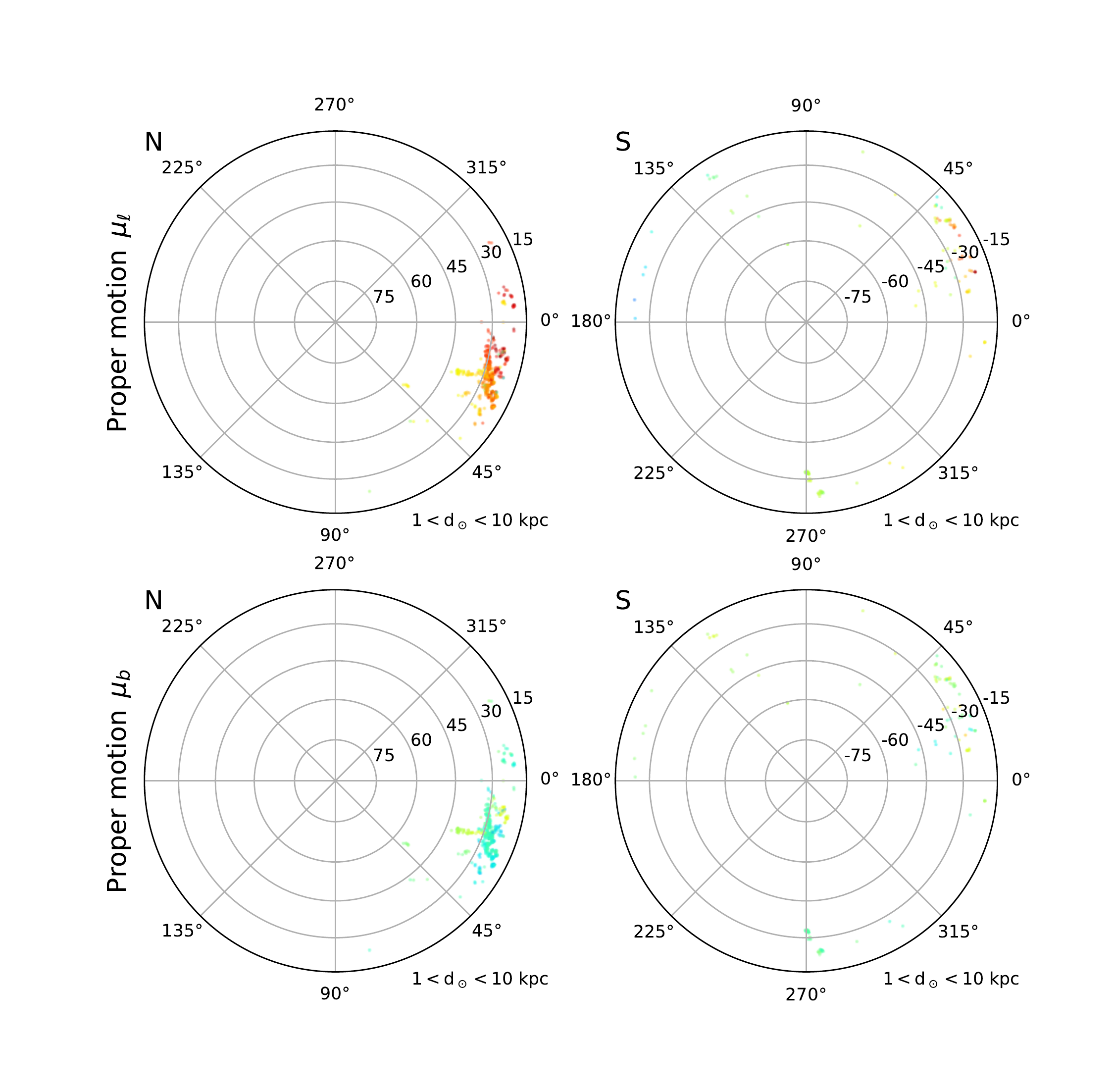}
\includegraphics[angle=0, viewport= 45 45 657 650, clip, height=9cm]{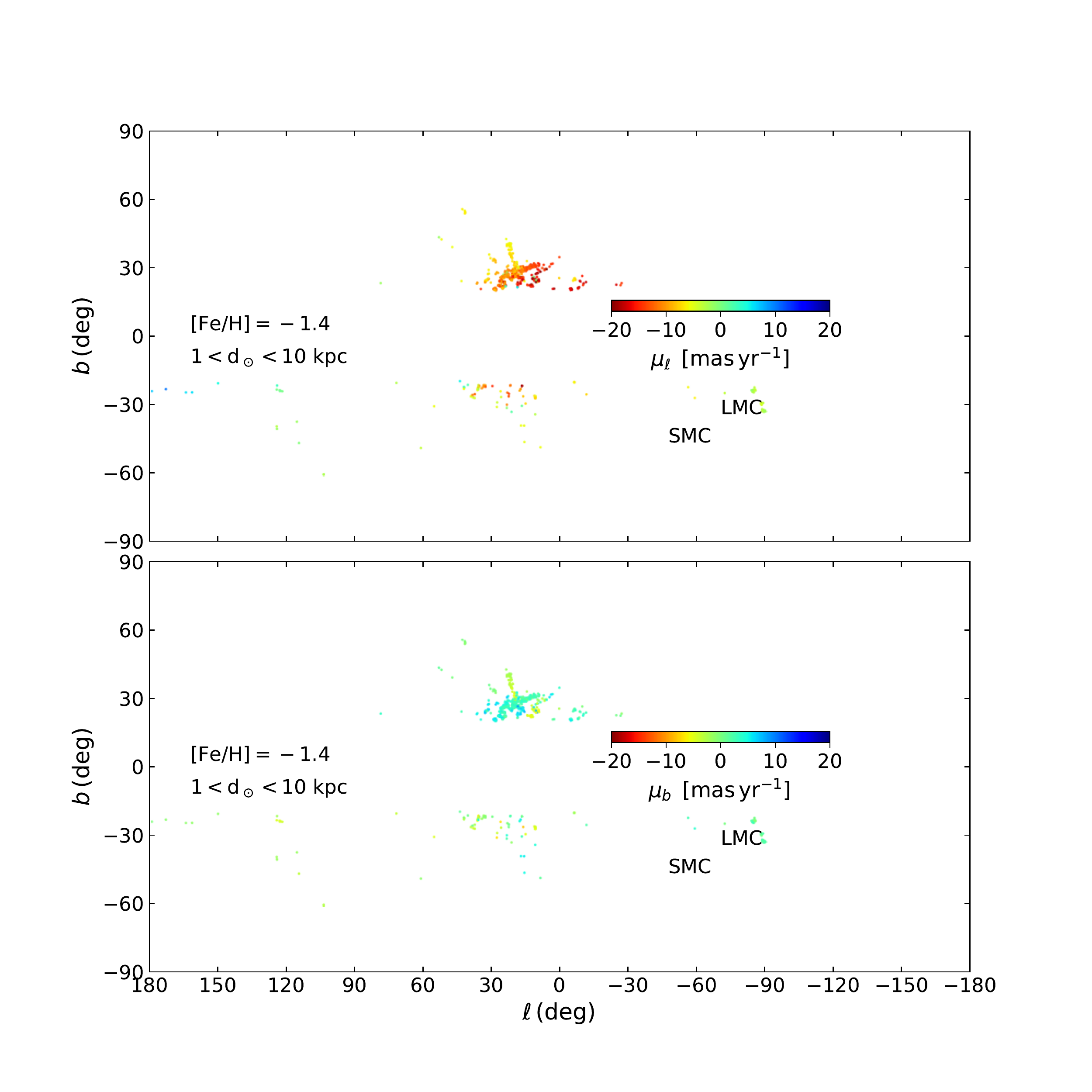}
}
\hbox{
\includegraphics[angle=0, viewport= 55 23 600 589, clip, height=9cm]{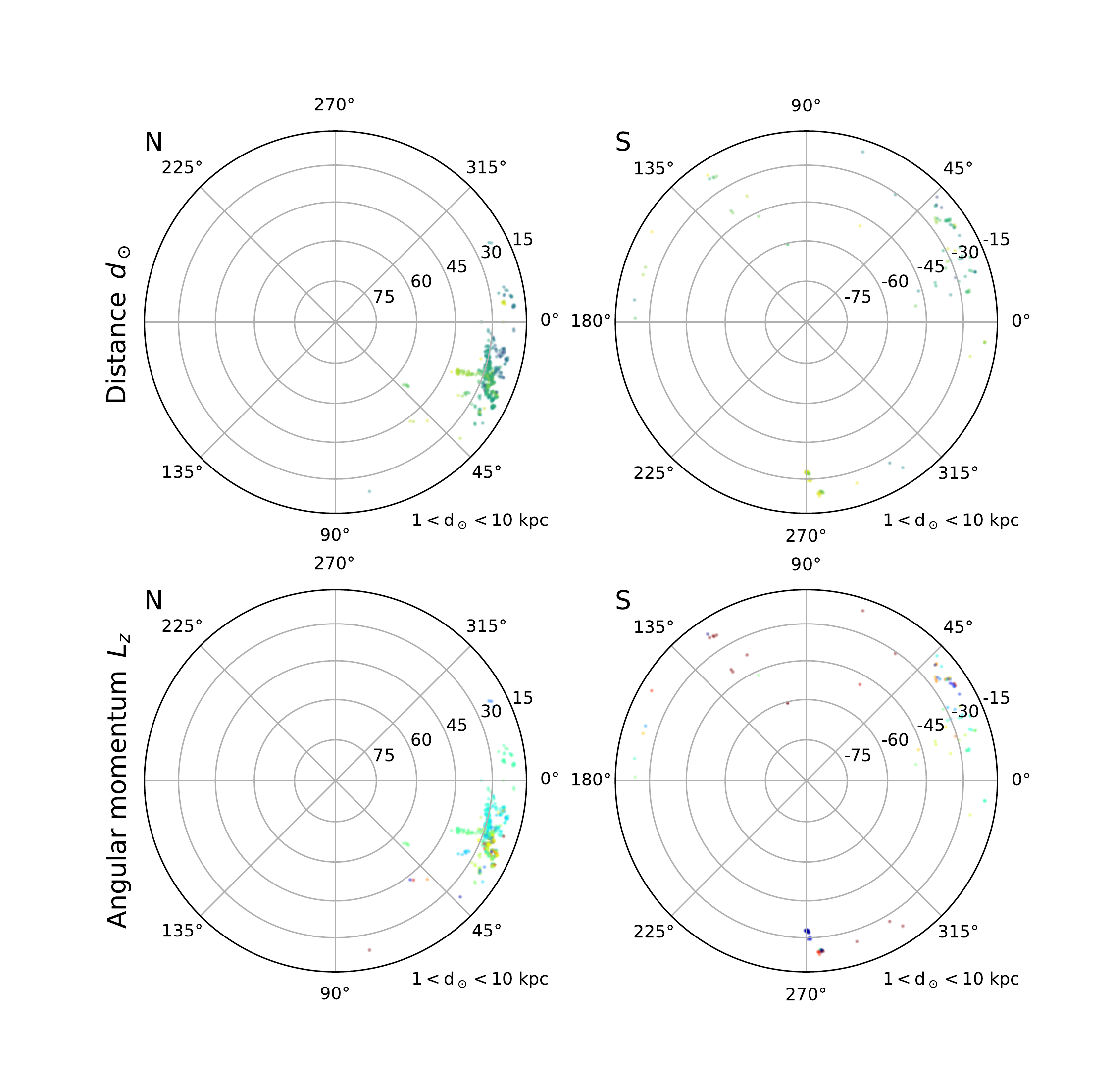}
\includegraphics[angle=0, viewport= 45 45 657 650, clip, height=9cm]{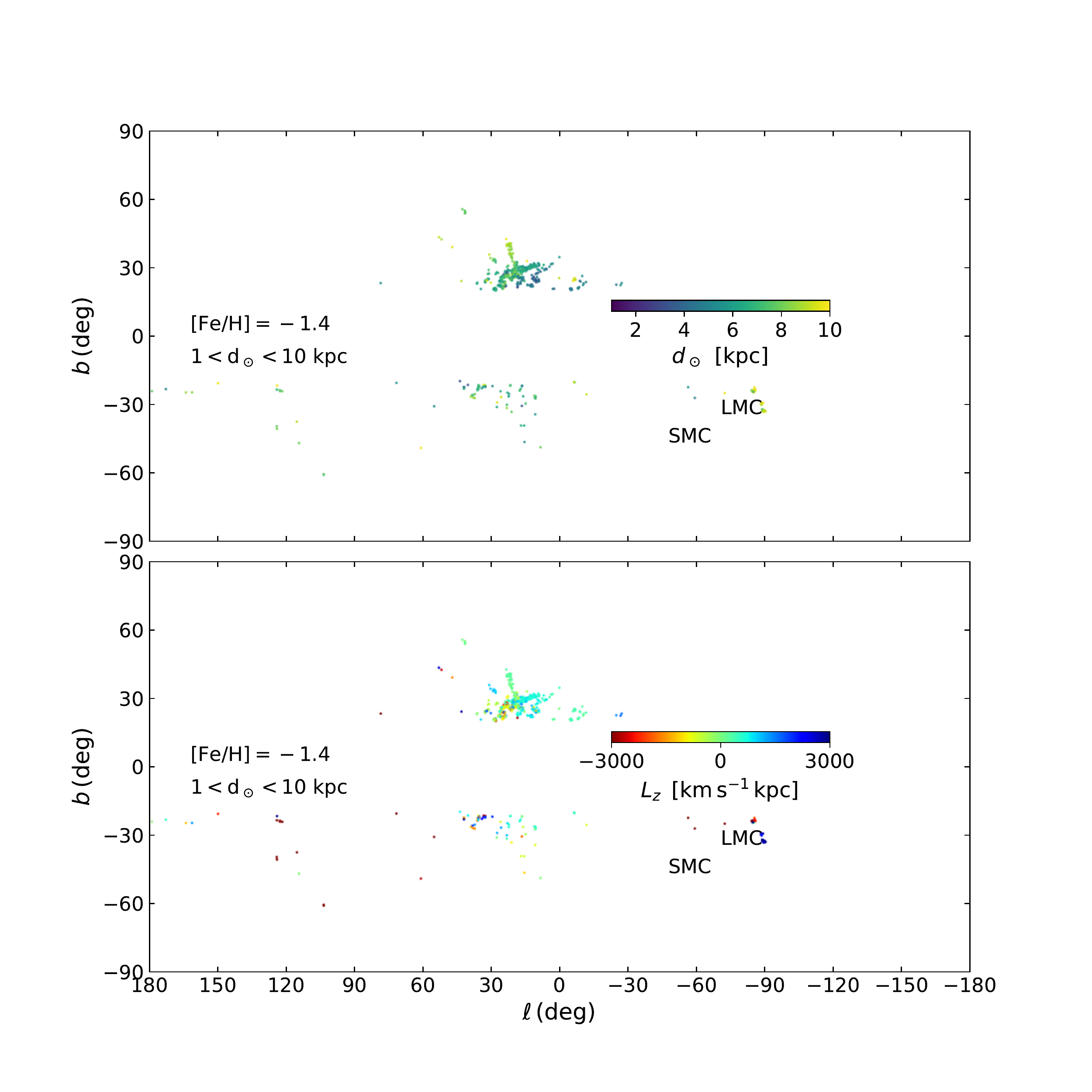}
}
}
\end{center}
\caption{As Figure~\ref{fig:FeH_m2.0_GUMS}, but for ${\rm [Fe/H]=-1.4}$  (cf. Figure~\ref{fig:FeH_m1.4}).}
\label{fig:FeH_m1.4_GUMS}
\end{figure*}

\end{document}